\pdfoutput=1
\newcommand*{\ATLASLATEXPATH}{latex/}
\documentclass[cernpreprint,texlive=2016,txfonts,UKenglish]{latex/atlasdoc}
\usepackage[backend=biber]{\ATLASLATEXPATH atlaspackage}
\usepackage{\ATLASLATEXPATH atlasbiblatex}

\usepackage{\ATLASLATEXPATH atlascontribute}

\usepackage{\ATLASLATEXPATH atlasphysics}
\graphicspath{{logos/}{figures/}}
\DeclareGraphicsExtensions{.pdf,.eps,.png,.jpg} % in order of preference

\usepackage{wxsec_2012-defs}

\AtlasTitle{Measurement of differential cross sections and $W^+/W^-$ cross-section ratios for $W$ boson production in association with jets at $\sqrt{s}=8$ TeV with the ATLAS detector}

\author{The ATLAS Collaboration}

 \PreprintIdNumber{CERN-EP-2017-213}

\AtlasJournalRef{JHEP 05 (2018) 077}
\AtlasDOI{10.1007/JHEP05(2018)077}

\AtlasAbstract{%
This paper presents a measurement of the $W$ boson production cross section and the $W^{+}/W^{-}$ cross-section ratio, both in association with jets, in proton--proton collisions at $\sqrt{s}=8$ TeV with the ATLAS experiment at the Large Hadron Collider. The measurement is performed in final states containing one electron and missing transverse momentum using data corresponding to an integrated luminosity of 20.2 fb$^{-1}$. Differential cross sections for events with at least one or two jets are presented for a range of observables, including jet transverse momenta and rapidities, the scalar sum of transverse momenta of the visible particles and the missing transverse momentum in the event, and the transverse momentum of the $W$ boson. For a subset of the observables, the differential cross sections of positively and negatively charged $W$ bosons are measured separately. In the cross-section ratio of $W^{+}/W^{-}$ the dominant systematic uncertainties cancel out, improving the measurement precision by up to a factor of nine. The observables and ratios selected for this paper provide valuable input for the up quark, down quark, and gluon parton distribution functions of the proton.
}

\hypersetup{pdftitle={ATLAS document},pdfauthor={The ATLAS Collaboration}}

\begin{document}

\maketitle

\tableofcontents

% List of contributors - print here or after the Bibliography.
%\PrintAtlasContribute{0.30}
%\clearpage

%-------------------------------------------------------------------------------
\section{Introduction}
\label{sec:intro}
%-------------------------------------------------------------------------------

With the large samples of proton--proton collision data available from the Large Hadron Collider (LHC), the measurement of the production of a \Wboson boson in association with jets allows precise tests of perturbative quantum chromodynamics (pQCD). In recent years, numerous theoretical advances have been made including calculations for up to five additional jets at next-to-leading-order (NLO)~\cite{Berger:2009ep,Berger:2010zx,Bern:2013gka} and calculations for one additional jet at next-to-next-to-leading-order (NNLO)~\cite{WjetNNLO,ZjetNNLO}, as well as merging approaches for NLO predictions of different jet multiplicities \cite{MiNLO, FxFx, FxFx2} and new parton shower approaches \cite{Dire,Vincia}. 
The theoretical predictions have undergone rigorous scrutiny using data from the ATLAS, CMS and LHCb experiments~\cite{STDM-2012-24,STDM-2011-08, CMSWjets2015, CMS-SMP-14-023, CMS-SMP-12-023, WjetsLCHb} with proton--proton collisions at the LHC and from the CDF and D\O\ experiments with proton--antiproton collisions at the Tevatron~\cite{WjetsCDF1,WjetsCDF2,WjetsD01,WjetsD02}. These results comprise a wide range of measurements of differential cross sections of observables, which are reconstructed from jets and leptonic decay products of the \Wboson boson. Detailed measurements of specific processes such as electroweak \Wboson boson production~\cite{STDM-2014-11}, small-angle emission of a \Wboson\ boson radiating from an energetic jet~\cite{STDM-2015-16} and production in association with heavy-flavour quarks~\cite{STDM-2012-14,STDM-2012-11,CMS-SMP-14-020,CMS-SMP-12-002,LHCbTtbarWbWc,LHCbWbc} complement these results. All of the studies mentioned here focus on jet production over a range of energy scales and attempt to probe pQCD to the statistical limits of the available data.

Using data corresponding to an integrated luminosity of \intLumi at $\sqrt{s}=\SI{8}{\tev}$, this paper presents the results for \Wjets production in final states containing one electron and missing transverse momentum, focusing on events with one or two additional jets. The data are measured for \Wboson production as well as for \Wplus and \Wminus production and the cross-section ratio of \WpmRatio\ as a function of the number of jets (\njs). For events with at least one jet, the differential cross sections are shown as a function of the scalar sum of the transverse momenta of electron, neutrino and jets (\HT), the transverse momentum (\pt) of the \Wboson\ boson, and the \pt\ and rapidity of the most energetic jet (leading jet). These observables are sensitive to higher-order terms in the prediction as well as the parton distribution functions (PDFs). For events with at least two jets, the differential cross sections are shown for \Wboson boson production and include distributions as a function of the \pt and rapidity of the second leading jet, the dijet angular separation, and the dijet invariant mass. These observables are sensitive to hard parton radiation at large angles and different matrix-element/parton-shower matching schemes.

The results for \Wjets production presented here are a useful test of jet production with energetic jets as well as jets with large rapidities. As in a previous ATLAS measurement using data at $\sqrt{s}=\SI{7}{\tev}$~\cite{STDM-2012-24}, the systematic uncertainties are larger than the statistical uncertainty of the data.
The new measurements are based on an independent dataset, at a higher centre-of-mass energy and with larger integrated luminosity. The analysis has improved event selections to reduce backgrounds from top quark production -- an important improvement since the increase in cross section with centre-of-mass energy is greater for top quark production than for \Wboson boson production. Several new sets of predictions and new measurements of the \pt of the \Wboson boson in addition to other observables provide additional information about pQCD.

In the ratio of \Wplus to \Wminus production, many of the experimental and theoretical uncertainties cancel out, making it a more precise test of the theoretical predictions. In addition, differential cross section measurements of \Wplus and \Wminus production and their ratio are sensitive to the PDFs for up and down quarks. 
The measurement of separate \Wplus and \Wminus cross sections as well as the \WpmRatio cross section ratio is new compared to the previous ATLAS measurement using data at \SI{7}{\tev}~\cite{STDM-2012-24}.
In previous measurements of \Wboson production for inclusive jet multiplicities~\cite{STDM-2010-16}, the \Wplus and \Wminus asymmetry probes the momentum fraction of the parton, $x$, in the range of $10^{-3}\lesssim x \lesssim 10^{-1}$. For events with at least one jet, a charge ratio measurement is sensitive to higher values of $x$, potentially accessing $x\sim0.1\mbox{--} 0.3 $~\cite{PDFconstrWratioTheoPaper}. 
The valence quark PDFs in this range are currently best constrained from fixed-target deep-inelastic scattering (DIS) experiments and the Tevatron \Wpm asymmetry measurements (see the discussion in Ref.~\cite{CT14PDFset}). The DIS measurements include effects from the nuclear target that require model-dependent corrections to obtain nucleon PDFs and the Tevatron results show tension between the different experiments as well as with the DIS results. 
It is therefore interesting to include new data, such as the measurement of the \Wboson boson cross sections and \Wplus and \Wminus cross-section ratios presented here, in PDF fits to improve the precision of valence quark and gluon PDFs at high $x$.

%-------------------------------------------------------------------------------
\section{ATLAS detector}
\label{sec:detector}
%-------------------------------------------------------------------------------

\newcommand{\AtlasCoordFootnote}{%
ATLAS uses a right-handed coordinate system with its origin at the nominal interaction point (IP)
in the centre of the detector and the $z$-axis along the beam pipe.
The $x$-axis points from the IP to the centre of the LHC ring,
and the $y$-axis points upwards.
Cylindrical coordinates $(r,\phi)$ are used in the transverse plane, 
$\phi$ being the azimuthal angle around the $z$-axis.
The pseudorapidity is defined in terms of the polar angle $\theta$ as $\eta = -\ln \tan(\theta/2)$. The rapidity, $y$, is defined as $\frac{1}{2}\ln[(E+p_z)/(E-p_z)]$, where $E$ denotes the energy of the jet and $p_z$ the momentum component of the jet along the beam direction.
Angular distance is measured in units of $\Delta R \equiv \sqrt{(\Delta\eta)^{2} + (\Delta\phi)^{2}}$.}

The ATLAS experiment~\cite{PERF-2007-01} at the LHC is a multi-purpose particle detector
with a forward-backward symmetric cylindrical geometry and a near $4\pi$ coverage in 
solid angle.\footnote{\AtlasCoordFootnote}
It consists of an inner tracking detector surrounded by a thin superconducting solenoid, providing a \SI{2}{\tesla} axial magnetic field, electromagnetic and hadronic calorimeters, and a muon spectrometer.
The inner detector covers the pseudorapidity range $|\eta| < 2.5$.
It consists of silicon pixel, silicon microstrip, and transition radiation tracking detectors.
Lead/liquid-argon (LAr) sampling calorimeters provide electromagnetic (EM) energy measurements
with high granularity.
A hadron (steel/scintillator-tile) calorimeter covers the central pseudorapidity range ($|\eta| < 1.7$).
The endcap and forward regions are instrumented with LAr calorimeters
for EM and hadronic energy measurements up to $|\eta| = 4.9$.
The muon spectrometer surrounds the calorimeters and includes a system of precision tracking chambers and fast detectors for triggering.
Three large air-core toroidal superconducting magnets, each with eight coils, provide the magnetic field for the muon system. 
In 2012, a three-level trigger system was used to select events.
The first-level trigger was implemented in hardware and used a subset of the detector information to reduce the accepted event rate to at most \SI{75}{\kilo\hertz}.
This was followed by two software-based trigger levels that
together reduced the accepted event rate to \SI{400}{\hertz} on average
depending on the data-taking conditions.

% This contains the introduction and the ATLAS detector description

%-------------------------------------------------------------------------------
\section{Simulated event samples}
\label{sec:MCsamples}
%-------------------------------------------------------------------------------

Simulated event samples are used for most of the background estimates, for the correction of the signal yield for detector effects and in comparison to the measured cross sections. The ATLAS detector simulation~\cite{SOFT-2010-01} is performed using GEANT4~\cite{Geant} and the simulated events are reconstructed and analysed using the same analysis chains as for data. Additional predictions that are only compared to the final measurements are described in \secref{theory}.

Samples of \Wen and \Zee events with associated jets were generated with \ALPGEN\ v2.14 \cite{Alpgen} and with \SHERPA v1.4.1 \cite{Sherpa1,Sherpa2}. The \ALPGEN\ event generator was also used to simulate \Wtaunu and \Ztautau production. For the \ALPGEN samples, events were produced with up to five additional partons in the final state from the matrix element. \PYTHIA v6.426 \cite{Pythia6} was used for the parton showering, hadronisation and underlying event, based on the \Perugia 2011C set of tuned parameters (tune) \cite{Perugia}, where the parton shower uses a dipole shower with a \pt-ordered evolution variable. For electromagnetic final-state radiation and the decay of $\tau$-leptons, \textsc{Photos} \cite{Photos} and \textsc{Tauola} \cite{Tauola} were used, respectively. Double counting of parton emissions between the \ALPGEN matrix element and the \PYTHIA parton shower calculations was removed through the \textsc{MLM} matching scheme \cite{Alpgen}. The proton structure is described by the \textsc{CTEQ6L1} PDF set \cite{CTEQ6L1}. The \ALPGEN samples include a matrix element calculation of \Wboson boson production in association with massive heavy-flavour partons, $\W+c$, $\W+c\bar{c}$ and $\W+b\bar{b}$ in addition to the light-flavour jet production. Overlap between heavy-flavour quarks originating from the matrix element and those originating from the parton shower was removed.
For the \SHERPA samples, events were produced with up to four additional partons in the final state from the matrix element and include a model of the parton shower, the hadronisation and the underlying event. The ME+PS@LO prescription~\cite{MEatNLO} is used to combine different parton multiplicities from the matrix element and the parton shower. The \SHERPA\ event generator uses the \textsc{CKKW} matching scheme \cite{CKKWmatching} and its own model of electromagnetic final-state radiation based on the Yennie--Frautschi--Suura method \cite{YFSmethod}. Massive $c$- and $b$-quarks are also included and the PDF set used is \textsc{CT10} \cite{CT10PDFset}.

The \ALPGENPYTHIA6 samples for \Wjets production provide the best description of data and are used as the main signal prediction throughout this measurement. The \SHERPA samples supply an alternative prediction and are used to estimate some of the systematic uncertainties.

Top quark pair production (\ttbar) was simulated with the \POWHEG-\textsc{Box} r2129 \cite{Powheg} event generator (referred to here as \POWHEG) interfaced to \PYTHIA v6.426 using the \Perugia 2011C tune and the \textsc{CT10} PDF set. The $h_\text{damp}$ parameter, which effectively regulates high-\pt emission in \POWHEG was set to the top quark mass of \SI{172.5}{\gev}. Single top quark production in the $s$-, $t$- and $Wt$- channels was modelled by \POWHEG and showered with \PYTHIA v6.426 (v6.427 for the $t$-channel) using the \Perugia 2011C tune. The PDF set is \textsc{CT10} (with a fixed 4-flavour scheme for $t$-channel production). Diboson processes ($WW$, $WZ$ and $ZZ$) were simulated using \HERWIG v6.520.2 \cite{Herwig} with the \textsc{AUET2} tune \cite{ATL-PHYS-PUB-2011-008} and the \textsc{CTEQ6L1} PDF set.

All simulated samples are normalised using their respective inclusive cross sections at higher orders in pQCD. The \Wboson and \Zboson predictions are scaled to the NNLO calculation obtained with \textsc{Dynnlo} v1.5 \cite{DYNNLO1,DYNNLO2} and the \textsc{MSTW2008} PDF set \cite{MSTW2008} (requiring $m_{\ell\ell}>\SI{60}{\gev}$ in case of \Zboson production). The production of top quarks is normalised using the prediction at NNLO+NNLL precision from the Top++2.0 program for \ttbar \cite{ttbarXsec1, ttbarXsec2, ttbarXsec3, ttbarXsec4, ttbarXsec5, ttbarXsec6, ttbarXsec7}, to the calculations in \refscite{singTopSChanXsec, singTopTChanXsec, singTopWtChanXsec} for single top quarks, and for diboson production to the NLO calculations in \refcite{DibosonNLO}.

The simulated events were overlaid with additional proton--proton interactions (\pileup) in the same and neighbouring crossings of proton bunches. These were generated with \PYTHIA v8.160 \cite{Pythia8} with the average number of interactions per bunch crossing matched to that measured in data. To achieve better agreement with data, the efficiencies for the electron triggering, reconstruction, identification, and isolation, as well as the efficiencies for the tagging or mis-tagging of heavy- and light-flavour jets, and the simulated vertex position were corrected in the simulated events.

%-------------------------------------------------------------------------------
\section{Data selection and event analysis}
\label{sec:selec}
%-------------------------------------------------------------------------------

The data used in this analysis were recorded during the 2012 proton--proton collision run at a centre-of-mass energy of \SI{8}{\TeV}. Crossings of proton bunches occurred every \SI{50}{ns} and the collisions achieved a peak instantaneous luminosity of $7.7\times 10^{33}\text{ cm}^{-2}\text{ s}^{-1}$. The mean number of simultaneous inelastic proton--proton interactions was $\avgmu=20.7$. After the application of data-quality requirements, the total integrated luminosity is \intLumi with an uncertainty of 1.9\% \cite{DAPR-2013-01}.

Events are selected for analysis by requiring that they satisfy a set of single-electron trigger criteria for an isolated electron with a transverse momentum above \SI{24}{\gev} or an electron with transverse momentum greater than \SI{60}{\gev}. Within this trigger algorithm the isolation momentum is defined as the sum of the transverse momenta of reconstructed charged-particle tracks with $\pt>\SI{1}{\gev}$ in a cone of size $\Delta R < 0.2$ around the electron (excluding the track of the electron). An electron trigger candidate is considered to be isolated if the isolation momentum is less than 10\% of the electron's transverse momentum. The threshold of the lower-\pt trigger is sufficiently low to ensure that electrons reconstructed with $\pt>\SI{25}{\gev}$ by the offline algorithms are selected with close to their maximum efficiency of about 96\% for central electrons. 
The higher-\pt trigger compensates for inefficiencies due to the isolation criteria applied. 

Events must have at least one reconstructed vertex with at least three associated tracks, each with a \pt greater than \SI{400}{\mev}. The vertex with the largest $\sum\pt^2$ of associated tracks is considered to be the primary vertex.

%-------------------------------------------------------------------------------
\subsection{Electron reconstruction and identification}
\label{sec:selec-el}
%-------------------------------------------------------------------------------

Electrons are reconstructed from energy clusters in the EM calorimeter that are matched to a track reconstructed in the inner detector. The electron is required to have $\pt>\SI{25}{\gev}$ and $\abseta<2.47$ (excluding the transition region between barrel and endcap calorimeters of $1.37<\abseta<1.52$) and match the online electron, which passed the trigger criteria. Each electron must satisfy a set of identification criteria in order to suppress misidentification of photons or jets. Electrons must pass the \emph{tight} selection, following the definition provided in \refcite{PERF-2016-01}. This includes requirements on the shower shape in the electromagnetic calorimeter, the leakage of the shower into the hadronic calorimeter, the number of hits measured along the track in the inner detector, the amount of transition radiation in the transition radiation tracker, and the quality of cluster--track matching as well as criteria to ensure that the reconstructed electron does not originate from a converted photon. A gaussian sum filter track refitting algorithm is used to improve the estimated electron track parameters. The electron is required to originate from the primary vertex by using the following criteria related to the electron track. The transverse impact parameter, $d_0$, must be smaller than five times its uncertainty ($|d_0|/\sigma_{d_0}<5$) and $|z_0\cdot \sin{\theta}|$ must be less than $\SI{0.5}{mm}$, where $z_0$ is the longitudinal impact parameter and $\theta$ is the polar angle of the electron with respect to the beam direction. 

In order to further suppress background from misidentified objects such as jets, the electron is required to be isolated using tracking-based and calorimeter-based criteria. The sum of the transverse momenta of tracks with $\pt>\SI{400}{\mev}$, excluding the electron track, in a radius of $\Delta R=0.3$ around the electron must be smaller than 7\% of the electron's \pt. Furthermore, the sum of transverse energies of topological clusters~\cite{PERF-2014-07} lying within a radius of $\Delta R<0.3$ around the electron centre and excluding the core area, must be smaller than 14\% of the electron's \pt. The calorimeter-based isolation is corrected for two effects: soft energy deposits in the isolation cone due to \pileup, using an ambient energy density approach~\cite{STDM-2011-05}, and for high-energy electrons, the energy leakage of the electron's energy from the core into the surrounding isolation cone.

%-------------------------------------------------------------------------------
\subsection{Jet selection}
\label{sec:selec-jets}
%-------------------------------------------------------------------------------

Jets are reconstructed using the \antikt algorithm \cite{AntiKtAlgo,FastJet} with a radius parameter $R=0.4$ and topological clusters of energy depositions in the calorimeter as input. The topological clusters are calibrated with the local cluster weighting method~\cite{LocalHadronic} to account for the hadronic and electromagnetic activity inside the clusters. Jets are then calibrated to the hadronic jet energy scale (JES), by applying \pt- and $\eta$-dependent factors that are determined from a combination of simulated events and \insitu methods~\cite{ATLAS-CONF-2015-037,ATLAS-CONF-2015-017,ATLAS-CONF-2015-057}.
These factors include corrections for inactive material in the detector, out-of-cone effects, \pileup contributions estimated using a jet-area-based approach~\cite{PERF-2014-03}, as well as a global sequential correction~\cite{ATLAS-CONF-2015-002}. The latter corrects for differences between quark- and gluon-initiated jets and the punch-through of a jet into the muon system. Events with jets arising from detector noise or non-collision effects~\cite{DAPR-2012-01} are rejected.

Jets are required to have $\pt>\SI{30}{\gev}$ and a rapidity of $\absrap<4.4$. Jets from additional proton--proton interactions are suppressed by requiring that more than 50\% of the total summed scalar \pt of the tracks associated with the jet must originate from tracks that are associated with the primary vertex~\cite{ATLAS-CONF-2014-018}. This requirement is applied to jets that are within the acceptance of the tracking detectors, $\abseta<2.4$, and have a \pt lower than \SI{50}{\gev}. Less than 5\% of non-pile-up jets are misidentified by this criterion. To avoid double counting with the selected electron, jets within $\Delta R=0.2$ of the electron are removed.

Jets containing $b$-hadrons are identified using a neural-network-based algorithm (MV1) \cite{PERF-2012-04}, which exploits information from the track impact parameters, secondary vertex location and decay topology. The operating point used for this analysis corresponds to an overall 60\% efficiency for heavy-flavour jets in \ttbar events and a less than 2\% mis-tag rate for light-flavour jets in dijet events. The $b$-tagged jets must have $\pt>\SI{20}{\gev}$ and $\abseta<2.5$.

%-------------------------------------------------------------------------------
\subsection{Event selection}
\label{sec:selec-W}
%-------------------------------------------------------------------------------

Events must contain one electron satisfying the selection criteria specified above. If the event contains a second electron that satisfies the \emph{medium} identification criteria and has $\pt>\SI{20}{\gev}$ and $\abseta<2.47$ (excluding $1.37<\abseta<1.52$), the event is rejected. This reduces the contribution from \Zboson boson production. To remove events where a jet is near the electron, the selected electron must be separated from any jet by $\Delta R>0.4$, otherwise the event is not considered. To suppress the background from \ttbar events, events with at least one $b$-tagged jet are also rejected. The application of a $b$-tagged jet veto reduces the \ttbar background for events with three or more associated jets by more than a factor of about two compared to the previous ATLAS measurement~\cite{STDM-2012-24}.

Events are required to have a missing transverse momentum (\met) and a transverse mass (\mt) consistent with the decay of a \Wboson boson. The missing transverse momentum~\cite{PERF-2014-04} is calculated as the negative vector sum of the transverse momenta of calibrated electrons, photons~\cite{PERF-2013-04}, hadronically decaying $\tau$-leptons~\cite{PERF-2014-06}, jets and muons~\cite{PERF-2014-05}, as well as additional low-momentum tracks which are associated with the primary vertex but are not associated with any other \met\ component. The transverse mass is defined as $\mt=\sqrt{2\pt^e\pt^\nu\left(1-\cos{\left(\phi^e-\phi^\nu\right)}\right)}$, where $\pt^\nu$ and $\phi^\nu$ of the neutrino correspond to that from the vector of the missing transverse energy (\metvec). Events in this analysis must have $\met>\SI{25}{\gev}$ and $\mt>\SI{40}{\gev}$. The set of selection criteria defines the \sr for this measurement.

The transverse momentum of the \Wboson\ boson is defined as the absolute value of the vectorial sum of the transverse momentum component of the selected electron and \metvec. The measurement of \Wplus and \Wminus production is performed by selecting events according to the charge of the electron.

%-------------------------------------------------------------------------------
\subsection{Background estimation}
\label{sec:bkg}
%-------------------------------------------------------------------------------

The major backgrounds to \Wboson boson production with decays into the electron plus neutrino final state are \Wtaunu, \Zee, \Ztautau, \ttbar (mainly $\ttbar\rightarrow\bbbar q\bar{q}' e\nu$), single-top-quark, diboson ($WW$, $WZ$, $ZZ$), and \mj events. Most of these background events contain an isolated, energetic electron in the final state. In the case of \Wtaunu\ and \Ztautau, an electron can be present in the final state via $\tau\rightarrow\nu_\tau \bar{\nu}_e e$. For the \mj background, an electron can be identified in the final state via three main modes: a light-flavour jet is misidentified as an electron, a bottom- or charm-hadron within a jet decays into an electron or an electron from a photon conversion passes the selection. In all cases, the event must also contain \met from either a mismeasurement of the deposited energy or from neutrinos in heavy-flavour decays.  

For events with less than four jets, the largest background is \mj events, whereas for five jets and above, \ttbar events dominate. An overview of the background contributions is given in \tabref{DetLevel_exclNjet}. For events with one (two)~jets, the multijet background constitutes 8\% (15\%) of the total number of events and all other backgrounds are less than 6\% (10\%). The use of tracks in the \met calculation to estimate the low-momentum contributions, instead of using soft energy deposits in the calorimeter, substantially suppresses the \mj background, in particular for one-jet events. At high jet multiplicities, the number of \Wjets events is less than the sum of all backgrounds, and for seven or more jets, the \ttbar background alone is larger than the signal. However, compared to previous ATLAS \Wjets publications, which did not include a veto on $b$-tagged jets, the \ttbar background is reduced from more than 60\% of events with five jets to less than 30\%. 

\begin{table}[tbp]
\robustify\bfseries
\centering
\caption{Signal and background contributions in the \sr for different jet multiplicities as percentages of the total number of predicted events, as well as the total numbers of predicted and observed events. The uncertainty in the total predicted number of events is statistical only. \label{tab:DetLevel_exclNjet}}
\resizebox{\textwidth}{!}{
\renewcommand{\arraystretch}{1.2}
\sisetup{retain-explicit-plus}
\begin{tabular}{lllllllll}
\toprule
\njs & 0 & 1 & 2 & 3 & 4 & 5 & 6 & 7 \\ 
\midrule
$W\rightarrow e\nu$ & 94 \% & 86 \% & 75 \% & 67 \% & 57 \% & 47 \% & 40 \% & 35 \% \\ 
\Mj & 3 \% & 8 \% & 15 \% & 16 \% & 16 \% & 16 \% & 14 \% & 14 \% \\ 
\ttbar & {<1} \% & {<1} \% & 1 \% & 6 \% & 16 \% & 27 \% & 36 \% & 43 \% \\ 
Single $t$ & {<1} \% & {<1} \% & 1 \% & 1 \% & 2 \% & 2 \% & 2 \% & 1 \% \\ 
$W\rightarrow\tau\nu$ & 2 \% & 2 \% & 2 \% & 2 \% & 2 \% & 1 \% & 1 \% & 1 \% \\ 
Diboson & {<1} \% & {<1} \% & 1 \% & 1 \% & 1 \% & 1 \% & {<1} \% & {<1} \% \\ 
$Z\rightarrow ee$ & {<1} \% & 3 \% & 5 \% & 6 \% & 6 \% & 6 \% & 5 \% & 5 \% \\ 
$Z\rightarrow\tau\tau$ & {<1} \% & {<1} \% & {<1} \% & {<1} \% & {<1} \% & {<1} \% & {<1} \% & {<1} \% \\ 
\midrule
Total predicted & \multicolumn{1}{r}{\num{54310000}} & \multicolumn{1}{r}{\num{7611700}} & \multicolumn{1}{r}{\num{2038000}} & \multicolumn{1}{r}{\num{478640}} & \multicolumn{1}{r}{\num{120190}} & \multicolumn{1}{r}{\num{30450}} & \multicolumn{1}{r}{7430} & \multicolumn{1}{r}{1735} \\ 
 & \multicolumn{1}{r}{$\pm$\num{22000}} & \multicolumn{1}{r}{$\pm 4000$} & \multicolumn{1}{r}{$\pm 1700$ }& \multicolumn{1}{r}{$\pm$\num{720}} & \multicolumn{1}{r}{$\pm$\num{320}} & \multicolumn{1}{r}{$\pm$\num{150}} & \multicolumn{1}{r}{$\pm$\num{63}} & \multicolumn{1}{r}{$\pm$\num{20}} \\ 
Data observed & \multicolumn{1}{r}{\num{56342232}} & \multicolumn{1}{r}{\num{7735501}} & \multicolumn{1}{r}{\num{2070776}} & \multicolumn{1}{r}{\num{486158}} & \multicolumn{1}{r}{\num{120943}} & \multicolumn{1}{r}{\num{29901}} & \multicolumn{1}{r}{7204} & \multicolumn{1}{r}{1641} \\ 
\bottomrule
\end{tabular}}
\end{table}

All backgrounds with the exception of the multijet background are estimated using simulations and are normalised to the integrated luminosity of the data using the cross sections as detailed in \secref{MCsamples}. For the \ttbar simulation, an additional normalisation factor of 1.086 is applied to account for an observed difference in the overall normalisation with respect to the data; this offset is also observed in other \ttbar measurements~\cite{TOPQ-2015-06}. 

The modeling of \ttbar production in the simulation is cross-checked using a \ttbar-enriched data sample, which is selected by requiring events with four or more jets, at least one $b$-tagged jet, and all other \sr selection criteria, and has a purity for \ttbar\ events of greater than 90\%. 
The background contributions are estimated using the same procedure as in the signal region. For the kinematic observables studied here, the \ttbar simulation agrees well with the data. The additional normalisation factor applied to the \ttbar simulation was determined with this data sample.

For the \mj background, a data-driven method is used to determine both the total number of events in the \sr and the shape of the background for each differential distribution. The number of \mj background events is estimated by fitting, for each jet multiplicity, the \met distribution in the data (without the \met\ requirement, but all other \sr requirements applied) to a sum of two templates: one for the \mj background and another which includes the signal and all other backgrounds. The normalisation of both templates is allowed to vary freely. The shape of the \mj template is obtained from data, while the second template is obtained from simulation. 
The multijet-enriched data sample used to extract the \mj template is acquired using a dedicated electron trigger, an inverted electron identification criterion, and an inverted isolation criterion. The electron trigger is equivalent to the one used for the \sr, but does not contain an isolation requirement. The inverted identification requires that the electron passes the \emph{medium} criteria but fails the \emph{tight} criteria, and the inverted isolation that the sum of the \pt of tracks in a cone of $\Delta R=0.3$ around the electron, excluding the electron track, is larger than 7\% of the electron's \pt.
To increase the number of events in the multijet-enriched sample the electron impact parameter criteria are not applied. The \mj-enriched data sample is statistically independent from the \sr and any contribution from the signal or other backgrounds to this sample is subtracted using simulation.

The \met\ fit is performed in the range of \SI{15}{\gev} to \SI{75}{\gev} for each jet multiplicity and separately for the \Wboson, \Wplus and \Wminus production selections. The fit results are used to determine the number of \mj events in the \sr for each selection. For events with six or more jets (five or more jets for the \Wminus event selection) where the statistical uncertainties in the \mj template are large, the \mj contribution is extracted from a fit of the \met distribution with these multiplicities combined. 
For the differential distributions, the shape of the \mj contribution is determined from the multijet-enriched data sample and scaled to the total number of \mj events as extracted from the fit. 

\begin{figure}[tbp]
    \centering
    \includegraphics[width=0.49\textwidth]{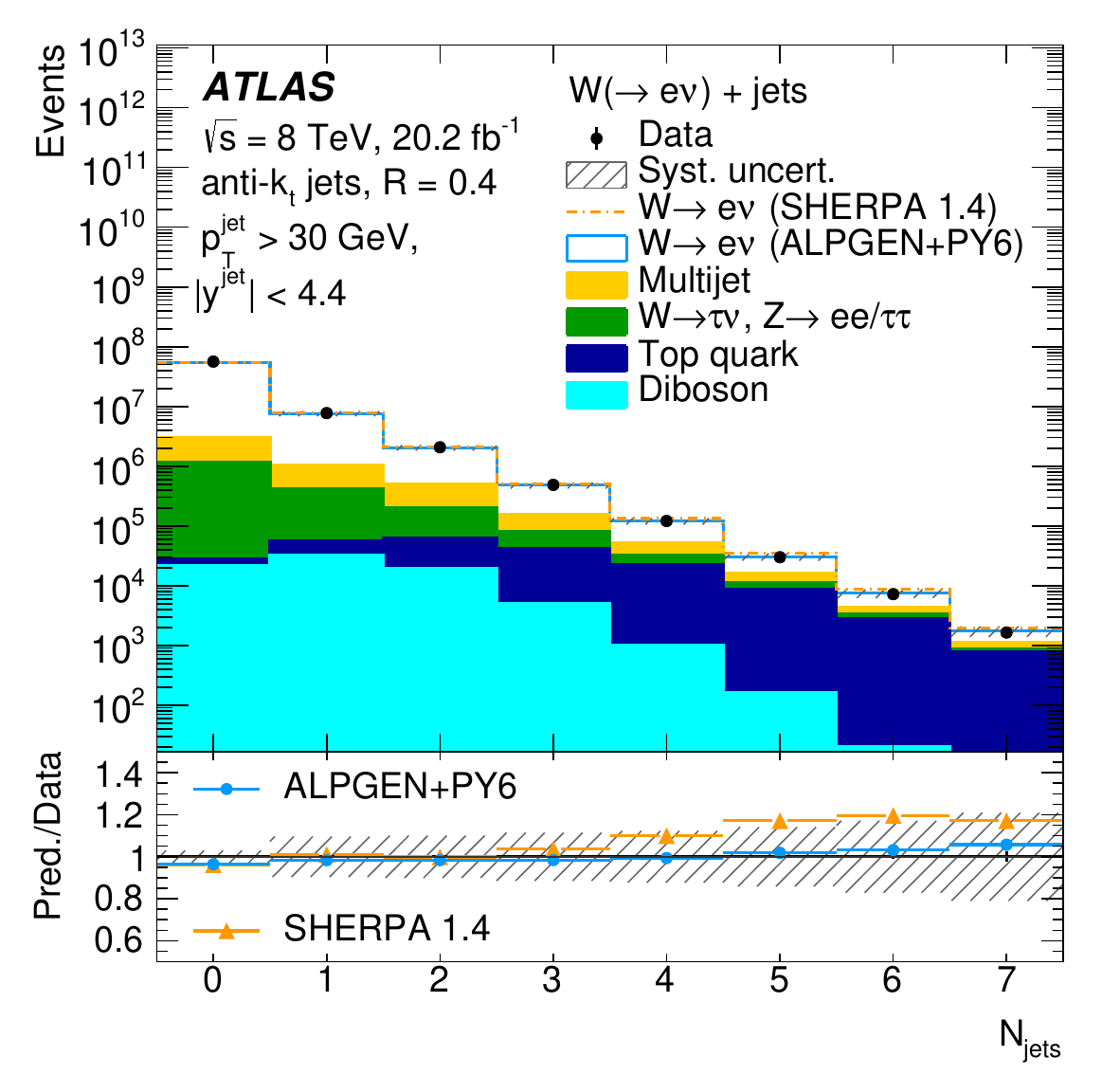}
    \includegraphics[width=0.49\textwidth]{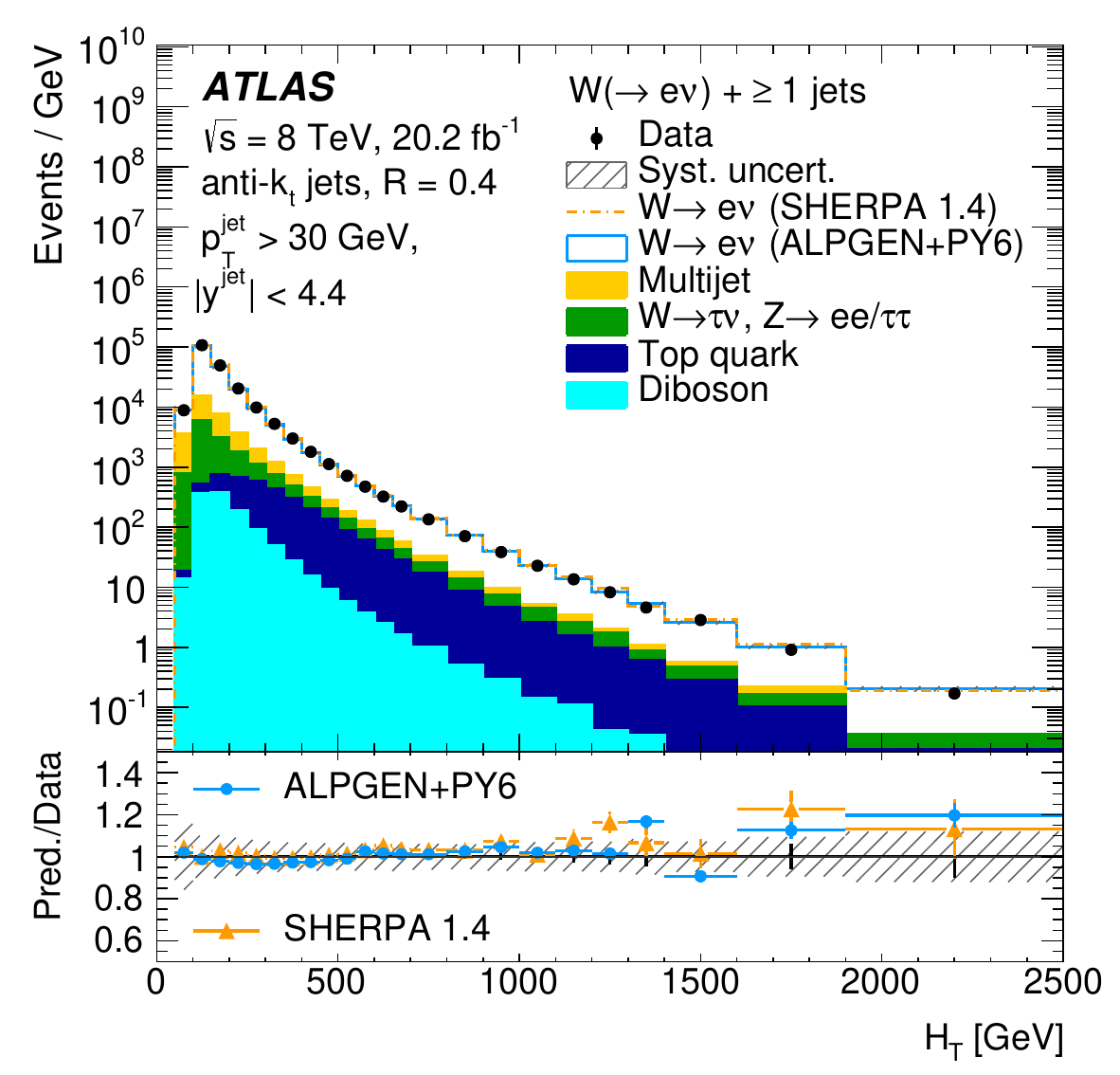}
    \newline
    \includegraphics[width=0.49\textwidth]{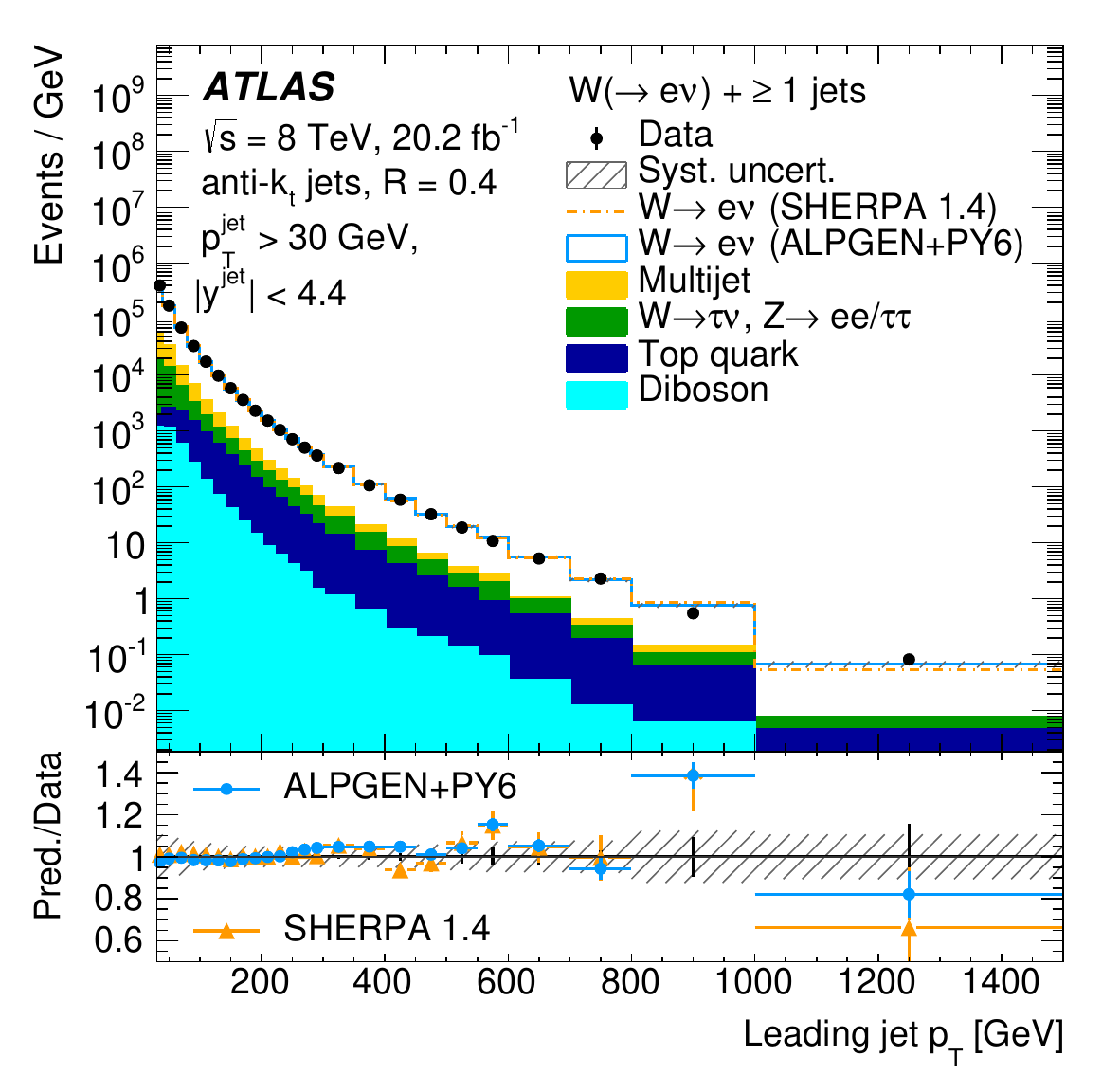}
    \includegraphics[width=0.49\textwidth]{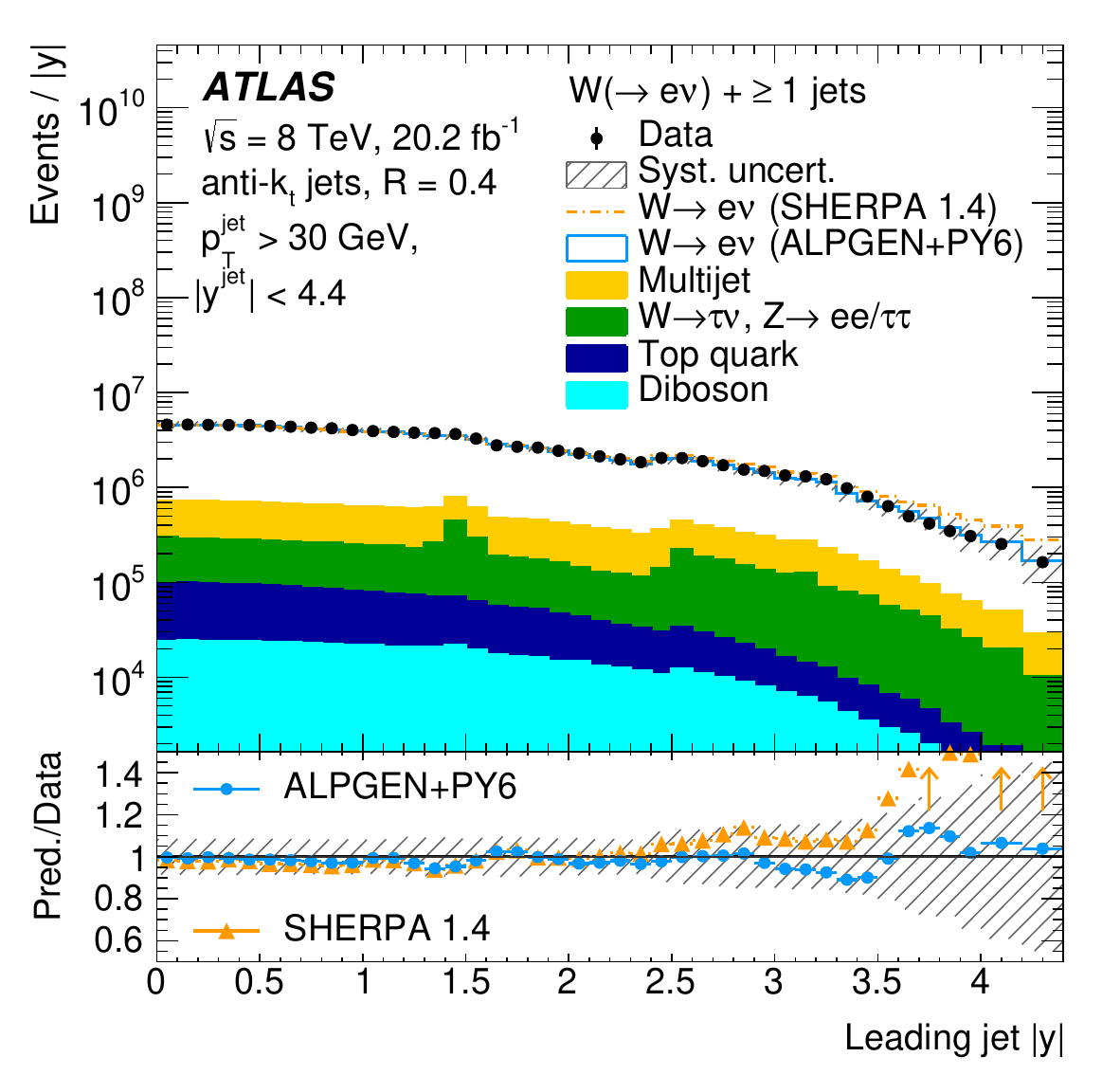}
    \caption{Distribution of events passing the \Wjets signal selection as a function of the exclusive jet multiplicity (upper left), the \HT (upper right), the leading jet's \pt (lower left), and the leading jet's rapidity (lower right). The lower panels display the ratio of the predictions for signal plus background to data using either \ALPGENPYTHIA6 (blue) or \SHERPA 1.4 (orange) as the signal simulation. The statistical uncertainty of the data is shown as black error bars and the total uncertainty in the prediction as the hatched band. The latter consists of the systematic uncertainties, including the uncertainty due to the luminosity, and the statistical uncertainties from the predictions and the data-driven \mj estimate.}
    \label{fig:detLevel}
\end{figure}

In \figref{detLevel}, the data are compared to the signal and background predictions as a function of the exclusive jet multiplicity, the \HT, and the transverse momentum and the rapidity of the leading jet. The data, in general, agree with the predictions within the experimental uncertainties.

% This contains the MC sample list, the event selection and the detector level results

%-------------------------------------------------------------------------------
\section{Correction for detector effects}
\label{sec:unfold}
%-------------------------------------------------------------------------------

The yield of \Wjets\ events is determined by subtracting the estimated background contributions from the event counts in data. Using simulated samples, the distributions are then corrected for detector effects to the fiducial phase space that is defined in \tabref{fidRegion}. Here, the electron definition is based on final-state electrons from the \Wboson boson decay and includes the contributions from photons, which are radiated within a $\Delta R=0.1$ cone around the electron direction (\emph{dressed} electron). The \met is determined from the transverse momentum of the neutrino from the \Wboson boson decay and is also used in the calculation of \mt.
Particle-level jets are obtained using an \akt algorithm with a radius parameter of $R=0.4$.  The jets are clustered using final-state particles (except muons and neutrinos) with a decay length of $c \tau > 10$~mm as input and the dressed electron is excluded as a jet. The jets are required to have $\pt>\SI{30}{\gev}$ and $\absrap<4.4$. If a jet is within $\Delta R=0.4$ from the selected electron, the event is not considered. 

\begin{table}[htbp]
\robustify\bfseries
\centering
\small{
\renewcommand{\arraystretch}{1.4}
\caption{Kinematic criteria defining the fiducial phase space for the \Wen final state in association with jets.\label{tab:fidRegion}}
\begin{tabular}{ll}
\toprule
 & Electron criteria \\
\midrule
Electron \pt & $\pt>\SI{25}{\gev}$ \\
Electron pseudorapidity & $\abseta<2.5$ \\
\midrule
 & \Wboson criteria \\
\midrule
Electron decay & Exactly one electron \\
Missing transverse momentum & $\met>\SI{25}{\gev}$ \\
Transverse mass & $\mt>\SI{40}{\gev}$ \\
\midrule
 & Jet criteria \\
\midrule
Jet \pt & $\pt>\SI{30}{\gev}$ \\
Jet rapidity & $\absrap<4.4$ \\
Jet--electron distance & $\Delta R(e,\text{jet})\geq0.4$ (otherwise event is removed) \\
\bottomrule
\end{tabular}}
\end{table}

The correction procedure uses the \ALPGENPYTHIA6 simulation of \Wjets production and corrects for selection efficiencies and resolution effects. Migrations between bins that occur during the reconstruction of events are corrected for using an iterative Bayesian unfolding method~\cite{BayesUnfold1,BayesUnfold2} with two iterations. In addition corrections for events that are outside the fiducial region but are reconstructed within the \sr\ or events that are not reconstructed due to detector inefficiencies are included. The correction procedure includes an extrapolation from the \sr, which has a veto on events with $b$-tagged jets, to the fiducial region, which does not. 
\Wboson boson production in association with $c$- and $b$-quarks is 18\% of the combined fiducial cross section for $\njs\geq1$. This reduces by only 2\% in the \sr from the $b$-tagged jet veto because, in events with one jet, contributions from $\Wboson+c$ production are larger and these contributions are only slightly affected by this veto. The extrapolation therefore has a small effect compared to other corrections, such as the one accounting for the electron identification efficiency.

For differential distributions, the unfolding is performed in two dimensions, one of which is always the jet multiplicity. In this way, migrations between jet multiplicity bins, which can be large, are considered. Migrations in \met\ compose a large part of the correction in and out of the fiducial region, in particular for zero-jet events, and are accounted for by the procedure. Other migrations, for example those in \mt\ are also included but are small. Differential cross sections for a given jet multiplicity, such as $\njs\geq1$, are obtained by summing over the contributing jet multiplicities in the two-dimensional result.

The \Wplus and \Wminus distributions are unfolded independently following the same procedure. The ratio of \Wplus to \Wminus cross sections is calculated from these unfolded distributions, taking correlations into account. All uncertainties of a statistical nature, such as the statistical uncertainty of the data, the statistical uncertainty of simulated samples used in the background estimate,  or the uncertainty from limited sample size of the signal simulation used in the unfolding are treated as uncorrelated between bins and between \Wplus and \Wminus production. All other systematic uncertainties are treated as fully correlated between bins and between the production of \Wplus and \Wminus bosons.

%-------------------------------------------------------------------------------
\section{Systematic uncertainties}
\label{sec:systs}
%-------------------------------------------------------------------------------

The dominant sources of systematic uncertainty in the cross-section measurement for events with at least one jet are the jet energy scale (JES) and the jet energy resolution (JER). The systematic uncertainties as a function of the number of jets in the \Wboson cross section and the \WpmRatio cross-section ratio measurements are summarised in \tabsref{sysXsec_inclNjet}{sysXsec_inclNjet_WpmRatio}. 

The JES systematic uncertainties are determined by a combination of data-driven \insitu techniques and simulation-based approaches \cite{ATLAS-CONF-2015-037,ATLAS-CONF-2015-017,ATLAS-CONF-2015-057,ATLAS-CONF-2015-002}. They are propagated through the analysis as 18 independent components and account for systematic uncertainties in the \insitu measurements, \pileup-corrections to the jet energies, jet punch-through to the muon system, effects due to the light quark or gluon origin of the jets, $b$-tagged jet energy calibration and other smaller effects. 
The uncertainty in the JES varies as a function of the jet \pt and $\eta$ and is approximately 3.5\% for central jets with $\pt > \SI{30}{\gev}$ and decreases to about 1\% for central jets with $\pt > 100$ \gev. For forward jets, the JES uncertainty is almost twice as large as for central jets, mainly due to the uncertainties in the jet-$\eta$-intercalibration~\cite{ATLAS-CONF-2015-017}. In the analysis, jet energies are shifted in simulated events by the size of the JES uncertainty component, and the event selection, including a recalculation of \met and \mt, is re-evaluated. The full analysis chain, which includes the background estimates and the unfolding, is repeated and the change in the cross section with respect to the nominal is taken as the systematic uncertainty. For a given source, the average of the up and down variations is taken as the symmetric uncertainty. 
The impact of the JES uncertainties on the cross section ranges from 8\% to 55\% for $\njs\geq1$ to $\njs\geq7$ but decreases for the \WpmRatio cross-section ratio to below 1\% and up to 17\% for $\njs\geq1$ to $\njs\geq6$. This method of propagating the systematic uncertainties is used for all other uncertainties except for uncertainties due the unfolding procedure itself. The total systematic uncertainty is the sum in quadrature of the individual uncertainties. 

The uncertainty of the JER is also determined through data-driven \insitu techniques and includes a dedicated estimate of effects from electronic noise in the calorimeter and pile-up~\cite{ATLAS-CONF-2015-037}. It is propagated through the analysis by smearing the energies of simulated jets, thereby degrading the jet resolution. For central jets, the JER uncertainty is small -- about 2\% for jets with a \pT of \SI{30}{\gev} -- but increases for jets in the forward region. In the \Wjets cross section, this translates to uncertainties of 9\% to 20\% for $\njs\geq1$ to $\njs\geq7$. In the \WpmRatio cross-section ratio, the impact of the JER uncertainty decreases to values of less than 1\% to 5\% for $\njs\geq1$ to $\njs\geq6$.

Additional experimental systematic uncertainties considered in this analysis include uncertainties in the $b$-tagged jet identification efficiencies~\cite{PERF-2012-04,ATLAS-CONF-2014-004,ATLAS-CONF-2014-046}, uncertainties due to the low-momentum tracks in the \met calculation \cite{PERF-2014-04}, and uncertainties in the electron energy scale, energy resolution and scale factors used to correct trigger, reconstruction, identification, and isolation efficiencies in the simulation \cite{PERF-2013-05,PERF-2016-01}. For the \Wplus and \Wminus cross sections and their ratio, the charge misidentification for electrons in the simulation is adjusted by randomly flipping the charge so that the overall misidentification rate matches that in the data. The uncertainty due to this correction is small. An uncertainty of 1.9\% \cite{DAPR-2013-01} in the integrated luminosity is applied to the signal predictions and all background estimates that are determined from simulation. The effect of the small relative uncertainty of the LHC proton beam energy~\cite{STDM-2012-20} is negligible and and is not included here.

The \mj background estimate is affected by uncertainties due to the choice of template and fit procedure. The uncertainty in the shape of the \mj template is estimated by varying separately both the inverted isolation criteria and the inverted identification of the electron used to select the multijet-enriched data sample. The influence of the signal template in the \met fit is determined by using the \SHERPA simulation instead of \ALPGENPYTHIA6 for the modelling of \Wjets production. The impact due to statistical uncertainties in the templates is evaluated by creating a thousand pseudo-data samples drawn from the templates and refitting the data with each. Uncertainties due to the fit procedure are estimated by varying the lower and upper bound of the fit range separately, as well as changing the binning used in the fit. The statistical uncertainty in the fit parameters is also included. The uncertainty in the \Wboson cross section from these sources ranges from less than 1\% to about 12\% for $\njs\geq0$ to $\njs\geq7$; the largest contributions to the uncertainty are due to the fit range variation, the modification of the inverted electron identification, the choice of \Wjets generator, and, at higher jet multiplicities, the statistical uncertainties. The uncertainty in the \WpmRatio cross-section ratio ranges from less than 1\% to 27\% for $\njs\geq0$ to $\njs\geq6$ and is larger than that of the \Wboson boson measurement due to statistical uncertainties from the fit as well as the inverted electron identification and the fit range uncertainties that do not fully cancel out in the ratio. 

Uncertainties from the background estimates that are derived using simulation include theoretical uncertainties in the cross section and the statistical uncertainty of the simulated samples. The theoretical uncertainties are evaluated for \ttbar and single top quark production by simultaneously varying the cross section by $\pm6.8\%$ \cite{ttbarXsec1, ttbarXsec2, ttbarXsec3, ttbarXsec4, ttbarXsec5, ttbarXsec6, ttbarXsec7, singTopSChanXsec, singTopTChanXsec, singTopWtChanXsec}, for diboson production ($WW$, $WZ$, $ZZ$) by simultaneously varying the cross section by $\pm7\%$~\cite{ATL-PHYS-PUB-2017-005} and for \Zboson production by varying the cross section by $\pm5\%$~\cite{ATL-PHYS-PUB-2017-006}. For the \ttbar estimate, the normalisation factor, as discussed in \secref{selec}, is also removed and the difference is taken as an uncertainty. Additional uncertainties in the modelling of the shape of the distributions are not considered. Backgrounds from single top quark, diboson and \Zboson boson production are small, and the impact of the cross-section uncertainties is minimal, therefore any modelling uncertainties are negligible. 
For the \ttbar background, the theoretical uncertainties only have a noticeable effect in the \njs\ distribution for events with 5--7 jets where \ttbar production is a significant contribution. The impact of \ttbar background modelling uncertainties is cross-checked by comparing to an alternative \ttbar prediction from \textsc{MC@NLO}+\HERWIG~\cite{MCatNLO}. The results from this prediction are well covered by other uncertainties, except for in events with $\njs\geq7$ where this prediction is known to have large differences from the data in the \ttbar-enriched data sample. The combined impact of the non-\mj background uncertainty sources ranges from less than $1\%$ to 22\% for $\njs\geq0$ to $\njs\geq7$ for the \Wboson cross section, and from $\ll1\%$ to 12\% for $\njs\geq0$ to $\njs\geq6$ in the \WpmRatio cross-section ratio. The dominant sources of uncertainty are those related to the \ttbar\ normalisation.

\begin{table}[tbp]
\robustify\bfseries
\centering
\caption{Relative systematic uncertainties in the measured \Wjets cross sections in percent as a function of the inclusive jet multiplicity. The uncertainty from $b$-tagging includes the uncertainties in the $b$-tagged jet identification and misidentification efficiencies as well as the impact of $\Wboson+c,c\bar{c},b\bar{b}$ cross sections in the extrapolation from the \sr to the fiducial region. \emph{Other backgrounds} summarises the impact of \Zboson and diboson cross sections as well as the statistical uncertainty in the background estimates. \emph{Other} combines uncertainties in the \pileup modelling and the impact of matching jets to the primary vertex. The luminosity uncertainty includes a global uncertainty of 1.9\%.}
\label{tab:sysXsec_inclNjet}
\small{
\renewcommand{\arraystretch}{1.4}
\sisetup{retain-explicit-plus}
\begin{tabular}{lSSSSSSSS}
\toprule
 & {Inclusive} & \multicolumn{1}{c}{$\geq\text{1 jet}$} & \multicolumn{1}{c}{$\geq\text{2 jets}$} & \multicolumn{1}{c}{$\geq\text{3 jets}$} & \multicolumn{1}{c}{$\geq\text{4 jets}$} & \multicolumn{1}{c}{$\geq\text{5 jets}$} & \multicolumn{1}{c}{$\geq\text{6 jets}$} & \multicolumn{1}{c}{$\geq\text{7 jets}$} \\
\midrule
Jet energy scale & 0.1 & 7.5 & 10 & 13 & 18 & 27 & 38 & 55 \\
Jet energy resolution & 0.5 & 8.8 & 9.9 & 12 & 14 & 15 & 18 & 22 \\
$b$-tagging & 0.1 & 0.5 & 1.5 & 3.8 & 8.3 & 15 & 23 & 33 \\
Electron & 1.1 & 1.4 & 1.4 & 1.5 & 1.8 & 2.2 & 2.6 & 3.3 \\
\met & 1.1 & 2.6 & 4.2 & 5.5 & 7.1 & 8.8 & 12 & 14 \\
\Mj background & 0.5 & 1.3 & 2.1 & 2.6 & 2.4 & 4.8 & 9.1 & 12 \\
Top quark background & <0.1 & 0.2 & 0.8 & 2.5 & 5.7 & 10 & 15 & 22 \\
Other backgrounds & <0.1 & 0.1 & 0.2 & 0.3 & 0.5 & 1.0 & 1.7 & 2.6 \\
Unfolding & 4.7 & 4.1 & 4.9 & 4.4 & 4.0 & 4.9 & 7.8 & 6.2 \\
Other & 0.3 & 0.8 & 1.0 & 2.1 & 4.6 & 8.7 & 14 & 21 \\
Luminosity & 1.9 & 2.1 & 2.3 & 2.6 & 3.1 & 3.9 & 4.8 & 6.0 \\
\midrule
Total systematic uncert. & 5.4 & 13 & 16 & 20 & 27 & 38 & 55 & 78 \\
\bottomrule
\end{tabular}}
\end{table}

\begin{table}[tbp]
\robustify\bfseries
\centering
\caption{Relative systematic uncertainties in the measured ($\Wplus+\text{jets}$)/($\Wminus+\text{jets}$) cross-section ratio in percent as a function of the inclusive jet multiplicity. The uncertainty from $b$-tagging includes the uncertainties in the $b$-tagged jet identification and misidentification efficiencies as well as the impact of $\Wboson+c,c\bar{c},b\bar{b}$ cross sections in the extrapolation from the \sr to the fiducial region. \emph{Other backgrounds} summarises the impact of \Zboson and diboson cross sections as well as the statistical uncertainty in the background estimates. \emph{Other} combines uncertainties in the \pileup modelling and the impact of matching jets to the primary vertex.}
\label{tab:sysXsec_inclNjet_WpmRatio}
\small{
\renewcommand{\arraystretch}{1.4}
\sisetup{retain-explicit-plus}
\begin{tabular}{lSSSSSSS}
\toprule
 & {Inclusive} & \multicolumn{1}{c}{$\geq\text{1 jet}$} & \multicolumn{1}{c}{$\geq\text{2 jets}$} & \multicolumn{1}{c}{$\geq\text{3 jets}$} & \multicolumn{1}{c}{$\geq\text{4 jets}$} & \multicolumn{1}{c}{$\geq\text{5 jets}$} & \multicolumn{1}{c}{$\geq\text{6 jets}$} \\
\midrule
Jet energy scale & <0.1 & 0.3 & 1.2 & 2.3 & 3.8 & 9.1 & 17 \\
Jet energy resolution & 0.1 & 0.7 & 1.6 & 2.5 & 2.6 & 3.0 & 4.6 \\
$b$-tagging & <0.1 & 0.2 & 0.5 & 1.5 & 4.2 & 9.4 & 17 \\
Electron & 0.1 & 0.1 & 0.1 & 0.1 & 0.4 & 0.9 & 1.5 \\
\met & 0.1 & 0.8 & 1.9 & 2.8 & 3.8 & 5.5 & 6.5 \\
\Mj background & 0.3 & 1.2 & 2.9 & 3.2 & 5.9 & 15 & 27 \\
Top quark background & <0.1 & 0.1 & 0.3 & 1.2 & 3.3 & 7.0 & 12 \\
Other backgrounds & <0.1 & 0.1 & 0.2 & 0.3 & 0.7 & 1.7 & 2.8 \\
Unfolding & 0.6 & 0.5 & 0.6 & 0.7 & 1.1 & 3.1 & 3.7 \\
Other & <0.1 & 0.1 & 0.3 & 0.9 & 2.4 & 6.4 & 12 \\
Luminosity & <0.1 & <0.1 & 0.1 & 0.2 & 0.5 & 1.1 & 1.8 \\
\midrule
Total systematic uncert. & 0.7 & 1.8 & 4.1 & 5.9 & 10 & 23 & 41 \\
\bottomrule
\end{tabular}}
\end{table}

In addition to the experimental uncertainties in the $b$-tagged jet identification efficiencies, a theoretical uncertainty in the cross section of \Wboson production in association with $c$- and $b$-quark jets is considered. This accounts for any mismodelling in the extrapolation from the \sr, which includes a veto of events with $b$-tagged jets, to the fiducial region, which has no such veto. The uncertainty in these cross sections is applied by scaling the $\Wboson+c$ contribution by a factor of 1.8 and the sum of the $\Wboson+c\bar{c}$ and $\Wboson+b\bar{b}$ contributions by a factor of 0.5. These factors are obtained by comparing the data to the signal and background predictions using a heavy flavour-enriched \Wjets\ data sample, which requires events with at least one $b$-tagged jet and one or two additional jets. The impact on the measured cross section is below 2\% for all jet multiplicities and in both the \Wboson cross section and the \WpmRatio cross-section ratio.

The uncertainties due to the unfolding result from imperfections in the modelling of \Wjets predictions as well as the size of the simulated sample used. The impact of the former is evaluated by repeating the unfolding using input from the \SHERPA generator instead of the \ALPGENPYTHIA6 generator and also by using input from \ALPGENPYTHIA6 where the true distribution in the unfolding matrix is reweighted to provide a better description of the data at reconstructed level. The dependence due to the size of the simulated sample is derived using pseudo-experiments and the spread of the results is taken as an uncertainty. The impact on the measured cross section ranges from 0.5\% to 3\%.

% This contains the unfolding and the systematics description

%-------------------------------------------------------------------------------
\section{Theoretical predictions}
\label{sec:theory}
%-------------------------------------------------------------------------------

The measured cross sections for \Wjets production are compared to a number of theoretical predictions at NNLO, NLO, and leading order (LO) in perturbative QCD, which are summarised in~\tabref{WjetsPred}. These predictions, with the exception of the NNLO results, are computed in the same phase space as the measurement, defined in~\tabref{fidRegion}. In general, the NNLO and NLO predictions include theoretical uncertainties due to the choice of scale and the PDFs, while the LO predictions include only statistical uncertainties. 

\subsection{NNLO predictions}
The \Wjets\ predictions at NNLO are calculated using the \njetii\ program~\cite{WjetNNLO,ZjetNNLO}, which uses the so-called N-jettiness subtraction technique to control the infrared singularities of the final-state partons. This calculation uses a renormalisation and factorisation scale choice of $\mu_\mathrm{o}=\sqrt{m_W^2+\Sigma (\pt^j)^2}$ and CT14 NNLO PDFs. All the kinematic selections listed in~\tabref{fidRegion} are applied except for the jet rapidity requirement, which is $|y| < 2.5$ for the leading jet for this calculation. In order to compare the \njetii\ results to the data, the ratio of events selected using a leading jet rapidity criterion of $|y| < 4.4$ to events using a criterion of $|y| < 2.5$ is estimated with the \ALPGENPYTHIA6 simulation as a function of each differential observable and applied as a correction to the \njetii\ results. The size of this correction is around 10\% to 15\% at low \pt of the \Wboson boson and of the jets as well as at low \HT and decreases to zero at around \SI{200}{\gev} to \SI{250}{\gev} in \pt (and at around \SI{500}{\gev} for \HT). For the differential cross section as function of the second leading jet's rapidity, the correction is approximately constant at 10\%. Uncertainties in this correction factor include statistical uncertainties from the \ALPGENPYTHIA6 sample and the change in the correction when using the \SHERPA 1.4.1 generator. The theoretical uncertainties in the NNLO prediction are obtained by multiplying and dividing $\mu_\mathrm{o}$ by a factor of two.

\begin{table}[tbp]
\robustify\bfseries
\centering
\caption{Summary of theoretical predictions, including the maximum number of partons at the highest order in \alphas used in this analysis, the PDF set used, if non-perturbative corrections (NPC) are applied and if a modelling of the parton shower (PS) is included and additional comments. The maximum number of partons in between parentheses is only used in the estimate of systematic uncertainties in the NPC. NLO electroweak (EW) corrections are applied to the prediction at NLO in \alphas only.}
\begin{tabular}{p{3.0cm}p{1.0cm}cp{2.0cm}p{0.9cm}p{0.9cm}p{2.5cm}}
\toprule
Program & Order & $N_\text{partons}^\text{max}$ & PDF set & NPC & PS & Comments  \\ 
 & in \alphas & at highest order & & & & \\ 
\midrule
\njetii & NNLO       & 1       & CT14  & \checkmark & & Not shown for \njs, \drjj\ and \mjj\ \\
 \addlinespace[2ex] \bhs  & NLO    & 1, 2 or 3 & CT10  & \checkmark & & \\
 \addlinespace[2ex] \MCFM  6.8               & NLO     & 1         & CT10 \newline+ 3 more & \checkmark & & \figref{res-ratio-1jet} only \\
 \addlinespace[2ex] \POWPYTHIA8 & NLO    &  1       & CT14 & & \checkmark & \figref{res-ratio-1jet} only \\
\addlinespace[2ex]\SHERPA 2.2.1  & NLO  & 2 & CT10 & & \checkmark & Including NLO EW corrections in \figref{res-ratio-1jet} \\
\addlinespace[2ex]\SHERPA 2.2.1  & LO  & 2 (3) & NNPDF 3.0 & & \checkmark & \\
\addlinespace[2ex]\ALPGENPYTHIA6  & LO     & 5         & CTEQ6L1 (LO) &  & \checkmark &  \\
\addlinespace[2ex]\ALPGENHERWIG  & LO     & 5         & CTEQ6L1 (LO) & & \checkmark & \\
\addlinespace[2ex]\SHERPA 1.4.1        & LO     & 4         & CT10 & & \checkmark &  \\
\bottomrule
\end{tabular}
\label{tab:WjetsPred}
\end{table}

\subsection{NLO predictions}
The \bhs\ predictions (abbreviated to BH+S in the figures) include NLO calculations for \Wjets production with up to five additional jets~\cite{Berger:2009ep,Berger:2010zx,Bern:2013gka}.  
The \bh\ program provides the NLO virtual matrix element corrections while \SHERPA calculates the tree-level diagrams and provides the phase-space integration. 
Focusing on events with one or two jets, only calculations at NLO for $\Wboson+1\text{-jet}$, $\Wboson+2\text{-jets}$, and $\Wboson+3\text{-jets}$ production are used for the corresponding measured jet multiplicity. These predictions use the CT10 NLO PDF set and the choice of renormalisation and factorisation scale is $\HT'/2$, where $\HT'$ is the scalar sum of the transverse momenta of the \Wboson boson and the jets. The theoretical uncertainties considered include uncertainties due to the PDF error set and uncertainties due to the choice of scale, which are evaluated by independently varying the renormalisation and factorisation scales up and down by a factor of two. For $\Wboson+1$-jet production, the \bhs matrix elements are also used in the exclusive sums approach~\cite{AlcarazMaestre:2012vp}, in which NLO information from $\Wboson+2$-jet production is utilised. Through this approach, additional contributions from higher multiplicity final states can be included in contrast to the standard fixed-order prediction. This is useful for observables that are sensitive to higher parton multiplicities. 

The \MCFM calculation in this paper predicts \Wjets production with one jet at NLO~\cite{MCFM,MCFM2}, with a second jet, if present, at LO accuracy as the real emission correction in the NLO calculation. Renormalisation and factorisation scales are set to $\HT/2$. Four choices of PDF sets are shown: CT10, HERAPDF~1.5~\cite{HERAPDF}, MSTW 2008 and NNPDF 2.3~\cite{NNPDF}, which are all at NLO. These predictions include uncertainties due to the PDF error set, the value of \alphas and the choice of scales, which are evaluated in the same way as above. 

The \SHERPA 2.2.1 generator is used to calculate \Wjets production at NLO for up to two associated jets and at LO for a third jet. This calculation includes matching with a parton shower, hadronisation, and modelling of the underlying event. The PDF set used is CT10 and the scale is set to $\HT'/2$. These predictions include uncertainties due to the PDF error set and the choice of scale, which are evaluated in the same way as above. The corresponding LO prediction from the same \SHERPA version is given in addition for comparison. In the figures, the LO prediction is shown without any uncertainties. Sizeable NLO corrections to the cross section from electroweak (EW) emissions are expected especially at large transverse momentum of the produced \Wboson bosons in association with one or two jets~\cite{SherpaNLOEW}. The NLO EW corrections are determined with the same set-up as the NLO QCD-only \SHERPA 2.2.1 predictions.

The \POWHEG r2129 results (abbreviated to PWHG+PY8 in the figures) are calculated at NLO for \Wboson\ production in association with one jet~\cite{Powheg}. This is interfaced to the parton shower of \PYTHIA8 \cite{Pythia82} and combined using the MiNLO technique~\cite{MiNLO}. The CT14 PDF set~\cite{CT14PDFset} is used for the \POWHEG calculation, and the PDF set \textsc{CTEQ6L1} together with the tune \textsc{AZNLO} \cite{STDM-2012-23} for the parton shower. The \POWHEG predictions of the overall cross section are corrected by a factor of 1.1 for events with $\njs\geq1$, as indicated in the figures, to match the total integrated number of events in the data. Only statistical uncertainties are included.  

\subsection{LO predictions}

Predictions from the multi-leg LO generators \ALPGEN and \SHERPA (version 1.4.1) are compared to the data. The details of these predictions are described in~\secref{MCsamples}. In addition to the \ALPGEN predictions showered with \PYTHIA6 (abbreviated to ALPGEN+PY6 in the figures), a prediction using an alternative parton shower model from \HERWIG \cite{Herwig} with \JIMMY \cite{Jimmy} for the underlying event is shown. This prediction uses the same PDF as \ALPGENPYTHIA6, but a different tune: \textsc{AUET2} \cite{ATL-PHYS-PUB-2011-008}. Only statistical uncertainties are shown. Theoretical uncertainties are large for LO calculations.

\subsection{Non-perturbative corrections} 

The \njetii, \bhs, and \MCFM results do not include non-perturbative effects from hadronisation and the underlying event. These corrections are computed for each bin with \SHERPA 2.2.1~\cite{Sherpa2} combining matrix element calculations with up to two parton emissions at LO in pQCD. The calculation uses the NNPDF 3.0 PDF set and dynamic renormalisation and factorisation scales determined by the CKKW scale-setting procedure. The corrections are typically around 2--3\% and are applied to the predictions for all measured distributions. Statistical uncertainties in these corrections and the systematic uncertainty, defined by the envelope of variations of the starting scale of the parton shower, the recoil scheme, the mode of shower evolution and the number of emitted partons from the matrix element, are included in the respective theory uncertainties. For the \WpmRatio\ predictions, no non-perturbative corrections are required as these effects cancel out in the ratio. The impact of QED radiation, which is considered as part of the dressed-electron definition in the measured cross sections, on the parton-level theoretical predictions is investigated using \SHERPA 2.2.1 with the same set-up as the NLO \SHERPA predictions described above and found to be very small. No correction for this effect is applied.

% This contains the theory description

%\newpage

%-------------------------------------------------------------------------------
\section{Cross-section results}
\label{sec:Wresults}
%-------------------------------------------------------------------------------

The measured cross sections for \Wen\ production and the cross-section ratios of \WpmRatio, obtained from separate measurement of \Wplus and \Wminus production, are shown for the jet multiplicity distributions as well as for distributions with $\njs \ge 1$. For distributions with $\njs \ge 2$, only the cross sections for \Wen\ production are shown. All results are compared to the set of predictions discussed in \secref{theory}.

\subsection{Jet multiplicity distribution}
\label{sec:Wresults-njet}

\begin{figure*} \centering
\begin{tabular}{cc}
\includegraphics[width=0.45\textwidth]{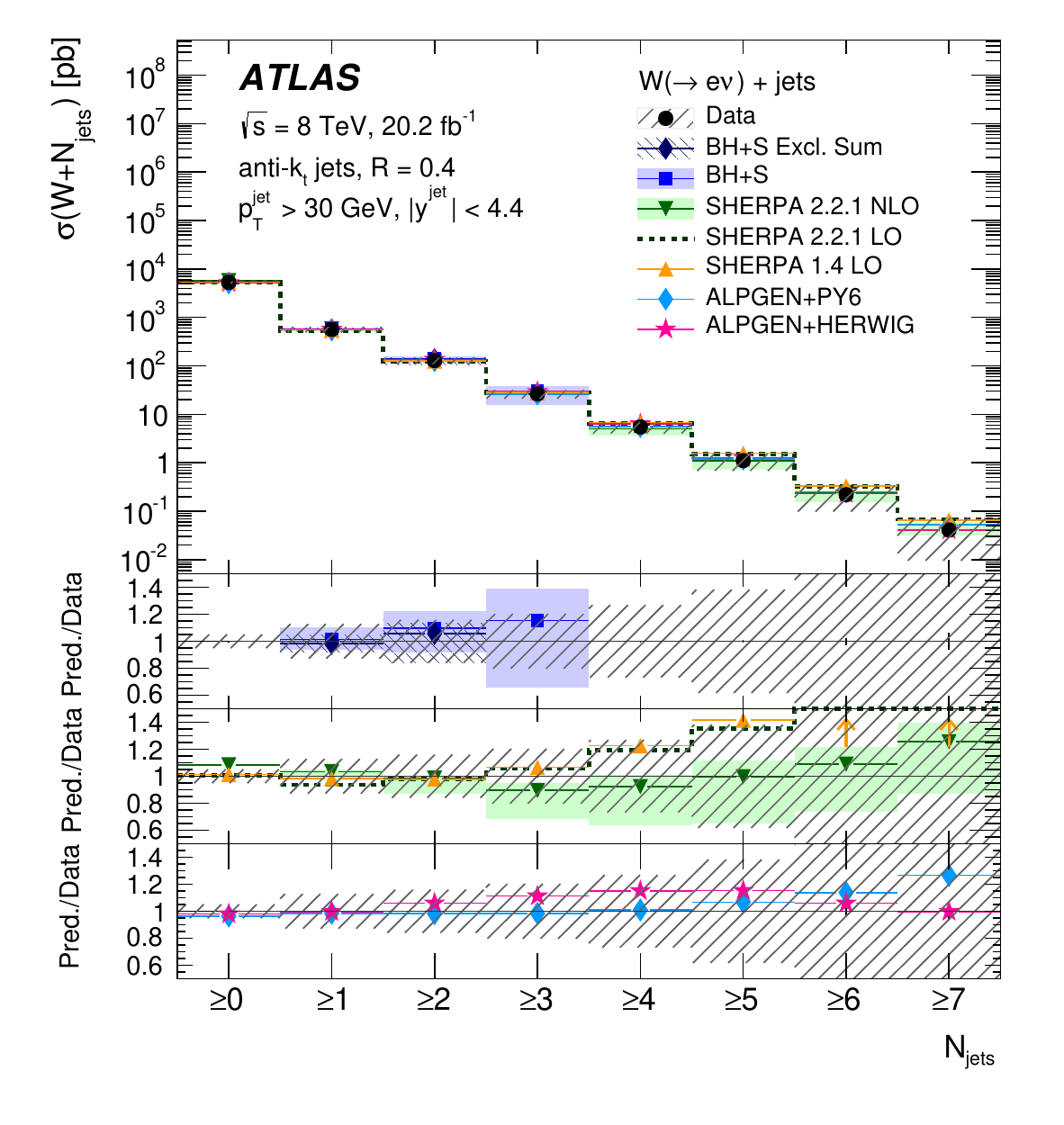} & 
\includegraphics[width=0.45\textwidth]{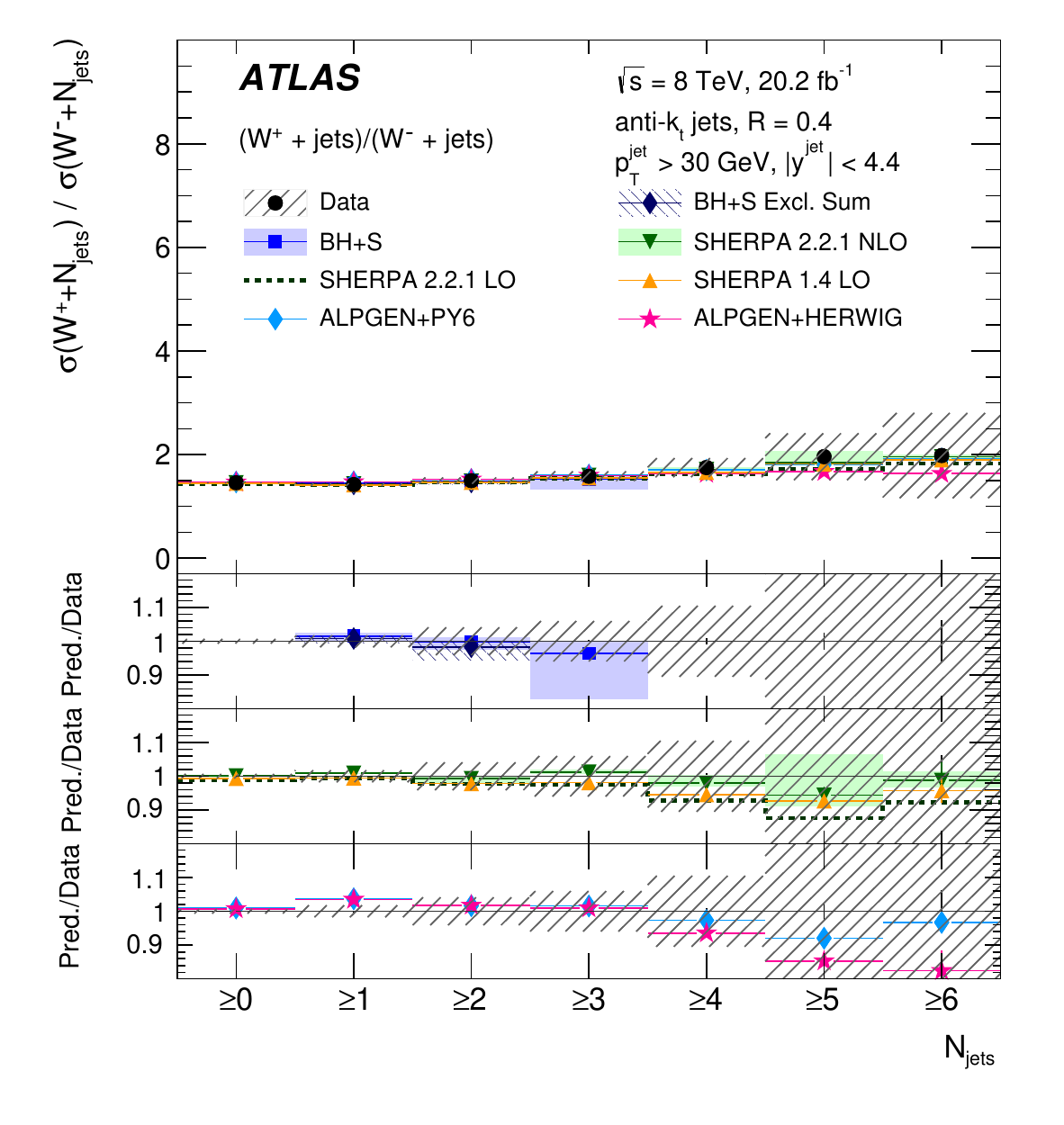} \\
\end{tabular}
\caption{Cross section for the production of \Wboson bosons (left) and the \WpmRatio\ ratio (right) for different inclusive jet multiplicities. \captionCommon }
\label{fig:res-njet}
\end{figure*}

The cross section for \Wboson production and the ratio of \WpmRatio for different inclusive jet multiplicities are shown in \figref{res-njet}. Overall the data agree with the predictions within the experimental uncertainties. At higher multiplicities, the LO \SHERPA\ predictions start to diverge from the data, while the NLO \SHERPA predictions provide a much better description of the data. The \ALPGEN\ predictions are shown for two different parton shower models, both of which are consistent with the data within the experimental uncertainties. The trends for all predictions are the same for the distributions of the \Wplus and \Wminus\ cross sections as well as the exclusive jet multiplicities (see Appendix~\ref{app:more-xsecs}). For the ratio of \WpmRatio, agreement between the data and the predictions is much improved, indicating that theoretical mismodelling related to jet emission cancels out in the ratio. The \ALPGEN predictions, which perform very well for the cross-section measurement have an offset in the \WpmRatio cross-section ratio for events with one jet, which is outside of the experimental uncertainties. This is present for both parton shower models, thereby indicating a problem in the matrix element calculation or an incorrect $u/d$ ratio in the LO PDF.

\subsection{Distributions for $\njs \ge 1$}
\label{sec:Wresults-1jet}

\begin{figure*} \centering
\begin{tabular}{cc}
\includegraphics[width=0.45\textwidth]{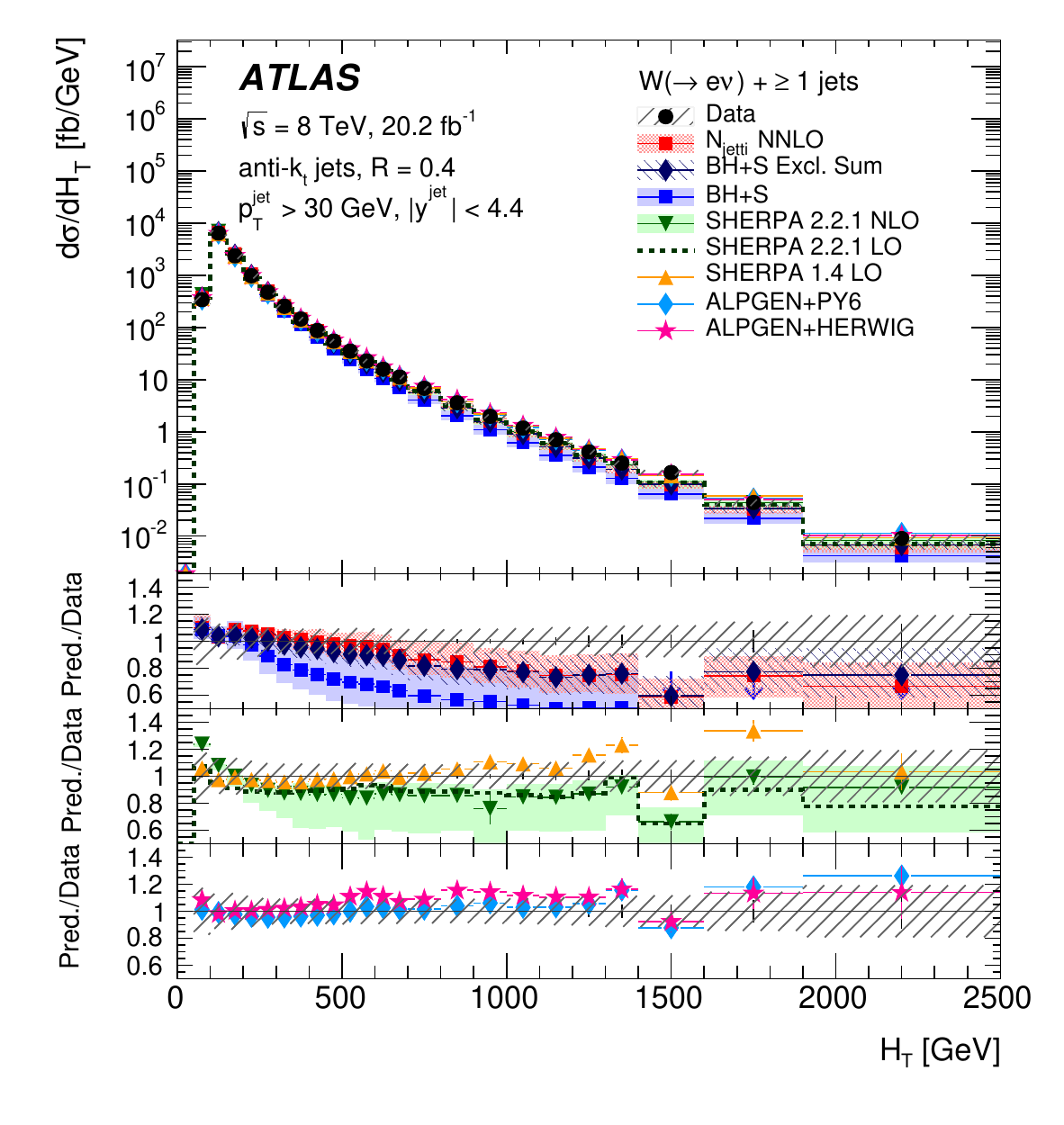} & 
\includegraphics[width=0.45\textwidth]{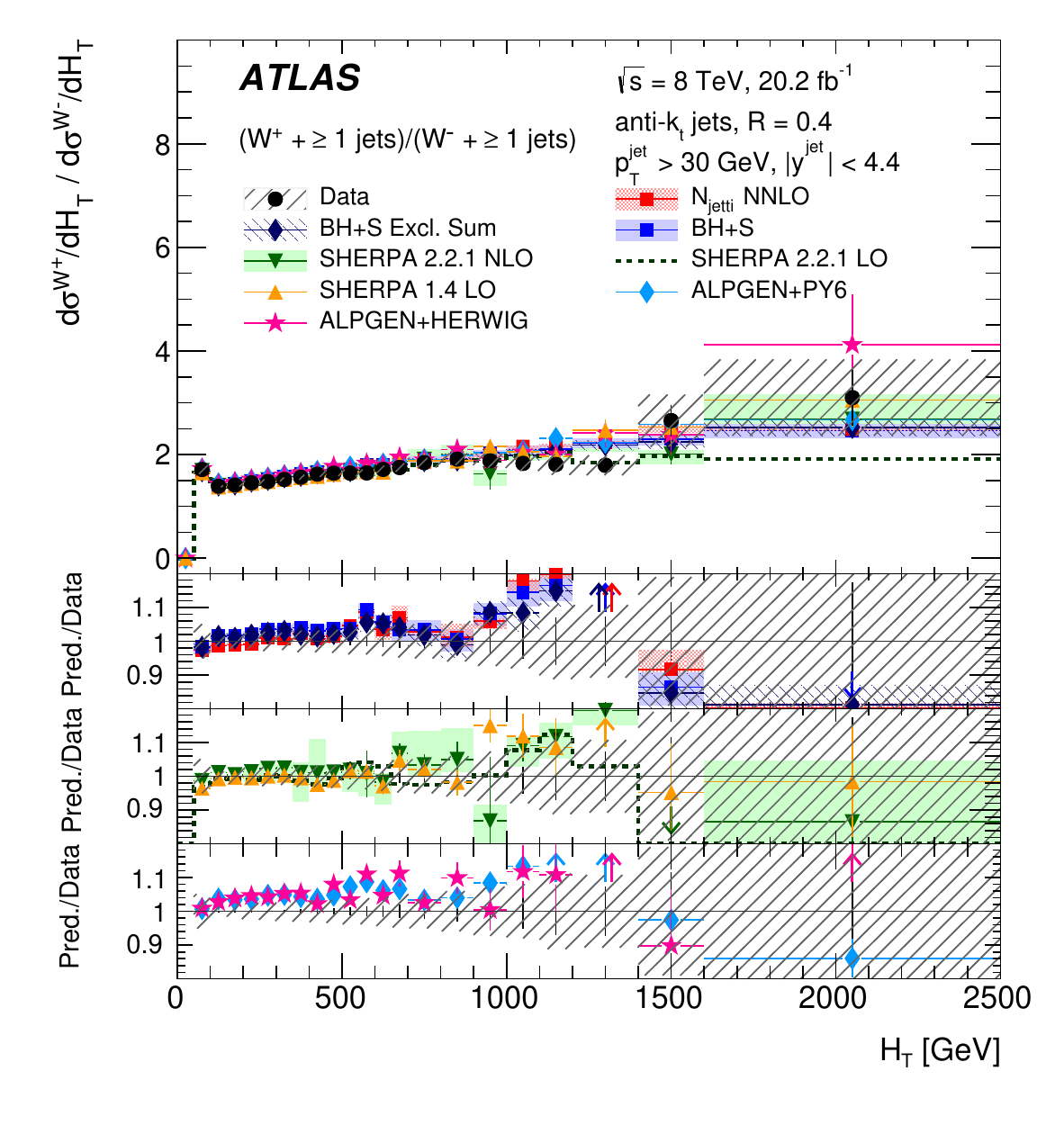} \\
\end{tabular}
\caption{Differential cross sections for the production of \Wboson bosons (left) and the \WpmRatio\ ratio (right) as a function of \htj\ for events with $\njs \ge 1$. \captionOverflow\  \captionCommon}
\label{fig:res-ht-1jet}
\end{figure*}

The differential cross section for \Wboson production and the ratio of \WpmRatio\ as a function of \htj\ are shown in \figref{res-ht-1jet} for $\njs \ge 1$. The \htj\ distribution is a very important test of pQCD as the higher values are sensitive to higher jet multiplicities and topologies such as $qq\rightarrow qq^\prime W$ (dijet production with a $W$ boson emitted from one of the initial or final state quarks). The LO predictions of \SHERPA\ and \ALPGEN, which both include multiple jets in the matrix element calculation describe the data best, although these predictions have large theoretical uncertainties. The \bhs\ predictions underestimate the data at large values of \htj. This is expected since, at these large values of \htj, contributions from additional jets are important, which are only partially present in this calculation. 
The predictions from the \bhs\ exclusive sums method and from the NNLO \njetii\ calculation, which include an additional jet emission at NLO, provide better agreement with the data. These effects cancel out to a large extent in the ratio of \WpmRatio. At the largest measured values of \HT, where the measured cross section is small, the total experimental uncertainty in the \WpmRatio cross-section ratio increases due to larger statistical uncertainties in the data and some systematic uncertainties that do not fully cancel out in the ratio. 

\begin{figure*} \centering
\begin{tabular}{cc}
\includegraphics[width=0.45\textwidth]{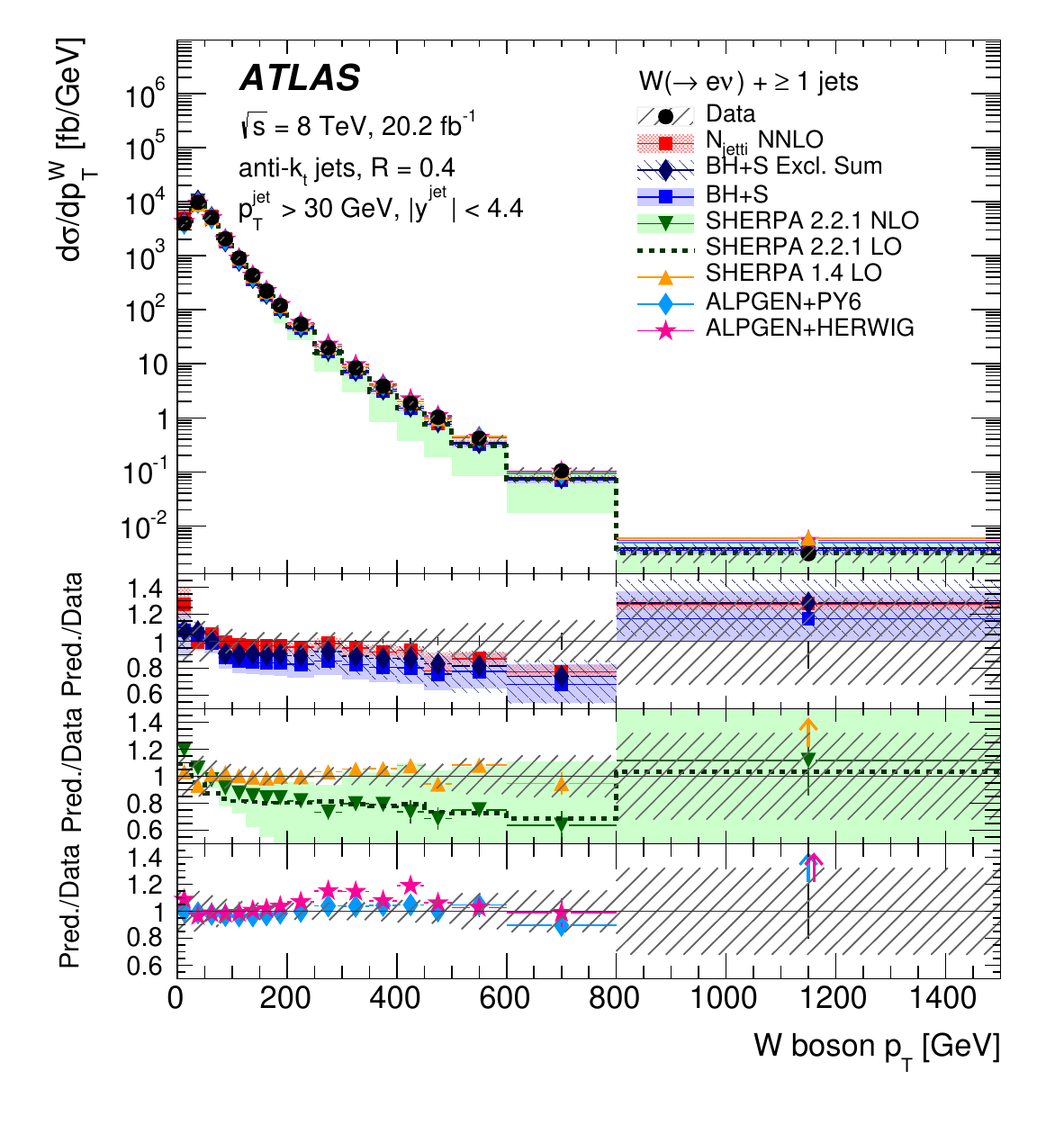} & 
\includegraphics[width=0.45\textwidth]{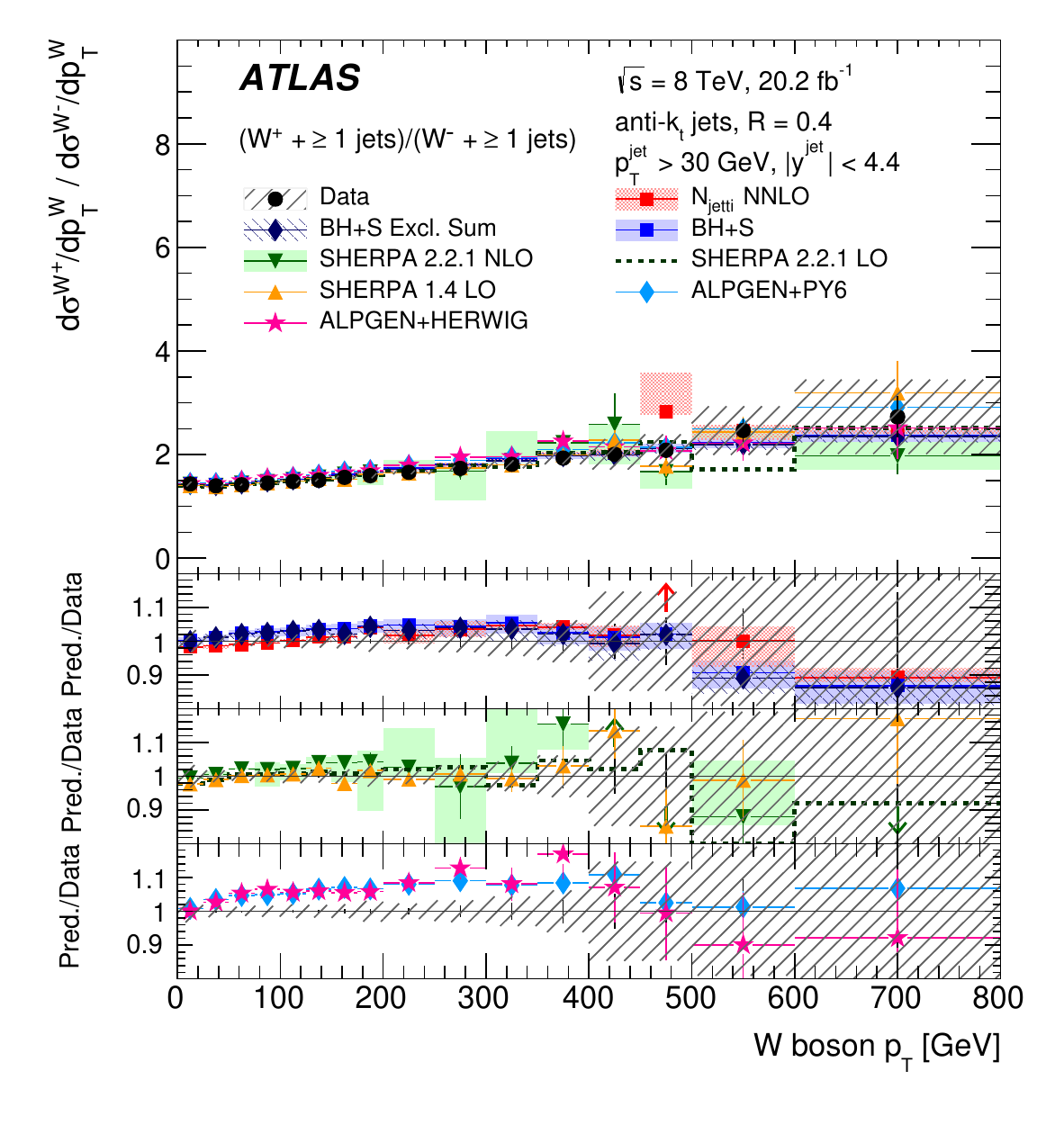} \\
\end{tabular}
\caption{Differential cross sections for the production of \Wboson bosons (left) and the \WpmRatio\ ratio (right) as a function of the \Wboson \pt for events with $\njs \ge 1$. \captionOverflow\ \captionCommon}
\label{fig:res-wpt-1jet}
\end{figure*}

The distribution of the \pt of the \Wboson boson is potentially sensitive to the parton distributions in the proton. For $\njs \ge 1$, \figref{res-wpt-1jet} shows the differential cross section as a function of the \pt of the \Wboson boson for \Wboson production and the cross-section ratio of \WpmRatio. For \Wboson production, there is good agreement between the data and most LO predictions as well as the \njetii\ NNLO calculation. Both the NLO and LO \SHERPA 2.2.1 predictions perform worse than LO \SHERPA\ 1.4.1.  
The \ALPGEN\ predictions vary slightly for different parton shower models, with \PYTHIA\ providing a better description of the data. In the ratio of \WpmRatio, where the experimental precision is greatly improved, neither of these predictions describe the data well. 
Most predictions (except \njetii\ NNLO and \SHERPA\ 1.4.1) overestimate the data between one to almost four times the experimental uncertainties. This effect is largest for \ALPGEN\ and consistent with the offset seen for \ALPGEN\ in \figref{res-njet} (right) for events with one jet.

\begin{figure*} \centering
\begin{tabular}{cc}
\includegraphics[width=0.45\textwidth]{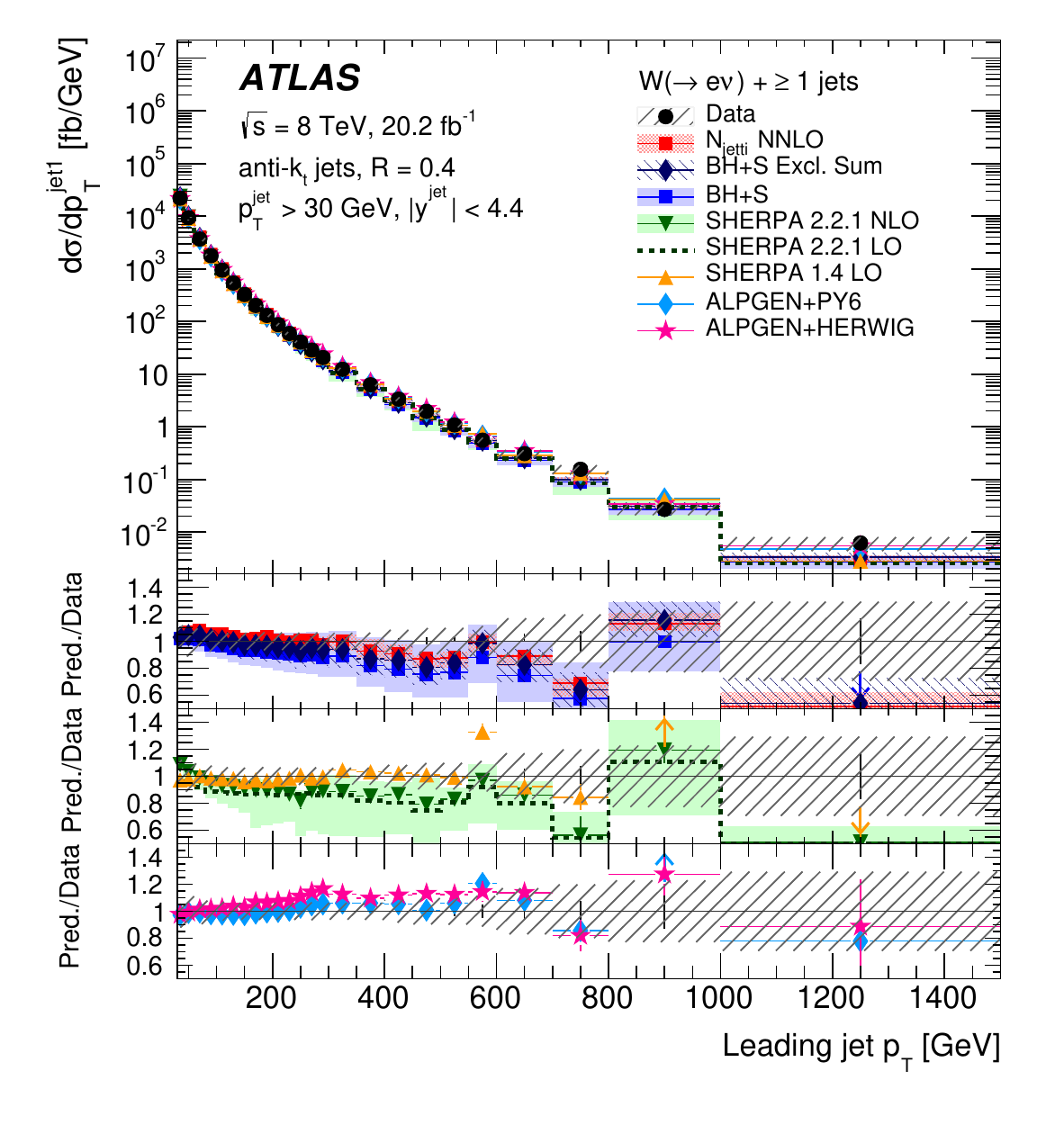} & 
\includegraphics[width=0.45\textwidth]{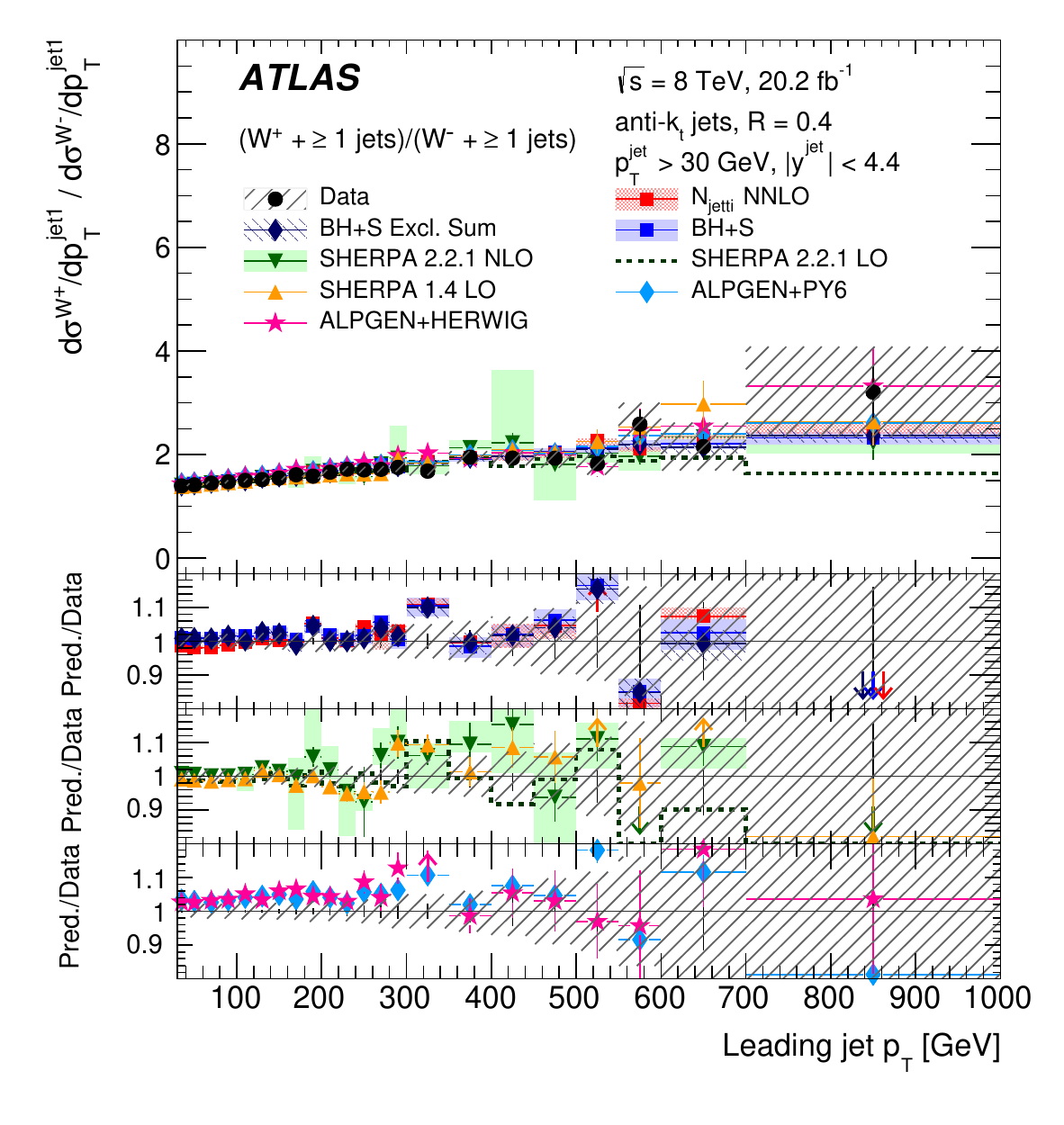} \\
\end{tabular}
\caption{Differential cross sections for the production of \Wboson bosons (left) and the \WpmRatio\ ratio (right) as a function of the leading jet \pt for events with $\njs \ge 1$. \captionOverflow\ \captionCommon}
\label{fig:res-leadjetpt-1jet}
\end{figure*}

\figref{res-leadjetpt-1jet} shows the differential cross section as a function of the leading jet \pt for events with $\njs \ge 1$ for \Wboson production and the ratio of \WpmRatio. The \njetii, \ALPGEN\ and LO \SHERPA 1.4.1 predictions show fair agreement with the data for both distributions. The \SHERPA\ 2.2.1 calculations for both NLO and LO as well as \bhs\ tend to predict a softer \pt distribution. These differences contrast with those observed in $W$+dijet production~\cite{STDM-2014-11} in the leading jet \pt for events with at least two jets. In that paper, the event selection requires a larger leading jet \pt\ and a dijet invariant mass greater than 500 \GeV. With this selection, the predictions tend to overestimate the cross section. This highlights how a given prediction can yield very different results in different phase spaces.

\begin{figure*} \centering
\begin{tabular}{cc}
\includegraphics[width=0.45\textwidth]{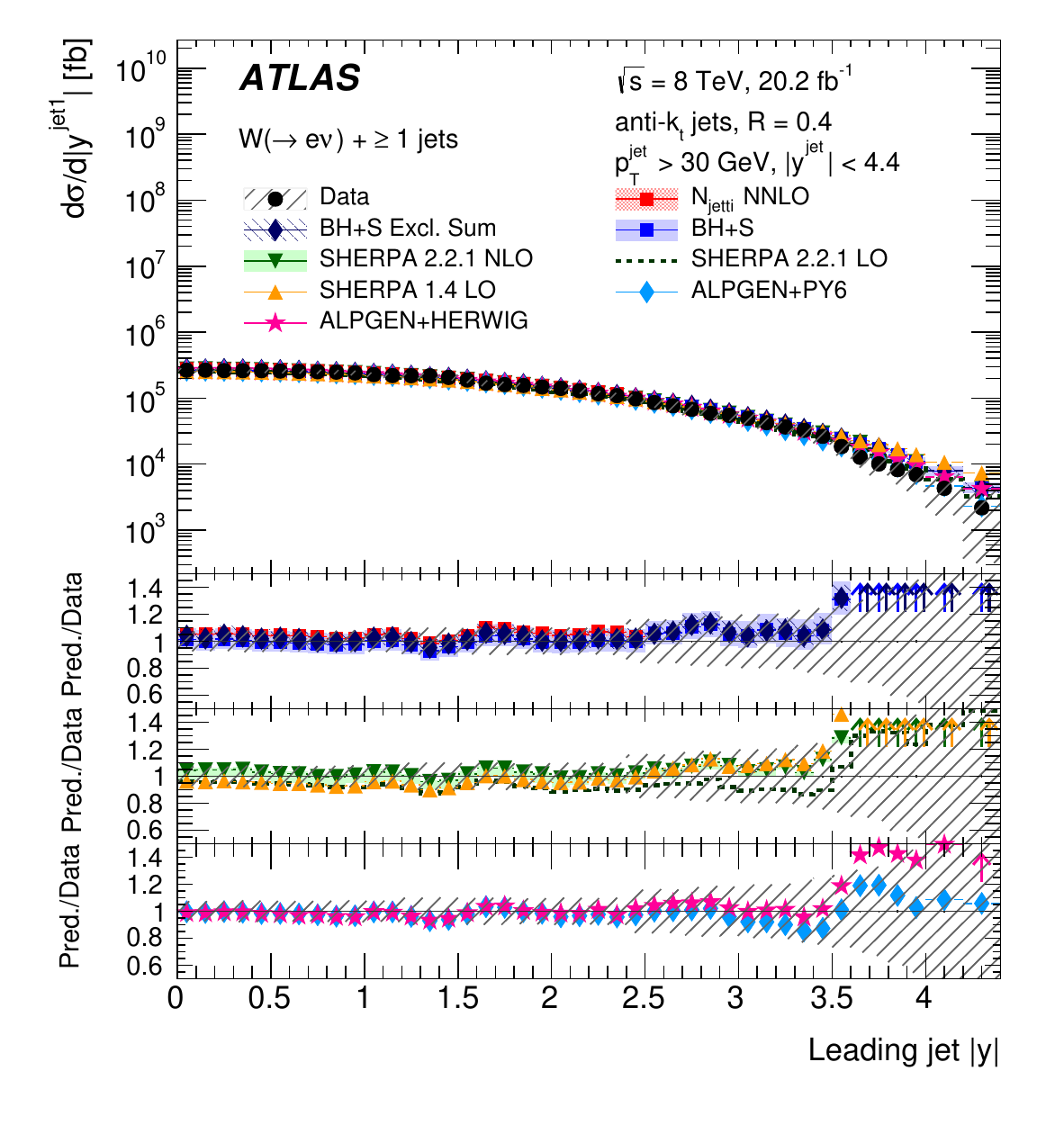} & 
\includegraphics[width=0.45\textwidth]{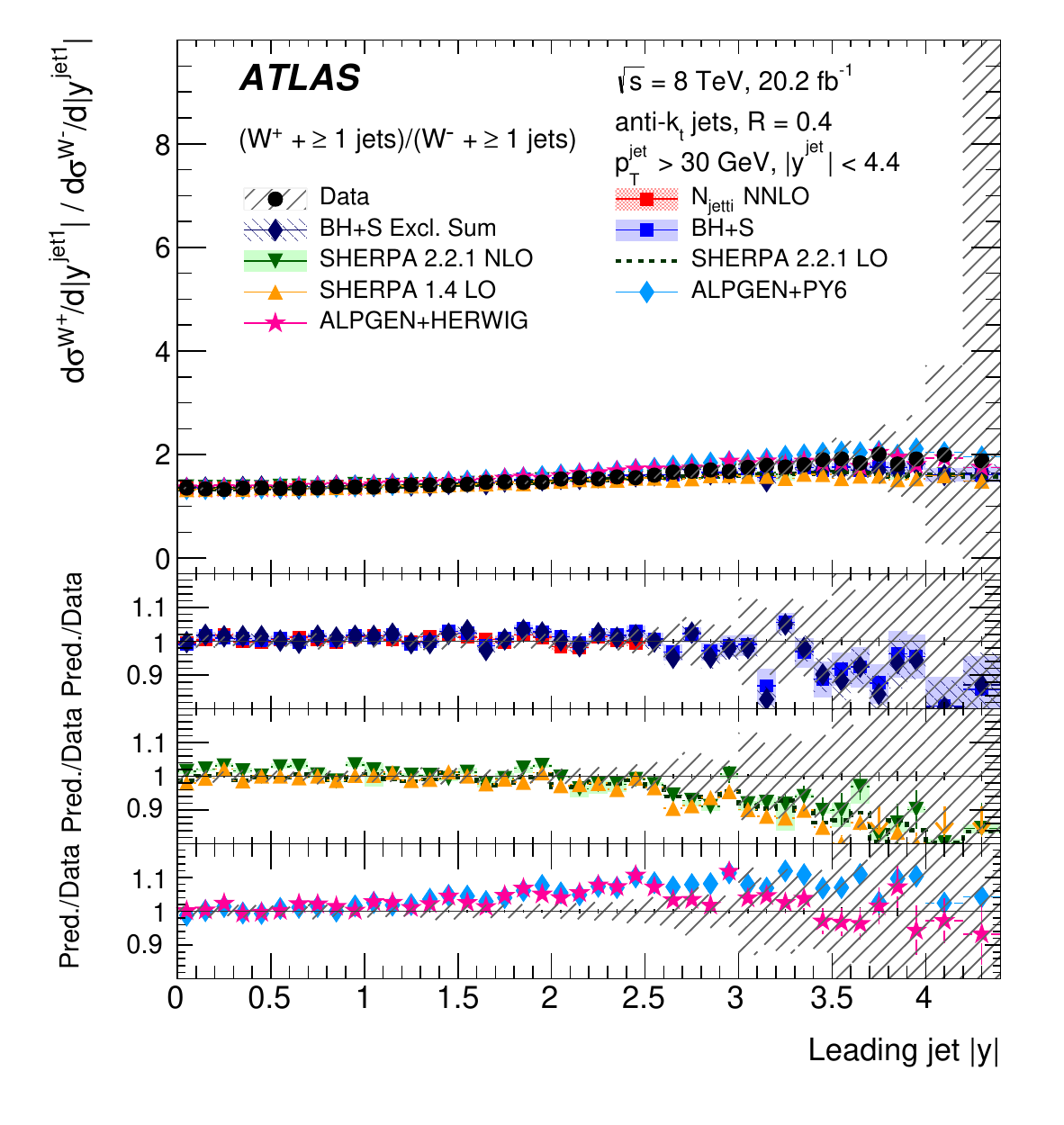} \\
\end{tabular}
\caption{Differential cross sections for the production of \Wboson bosons (left) and the \WpmRatio\ ratio (right) as a function of the leading jet rapidity for events with $\njs \ge 1$. \captionCommon}
\label{fig:res-leadjetrap-1jet}
\end{figure*}

The differential cross sections as a function of the leading jet rapidity for events with $\njs \ge 1$ for \Wboson production and the ratio of \WpmRatio\ are shown in \figref{res-leadjetrap-1jet}. In the forward region, the data turn down more sharply at $\absrap\approx 3.6$ compared to a smoothly falling distribution. The experimental uncertainties, which in this region are dominated by the difference between \ALPGEN and \SHERPA in the unfolding and the jet energy scale and resolution uncertainties, cover this effect. 
Most theory calculations predict a larger cross section for forward jets than that observed in the data and lie within 1--2 times the experimental uncertainties.
The parton shower model strongly influences the calculated cross section in the high rapidity region, as seen through the comparison of \ALPGENPYTHIA6\ and \ALPGENHERWIG. In addition, different PDF sets can affect the predicted cross section at high jet rapidities, but to a smaller extent (as can be seen by comparing with \figref{res-app-leadjetrap-1jet-ch} in \appref{more-xsecs}). The mismodelling in the forward region, however, largely cancels out in the ratio of \WpmRatio, resulting in good agreement with data. It can be noticed that \SHERPA underpredicts the ratio at high rapidities, and \ALPGEN overpredicts the data around $|y|\approx2.4$.


\begin{figure*} \centering
\begin{tabular}{cc}
\includegraphics[width=0.45\textwidth]{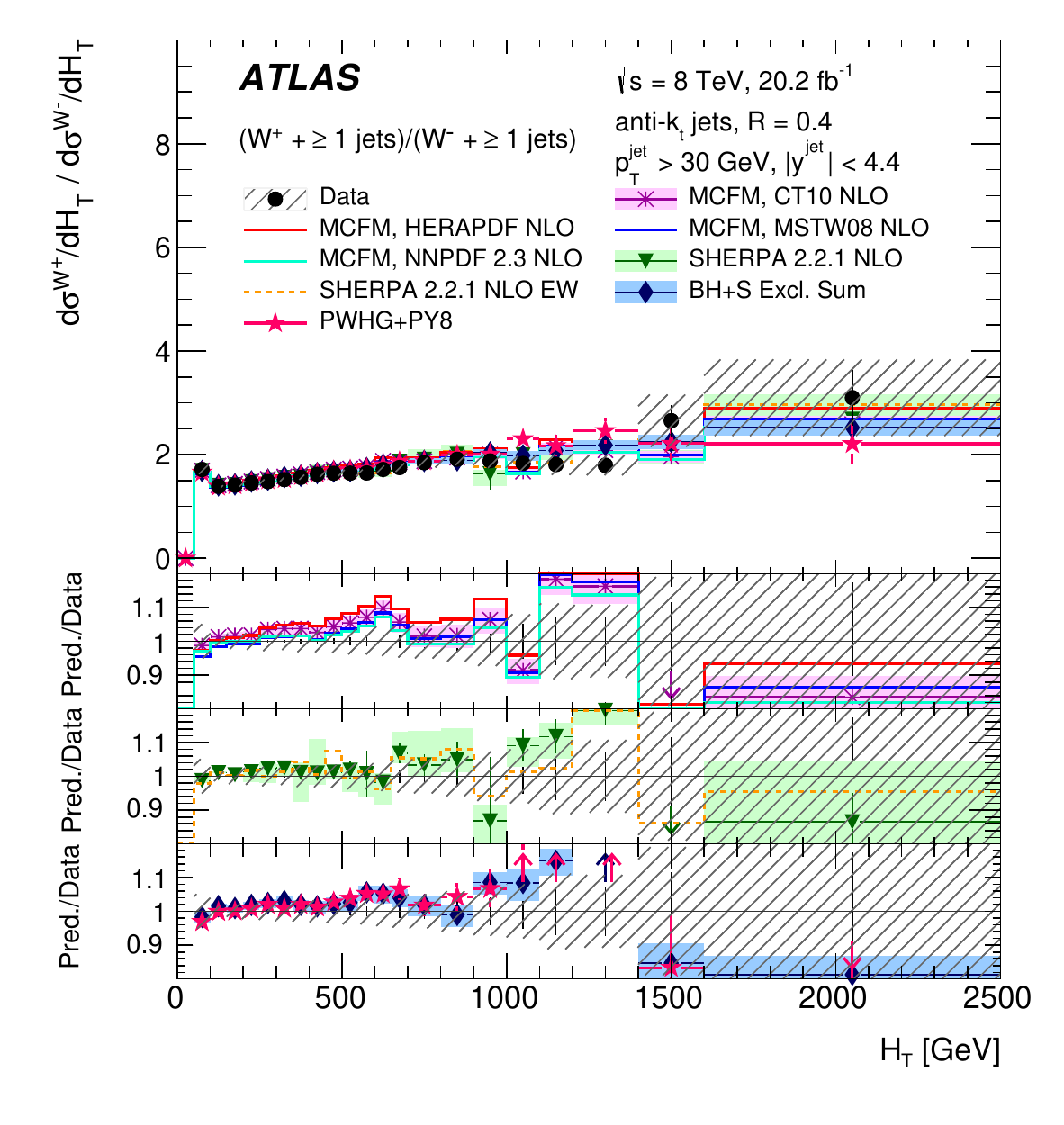} & 
\includegraphics[width=0.45\textwidth]{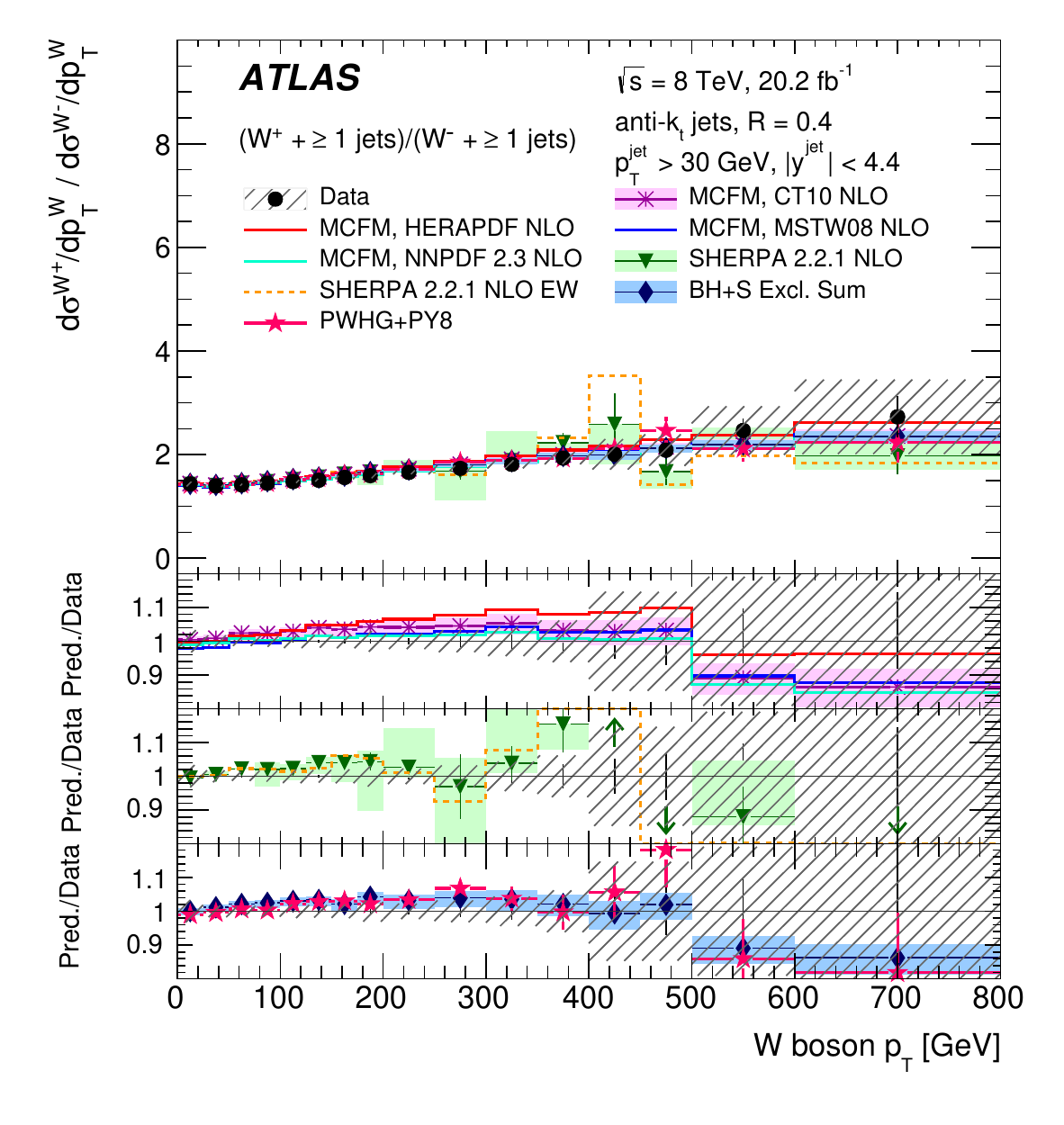} \\
\includegraphics[width=0.45\textwidth]{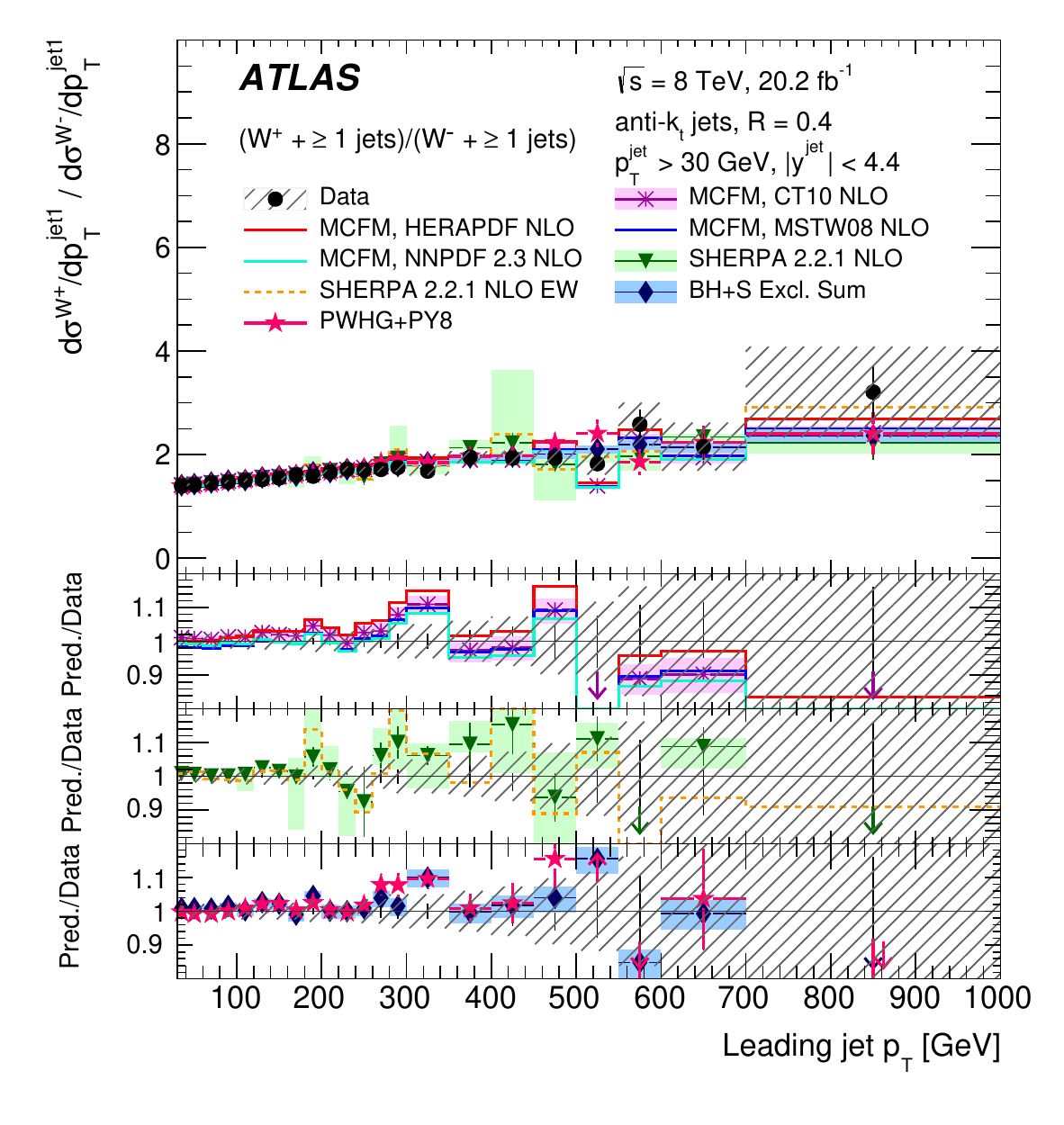} & 
\includegraphics[width=0.45\textwidth]{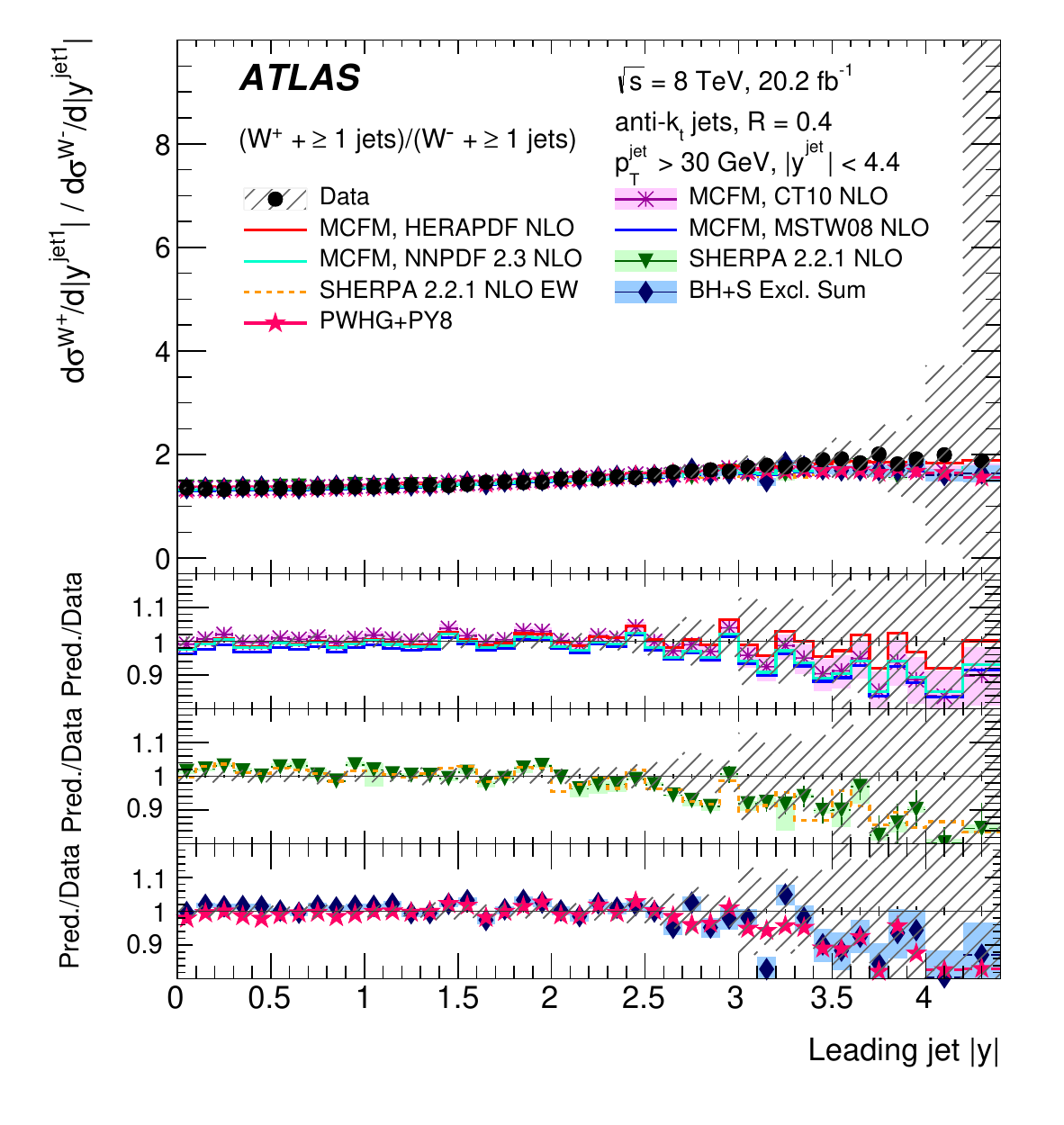} \\
\end{tabular}
\caption{\Wplus to \Wminus\ cross-section ratio as a function of \HT (top left), \Wboson \pt (top right), leading jet \pt (bottom left) and leading jet rapidity (bottom right) for events with $\njs \geq 1$. \captionCommon}
\label{fig:res-ratio-1jet}
\end{figure*}

The \WpmRatio cross-section ratios for the above four observables (\HT, \Wboson boson \pt, leading jet \pt and leading jet rapidity) are compared in \figref{res-ratio-1jet} to NLO \MCFM predictions with four different PDF sets: CT10, HERAPDF~1.5, MSTW~2008, and NNPDF~2.3. The theoretical uncertainties for the \MCFM prediction are displayed only for the CT10 PDF set. As seen in the figure, the \MCFM predictions vary depending on the PDF set used. These variations are largest for the \pt of the \Wboson boson and at forward jet rapidities. In the region of \SI{200}{\gev} to \SI{400}{\gev} in the distribution of the \pt of the \Wboson boson, where experimental uncertainties in the ratio are small (around 2\% to 6\%), the predictions from the four PDF sets differ by about 2\% to 5\% and are, in some cases, up to 2--3 times the experimental uncertainty away from the data. Similar trends are visible in the \HT distribution and the distribution of the leading jet \pt. These results should prove useful in global PDF fits as a counterpart to measurements of \Zboson boson production as a function of the \Zboson boson \pt ~\cite{STDM-2014-12, PDFimpactWjets, ZptTheoryPaper}.

Additional predictions from \POWHEG+\PYTHIA8 and \SHERPA 2.2.1, the latter including NLO electroweak corrections to the \Wplus and \Wminus cross sections, are also shown in \figref{res-ratio-1jet} for the \WpmRatio cross-section ratios. The description of the ratio by the \POWHEG+\PYTHIA8 predictions is similar to that from the \bhs\ exclusive sums method. The impact of the emission of a second jet calculated at NLO in \alphas in the latter is balanced by the \PYTHIA8 parton shower in the former. Corrections to the differential cross section from electroweak radiation calculated at NLO in $\alpha_\text{QED}$ grow for increasing values of \HT, leading jet \pt and the \pt of the \Wboson boson, resulting in a reduction of the predicted cross section of up to 30\% to 80\% depending on the distribution (the distributions are shown in \figsref{res-app-HT-1jet-ch}{res-app-leadjetpt-1jet-ch} in \appref{more-xsecs}). The inclusion of these corrections to \SHERPA 2.2.1 leads to a larger disagreement with the data. In the \WpmRatio cross-section ratio, differences due to these higher-order effects, both in \alphas and $\alpha_\text{QED}$, cancel out to a large extent.

\subsection{Distributions for $\njs \ge 2$}
\label{sec:Wresults-2jet}

\begin{figure*} \centering
\begin{tabular}{cc}
\includegraphics[width=0.45\textwidth]{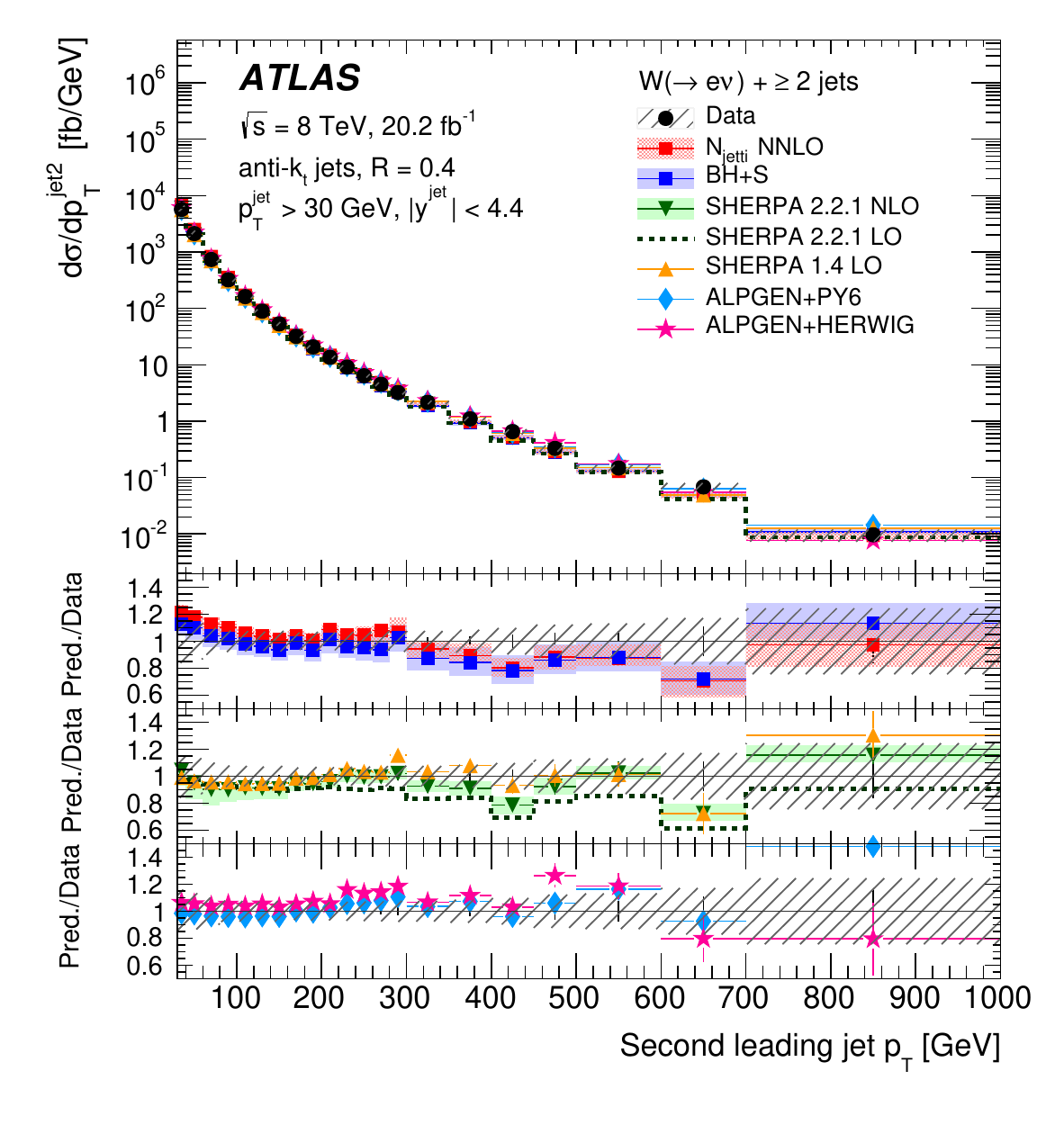} & 
\includegraphics[width=0.45\textwidth]{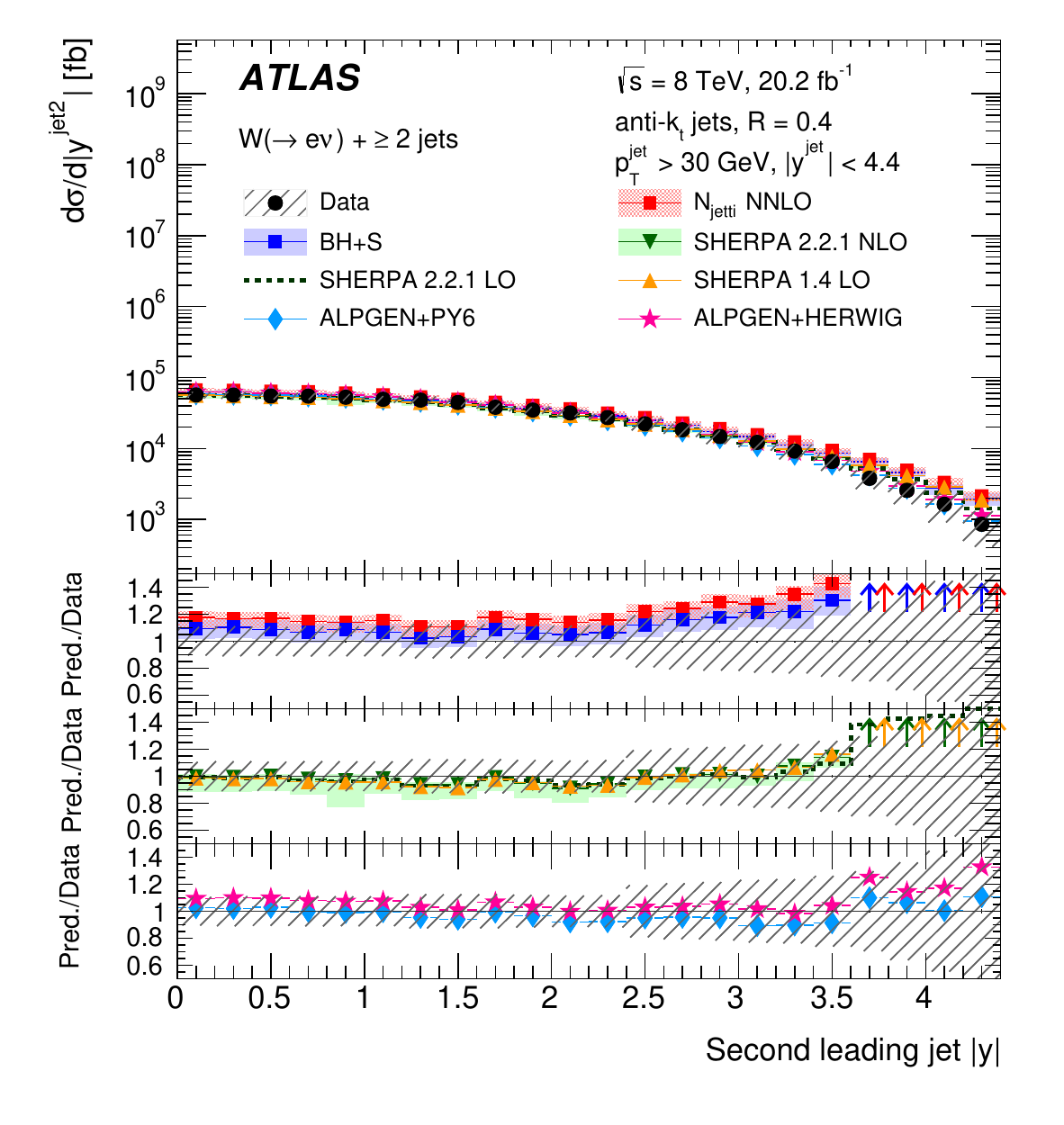} \\
\end{tabular}
\caption{Differential cross sections for the production of \Wjets\ as a function of second leading jet \pt (left) and rapidity (right) for events with $\njs \ge 2$. \captionOverflow\ \captionCommon}
\label{fig:res-jet-2jet}
\end{figure*}

For events with at least two jets, the differential cross sections as a function of the second leading jet \pt\ and rapidity are shown for \Wboson production in \figref{res-jet-2jet}. Both fixed-order predictions, \njetii\ ($\Wboson+1\text{~jet}$ at NNLO) and \bhs\ ($\Wboson+2\text{~jets}$ at NLO), predict the second jet at NLO. Both have a similar level of agreement with respect to the data for the second leading jet's \pt distribution. 
The second leading jet's rapidity distribution is modelled well by most predictions up to a rapidity of $|y|\approx2.5$, similar to the modelling of the rapidity for the leading jet. At large jet rapidities, with the exception of \ALPGEN, all other calculations tend to predict larger cross sections. 

\begin{figure*} \centering
\begin{tabular}{cc}
\includegraphics[width=0.45\textwidth]{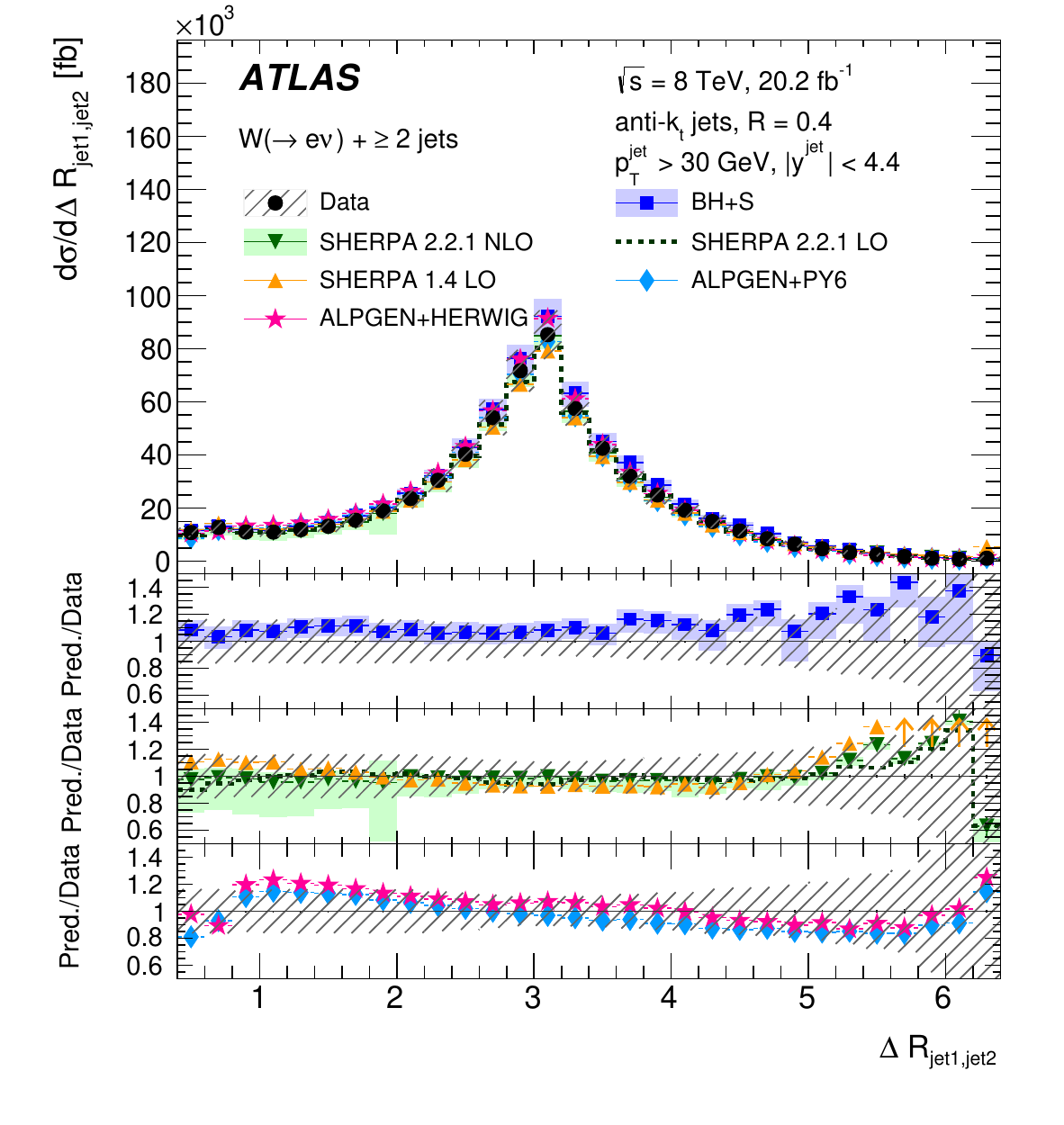} & 
\includegraphics[width=0.45\textwidth]{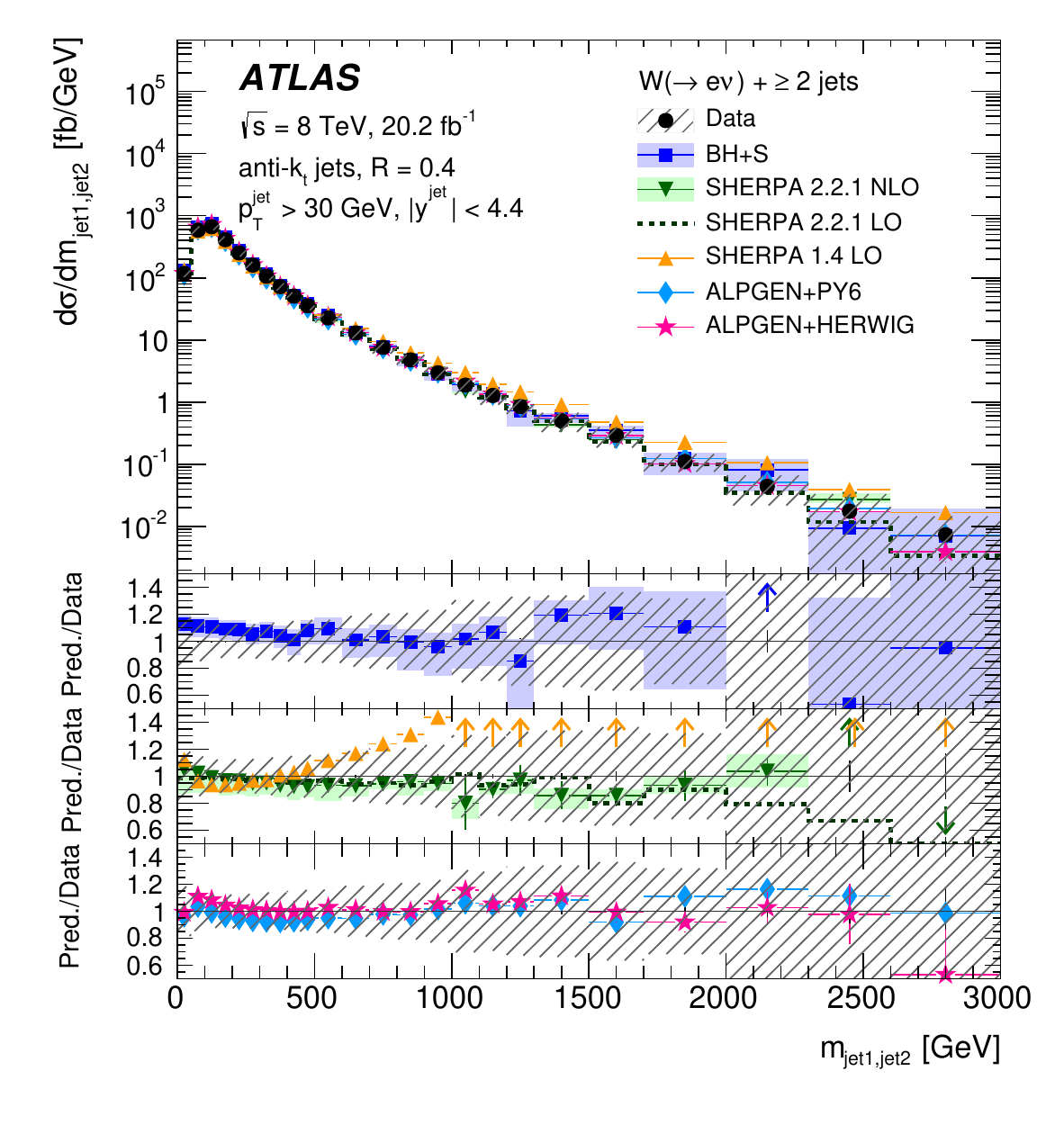} \\
\end{tabular}
\caption{Differential cross sections for the production of \Wjets\ as a function of \drjj\ (left) and dijet invariant mass (right) for events with $\njs \ge 2$. \captionOverflowBoth\ \captionCommon}
\label{fig:res-ang-2jet}
\end{figure*}

The cross sections as a function of the dijet angular separation (\drjj) and of the dijet invariant mass (\mjj) are shown in \figref{res-ang-2jet} for \Wboson production. These observables test hard parton radiation at large angles and matrix-element/parton-shower matching schemes. Jet production in the forward region can also be very sensitive to the tuning of the underlying event's contribution. \bhs\ describes the data well for both distributions even at large dijet invariant masses, with a cross section slightly higher than in the data at low invariant mass. This leads to the small observed offset in the \drjj\ distribution, which is dominated by this low \mjj\ region. The \SHERPA\ 1.4.1 generator predicts too many events at large angular separations and high dijet invariant masses. As a result, this prediction greatly overestimates the data in the highest bin of the \drjj\ distribution, which includes all higher values beyond the shown  \drjj\ range. Both LO and NLO \SHERPA 2.2.1 describe the data better, in particular, the dijet invariant mass distribution. In \SHERPA 1.4.1, considerable improvement in the description of this observable is found when requiring a larger \pt\ of the leading jet. The \ALPGEN\ predictions describe large invariant masses well, but deteriorate for small and large angular separations between the leading two jets. Differences between the \ALPGEN\ predictions with two different parton shower models are small for both distributions. For electroweak production of the \Wboson boson, which becomes larger for dijet invariant masses above 1 \TeV, a dedicated measurement with an optimised selection has been performed using data at $\sqrt{s}=\SI{7}{\tev}$ and $\sqrt{s}=\SI{8}{\tev}$~\cite{STDM-2014-11}.

The cross sections for all distributions shown in this paper are available in HepData~\cite{Hepdata}.
The additional jet multiplicities for the displayed observables, the cross section and \WpmRatio cross-section ratio as a function of the pseudorapidity of the electron, and the separate \Wplus and \Wminus cross section distributions are given in Appendix~\ref{app:more-xsecs}. 

\FloatBarrier

%-------------------------------------------------------------------------------
\section{Conclusion}
\label{sec:conclusion}
%-------------------------------------------------------------------------------

This paper presents measurements of \Wboson\ boson production cross sections and the \WpmRatio\ cross-section ratios, using data corresponding to an integrated luminosity of 20.2~fb$^{-1}$ of proton--proton collisions at $\sqrt{s}=8$~TeV recorded by the ATLAS detector at the LHC. The selected differential distributions focus on \Wboson production in association with one or two jets and are sensitive tests of perturbative QCD, the modelling of the parton shower and the parton structure of the proton. The \WpmRatio\ cross-section ratio can be measured to high precision as many of the experimental and theoretical uncertainties cancel out. 

Overall, the measured distributions show that NNLO and NLO predictions are able to describe the data. However, at high transverse momenta, large jet rapidities, or large dijet angular separations, many of these predictions underestimate or overestimate the cross sections. In many places, multi-leg LO generators, such as \ALPGEN\ and \SHERPA, which consider a larger number of parton emissions from the matrix element calculation, model the data best, although with large theoretical uncertainties.
There is, however, no single prediction that is able to describe all distributions well. The \htj, jet rapidity, and dijet invariant mass distributions are in general the least well described, suggesting that better modelling of events with energetic jets as well as jets with large rapidities is needed. 
In the \WpmRatio\ cross-section ratios, additional features in the description of the data by the predictions emerge; agreement for \ALPGEN\ worsens, but in many places improves for others. The choice of parton distribution functions can, in some cases, modify the predicted \WpmRatio\ cross-section ratio by about the experimental uncertainty.

The presented measurements will allow a better understanding of perturbative QCD and the parton distribution functions of the proton.

%-------------------------------------------------------------------------------
\section*{Acknowledgements}
%-------------------------------------------------------------------------------

% Acknowledgements for papers with collision data
% Version 02-Nov-2020

% Standard acknowledgements start here
%----------------------------------------------

We thank CERN for the very successful operation of the LHC, as well as the
support staff from our institutions without whom ATLAS could not be
operated efficiently.

We acknowledge the support of ANPCyT, Argentina; YerPhI, Armenia; ARC, Australia; BMWFW and FWF, Austria; ANAS, Azerbaijan; SSTC, Belarus; CNPq and FAPESP, Brazil; NSERC, NRC and CFI, Canada; CERN; ANID, Chile; CAS, MOST and NSFC, China; COLCIENCIAS, Colombia; MSMT CR, MPO CR and VSC CR, Czech Republic; DNRF and DNSRC, Denmark; IN2P3-CNRS and CEA-DRF/IRFU, France; SRNSFG, Georgia; BMBF, HGF and MPG, Germany; GSRT, Greece; RGC and Hong Kong SAR, China; ISF and Benoziyo Center, Israel; INFN, Italy; MEXT and JSPS, Japan; CNRST, Morocco; NWO, Netherlands; RCN, Norway; MNiSW and NCN, Poland; FCT, Portugal; MNE/IFA, Romania; JINR; MES of Russia and NRC KI, Russian Federation; MESTD, Serbia; MSSR, Slovakia; ARRS and MIZ\v{S}, Slovenia; DST/NRF, South Africa; MICINN, Spain; SRC and Wallenberg Foundation, Sweden; SERI, SNSF and Cantons of Bern and Geneva, Switzerland; MOST, Taiwan; TAEK, Turkey; STFC, United Kingdom; DOE and NSF, United States of America. In addition, individual groups and members have received support from BCKDF, CANARIE, Compute Canada, CRC and IVADO, Canada; Beijing Municipal Science \& Technology Commission, China; COST, ERC, ERDF, Horizon 2020 and Marie Sk{\l}odowska-Curie Actions, European Union; Investissements d'Avenir Labex, Investissements d'Avenir Idex and ANR, France; DFG and AvH Foundation, Germany; Herakleitos, Thales and Aristeia programmes co-financed by EU-ESF and the Greek NSRF, Greece; BSF-NSF and GIF, Israel; La Caixa Banking Foundation, CERCA Programme Generalitat de Catalunya and PROMETEO and GenT Programmes Generalitat Valenciana, Spain; G\"{o}ran Gustafssons Stiftelse, Sweden; The Royal Society and Leverhulme Trust, United Kingdom.

The crucial computing support from all WLCG partners is acknowledged gratefully, in particular from CERN, the ATLAS Tier-1 facilities at TRIUMF (Canada), NDGF (Denmark, Norway, Sweden), CC-IN2P3 (France), KIT/GridKA (Germany), INFN-CNAF (Italy), NL-T1 (Netherlands), PIC (Spain), ASGC (Taiwan), RAL (UK) and BNL (USA), the Tier-2 facilities worldwide and large non-WLCG resource providers. Major contributors of computing resources are listed in Ref.~\cite{ATL-SOFT-PUB-2020-001}.

%----------------------------------------------
% Created with Glance <Atlas.Glance@cern.ch>

%-------------------------------------------------------------------------------
\clearpage
\appendix
\part*{Appendix}
\addcontentsline{toc}{part}{Appendix}
%-------------------------------------------------------------------------------

\section{Additional cross-section distributions}
\label{app:more-xsecs}
\FloatBarrier

This appendix includes cross-section results for additional jet multiplicities, the differential cross section and the \WpmRatio cross-section ratio as a function of the pseudorapidity of the measured electron in the presence of any jet and of at least one jet, as well as the separate \Wplus and \Wminus cross sections for all \WpmRatio cross-section ratios, where they have not been shown earlier.

\subsection{Jet multiplicity and distributions for events with $\njs \ge 2$}
\label{app:2-jet}

\begin{figure*} \centering
\includegraphics[width=0.45\textwidth]{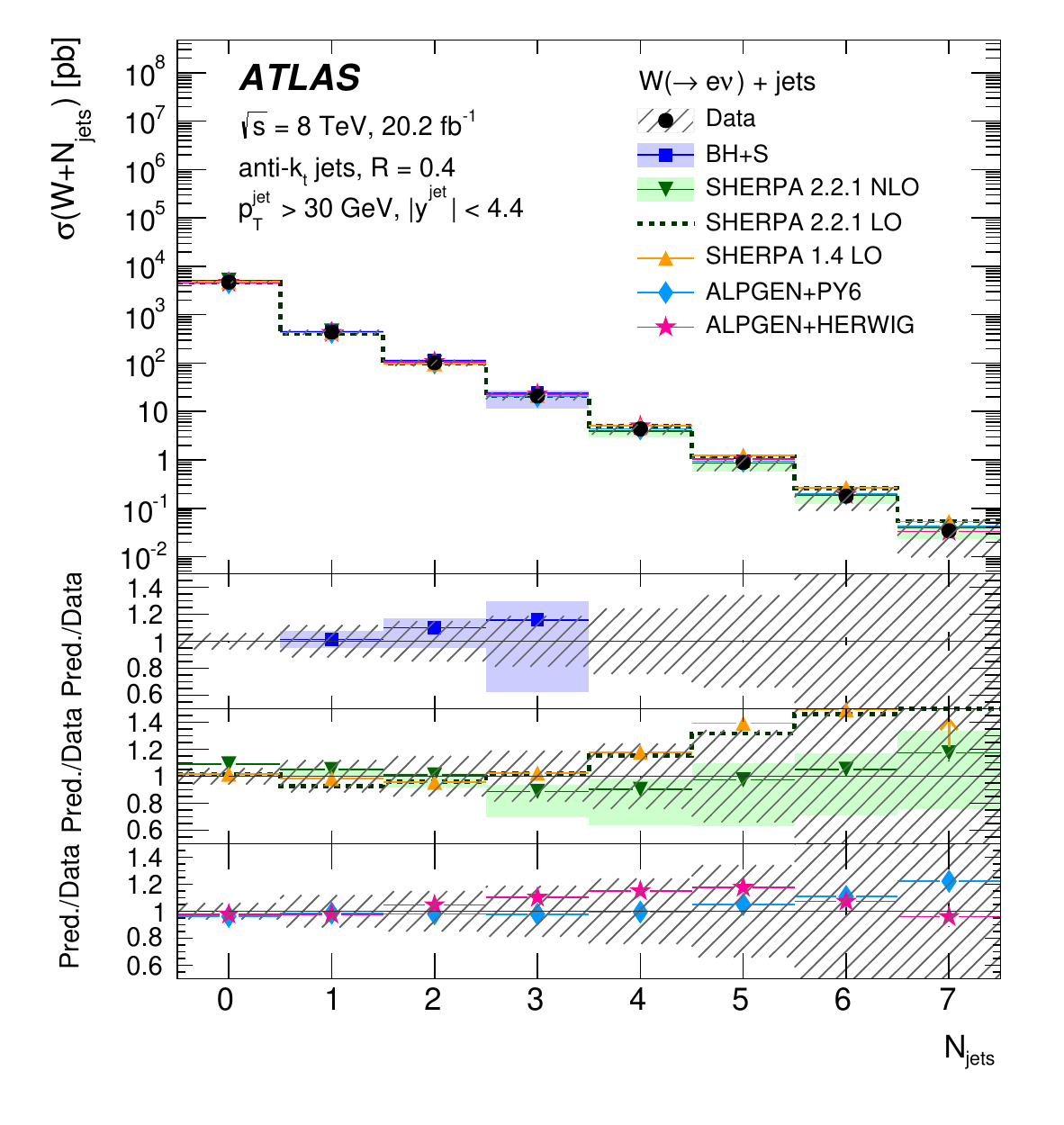} 
\caption{Cross section for the production of \Wjets\ as a function of exclusive jet multiplicity. \captionCommon}
\label{fig:res-njet-ex}
\end{figure*}

\begin{figure*} 
\centering 
\begin{tabular}{cc} 
\includegraphics[width=0.45\textwidth]{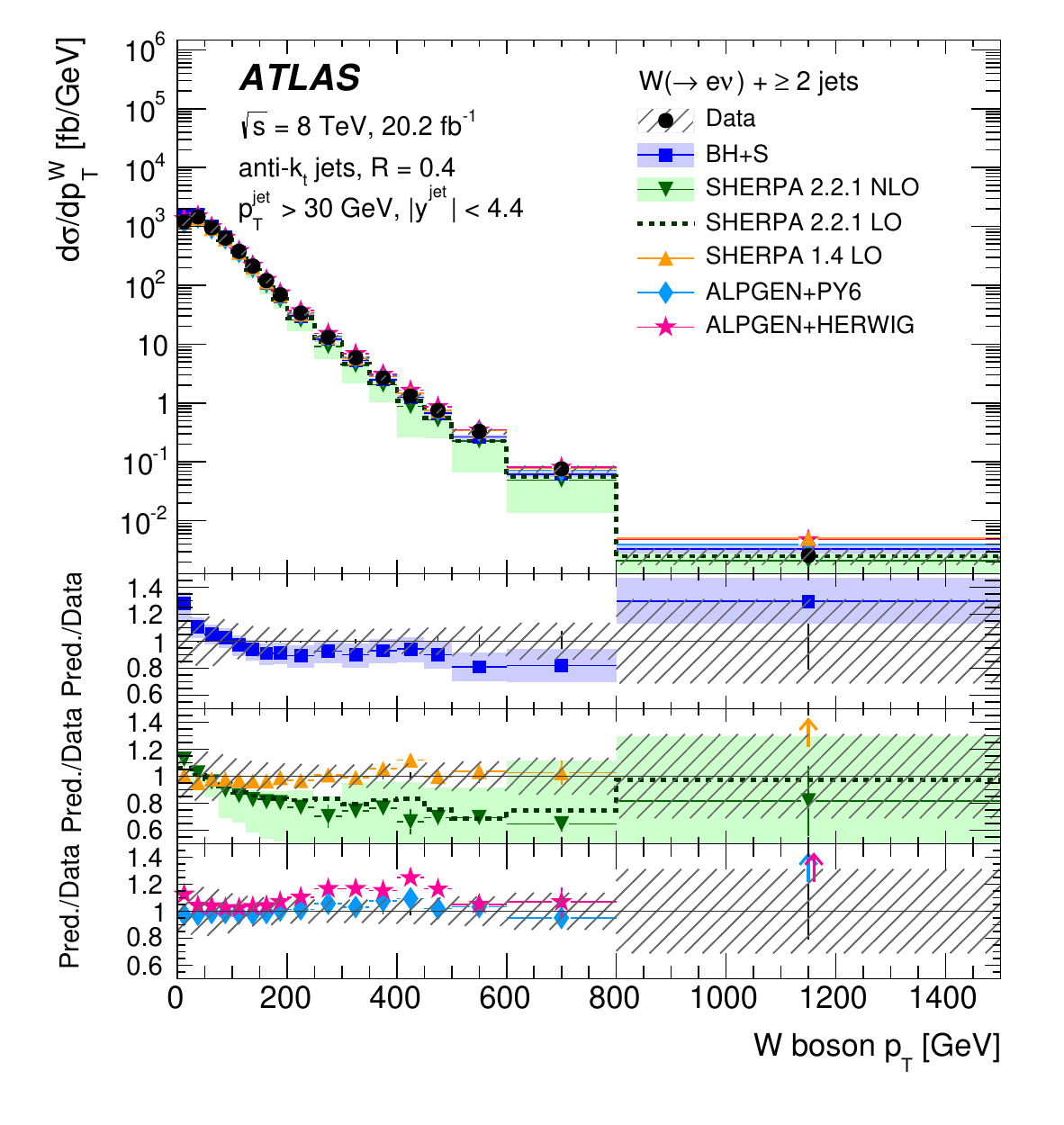} & 
\includegraphics[width=0.45\textwidth]{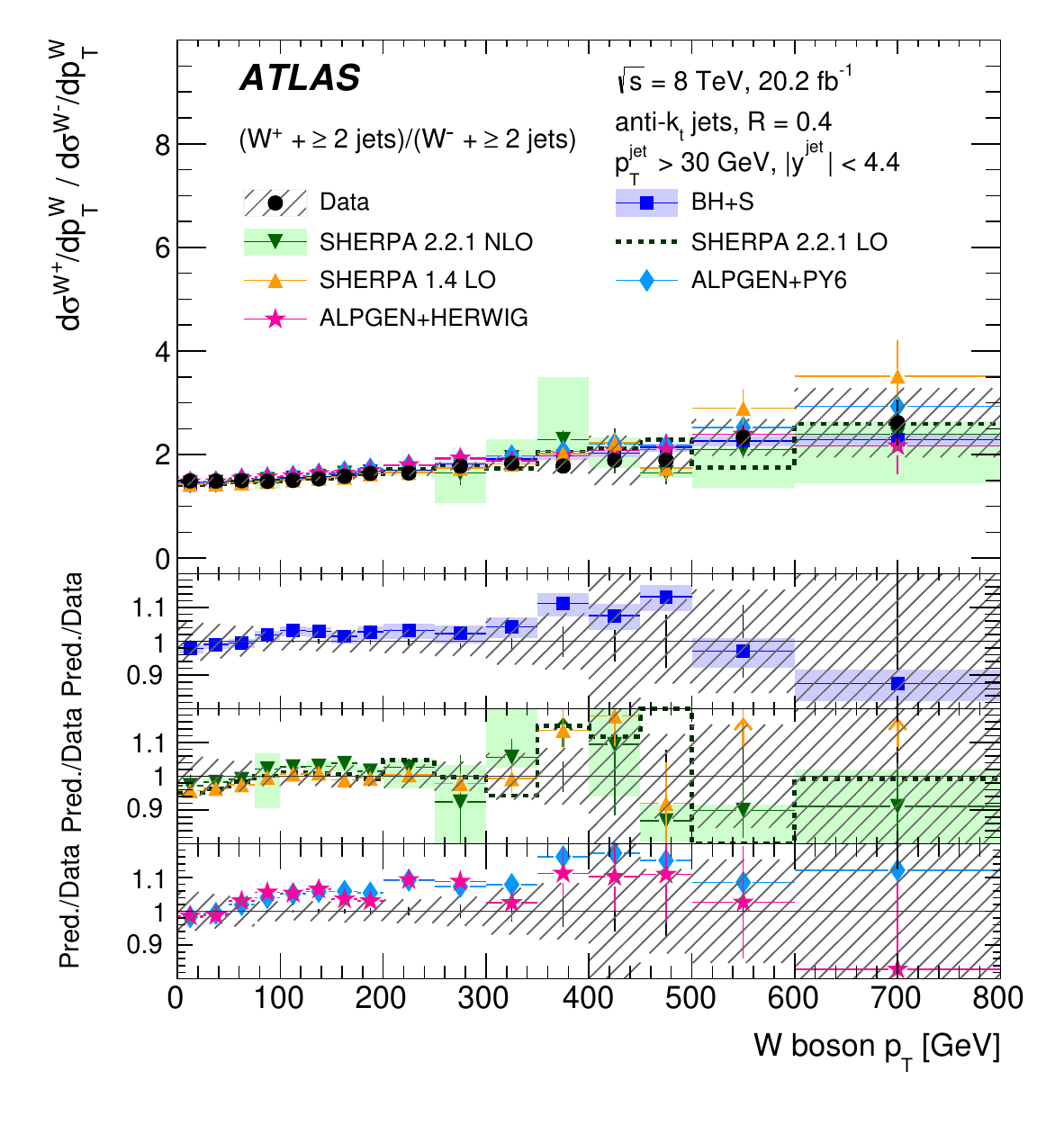} \\ 
\end{tabular} 
\caption{Differential cross sections for the production of \Wboson bosons (left) and the \WpmRatio\ ratio (right) as a function of the \Wboson \pt for events with $\njs \ge 2$. \captionOverflow\ \captionCommon} 
\label{fig:res-wpt-2jet} 
\end{figure*}

\begin{figure*} \centering
\begin{tabular}{cc}
\includegraphics[width=0.45\textwidth]{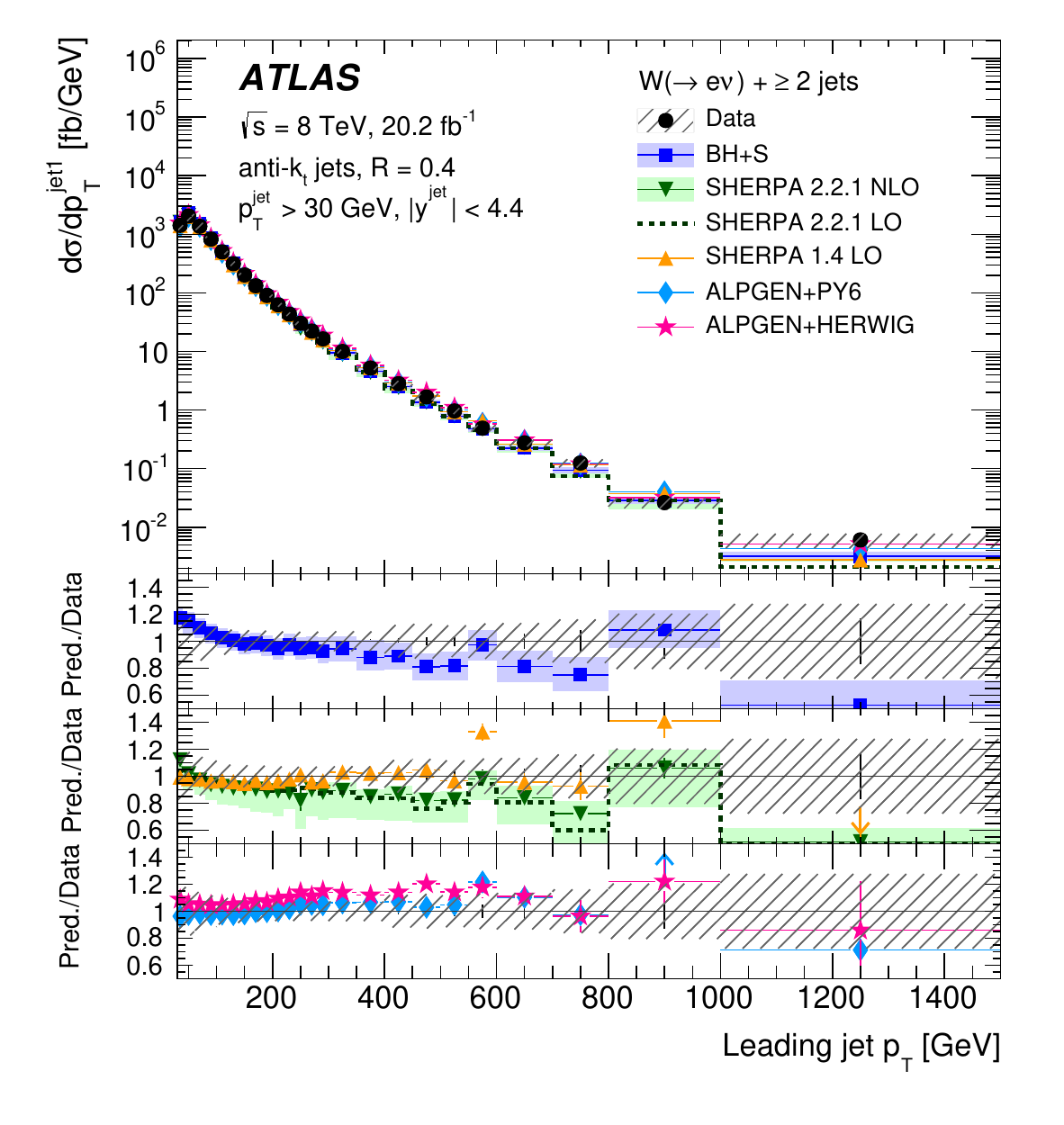} & 
\includegraphics[width=0.45\textwidth]{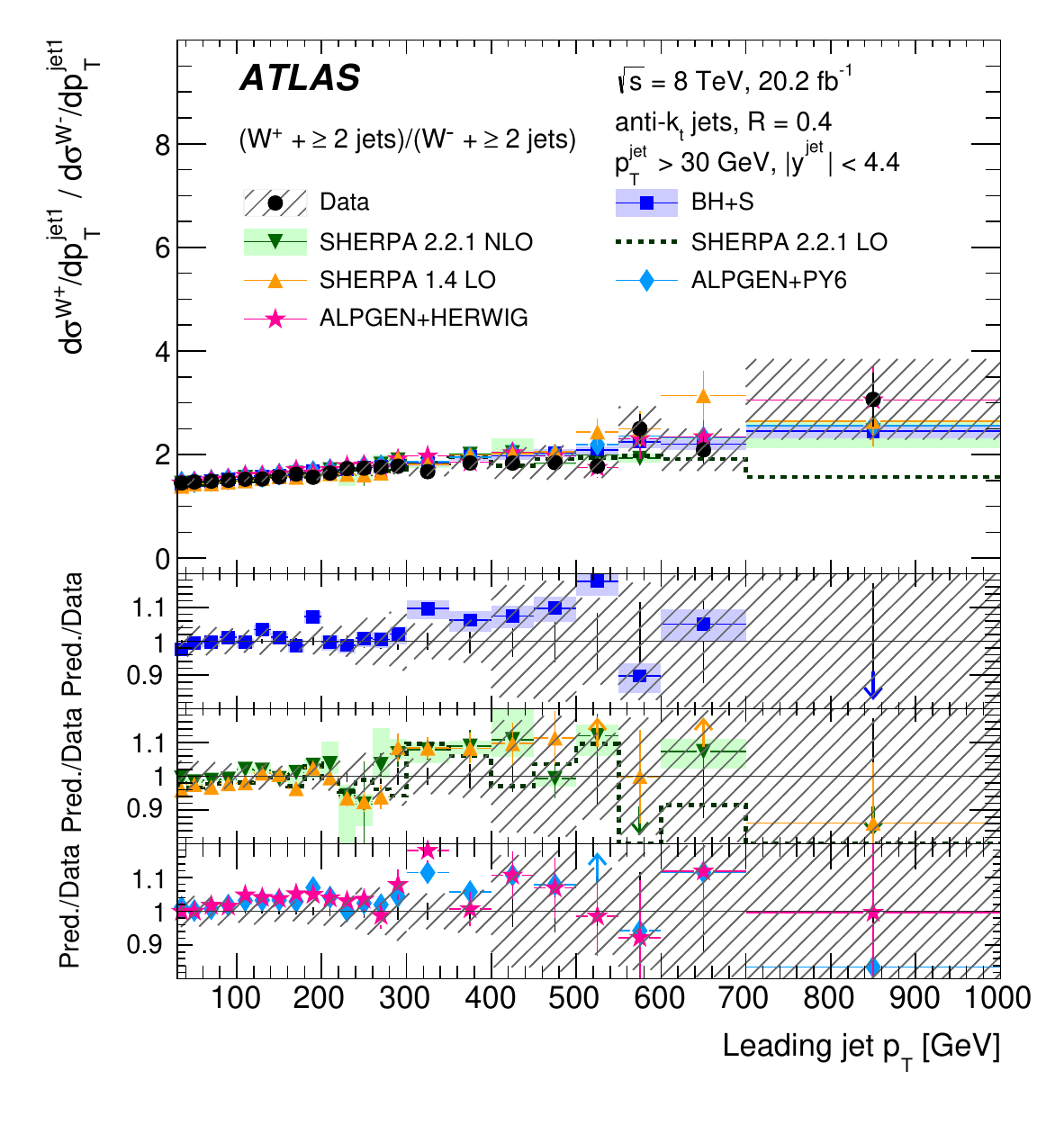} \\
\end{tabular}
\caption{Differential cross sections for the production of \Wboson bosons (left) and the \WpmRatio\ ratio (right) as a function of the leading jet \pt for events with $\njs \ge 2$. \captionOverflow\ \captionCommon }
\label{fig:res-leadjetpt-2jet}
\end{figure*}

The following additional multiplicities are measured:
\begin{itemize}
	\item The exclusive jet multiplicity distribution for \Wboson production (\figref{res-njet-ex}), and
	\item the \Wboson cross section and the \WpmRatio\ ratio as a function of 
\htj\ (\figref{res-ht-2jet}), 
\Wboson \pt (\figref{res-wpt-2jet}), and
the leading jet \pt (\figref{res-leadjetpt-2jet}) for events with $\njs \ge 2$.
\end{itemize}

\subsection{Pseudorapidity of the electron}

\begin{figure*} \centering
\begin{tabular}{cc}
\includegraphics[width=0.45\textwidth]{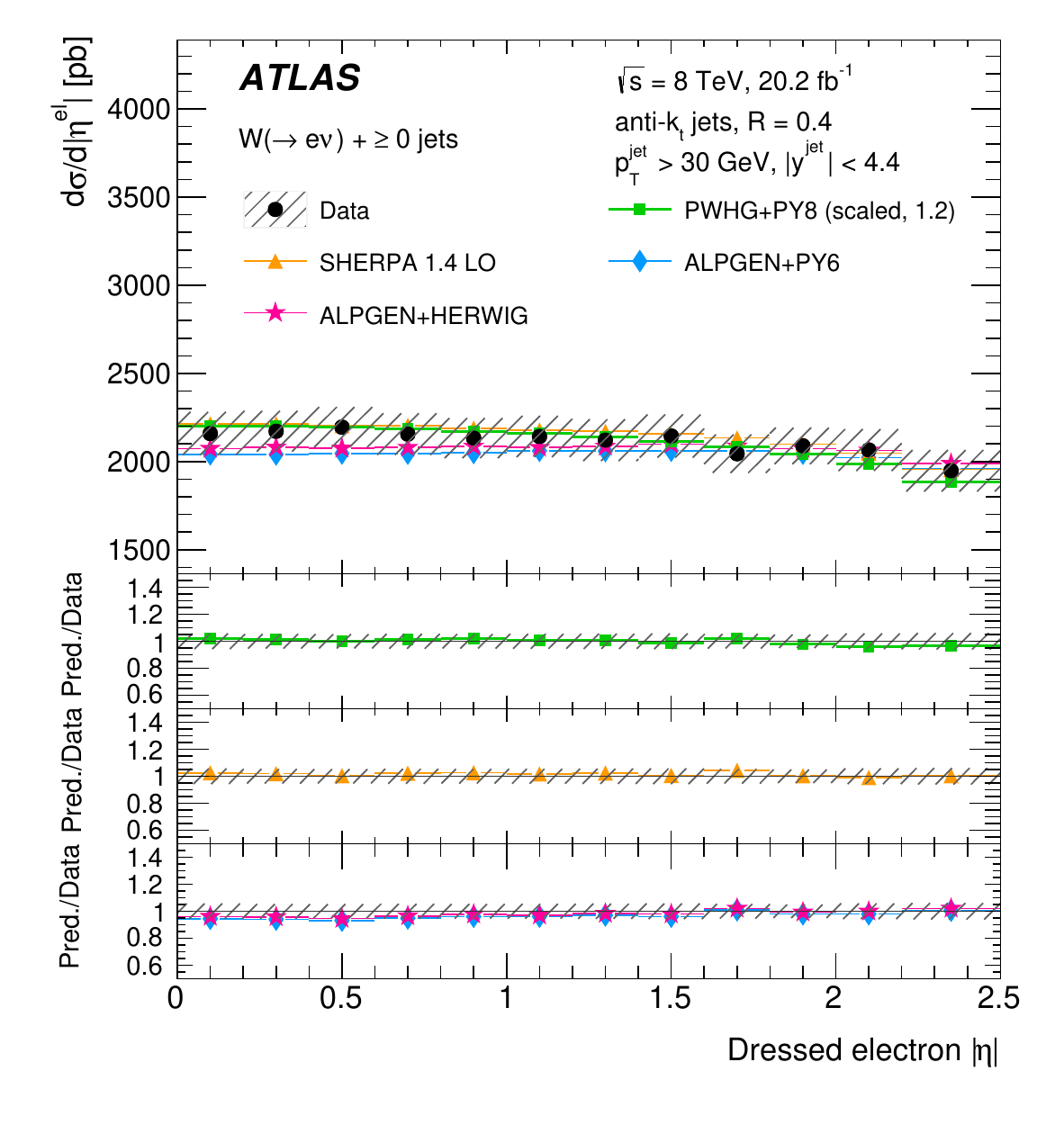} & 
\includegraphics[width=0.45\textwidth]{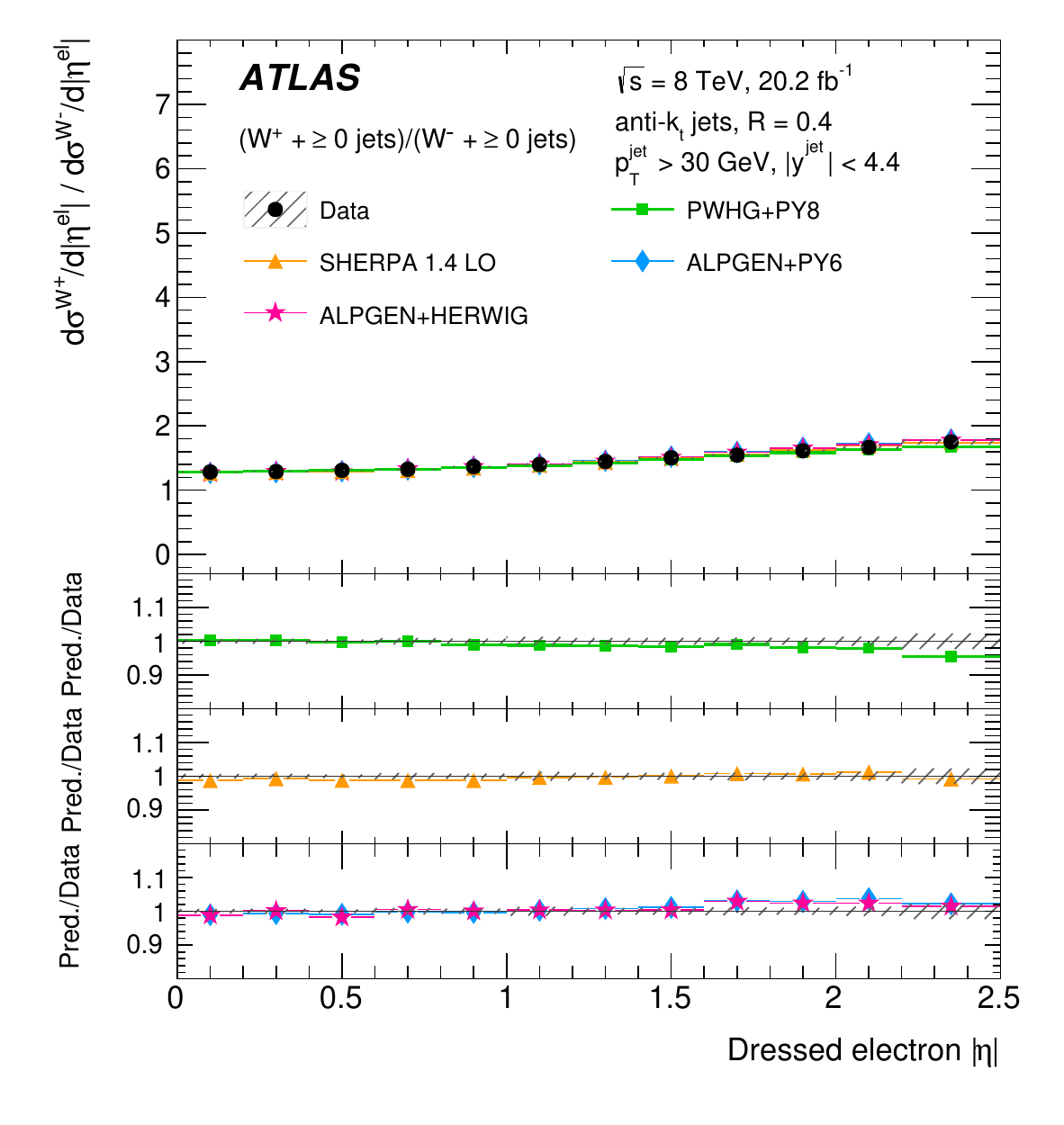} \\
\includegraphics[width=0.45\textwidth]{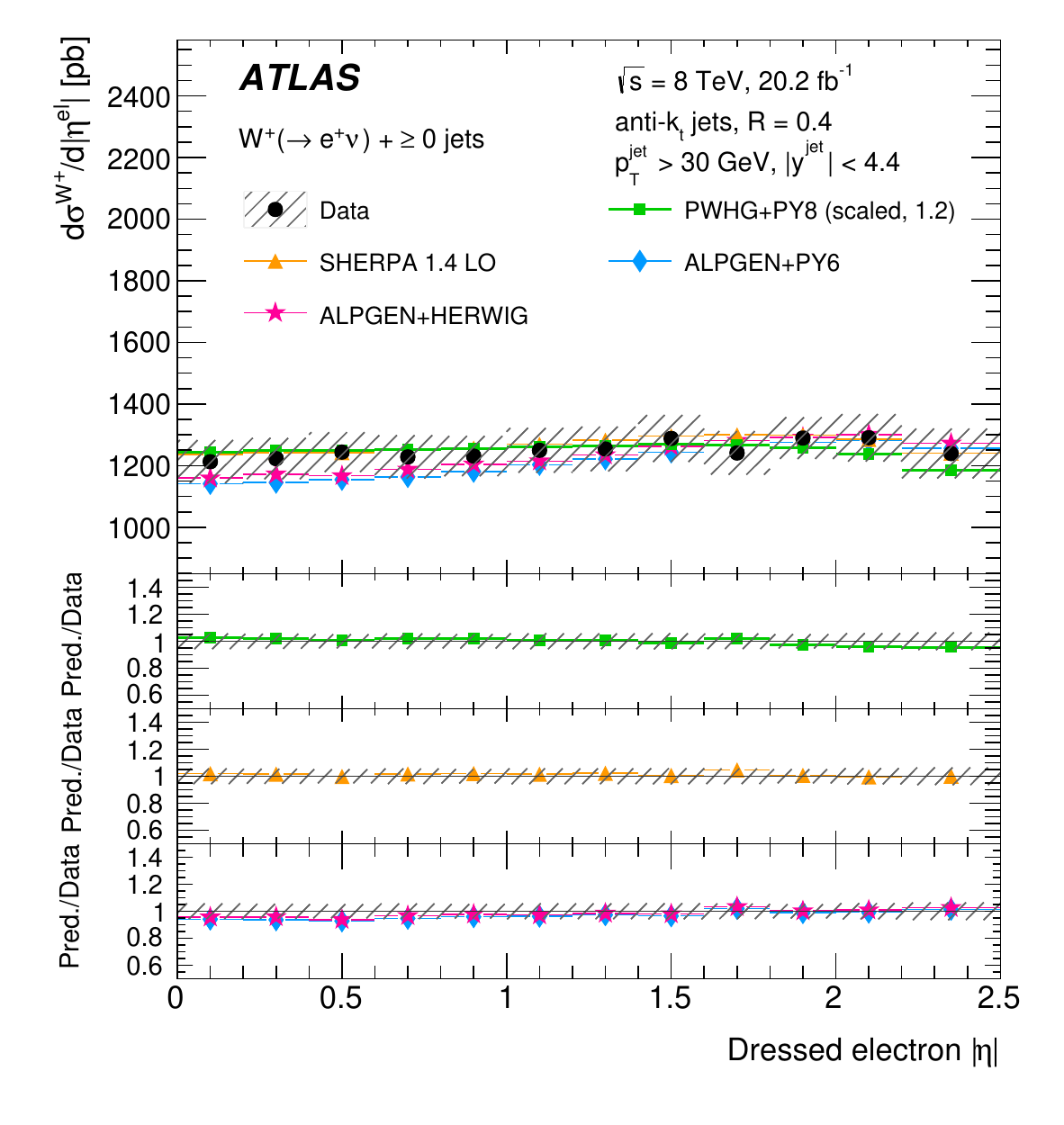} & 
\includegraphics[width=0.45\textwidth]{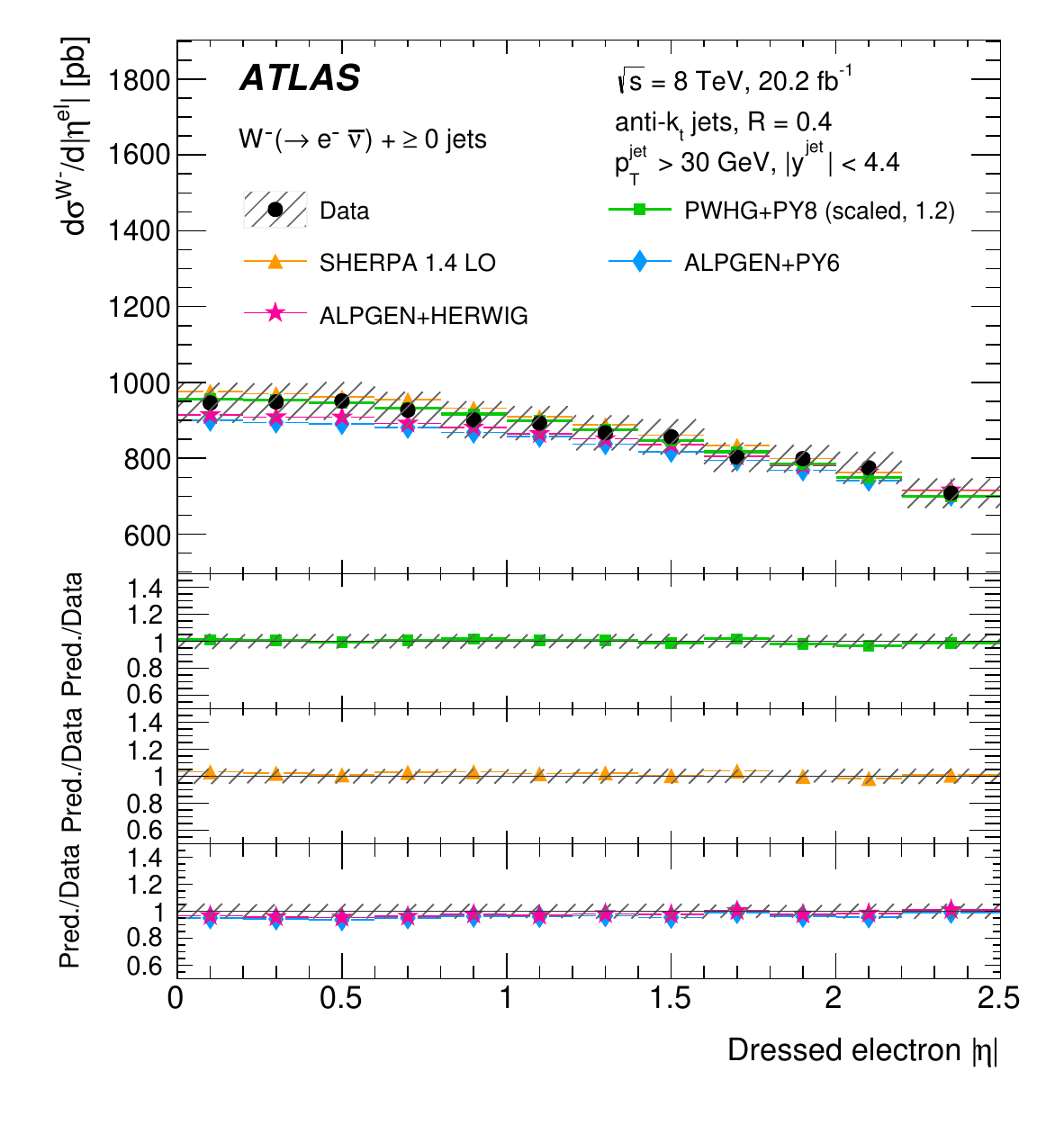} \\
\end{tabular}
\caption{Differential cross sections for the production of \Wboson bosons (top left), \Wplus\ (bottom left), \Wminus\ (bottom right) and the \WpmRatio\ ratio (top right) as a function of the electron $\eta$ for events with $\njs \ge 0$. \captionCommon}
\label{fig:res-eleta-0jet}
\end{figure*}

The \Wboson, \Wplus and \Wminus cross sections and the \WpmRatio\ cross-section ratio as a function of the electron $\eta$ for events with $\njs \ge 0$ and $\njs \ge 1$ are presented in \figsref{res-eleta-0jet}{res-eleta-1jet}.

\subsection{\Wplus and \Wminus cross sections}

\begin{figure*} \centering
\begin{tabular}{cc}
\includegraphics[width=0.45\textwidth]{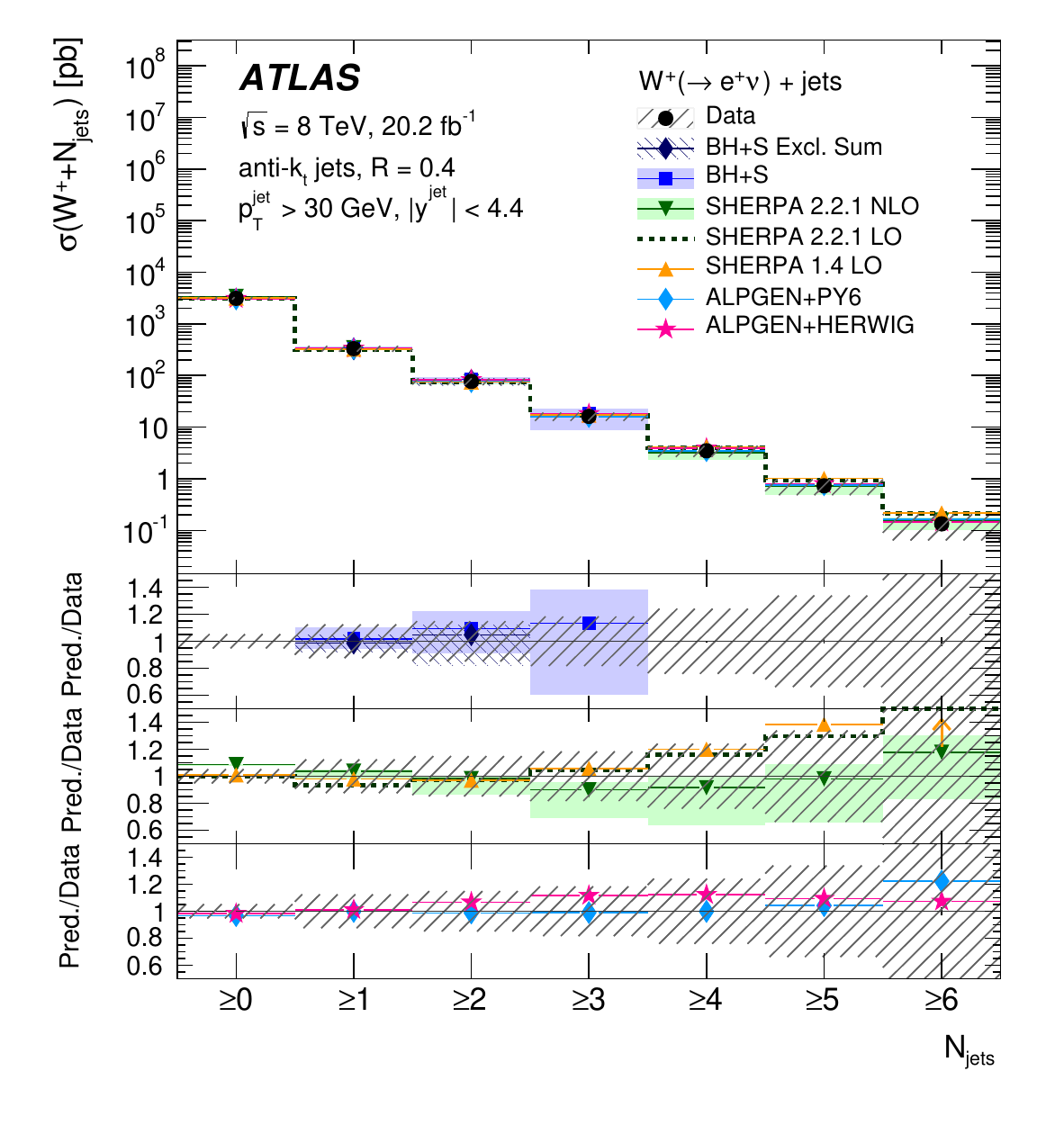} & 
\includegraphics[width=0.45\textwidth]{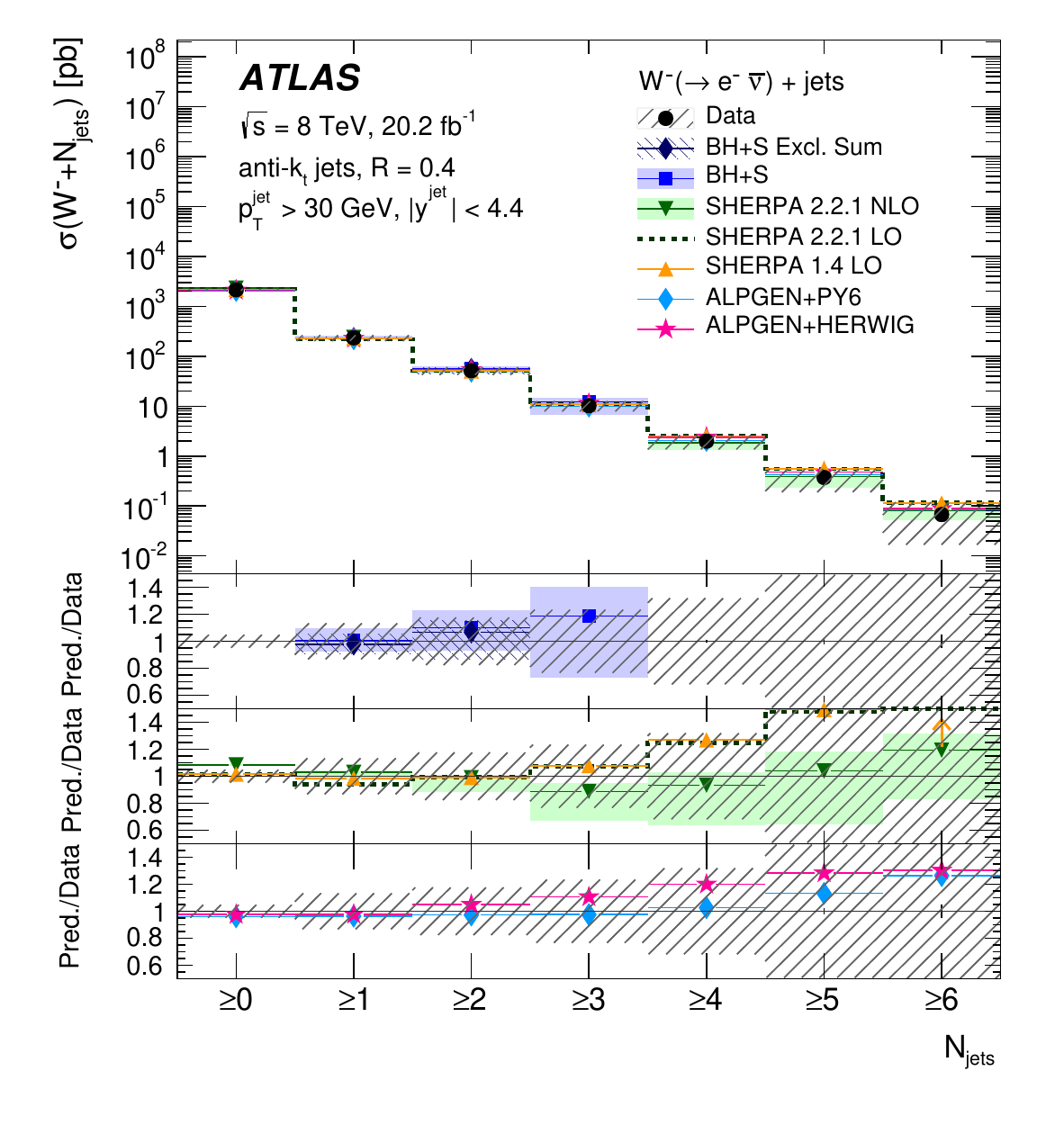} \\
\end{tabular}
\caption{Differential cross sections for the production of \Wplus\ (left) and \Wminus\ (right) as a function of the inclusive jet multiplicity. \captionCommon }
\label{fig:res-njet-ch}
\end{figure*}

\begin{figure*} \centering
\begin{tabular}{cc}
\includegraphics[width=0.45\textwidth]{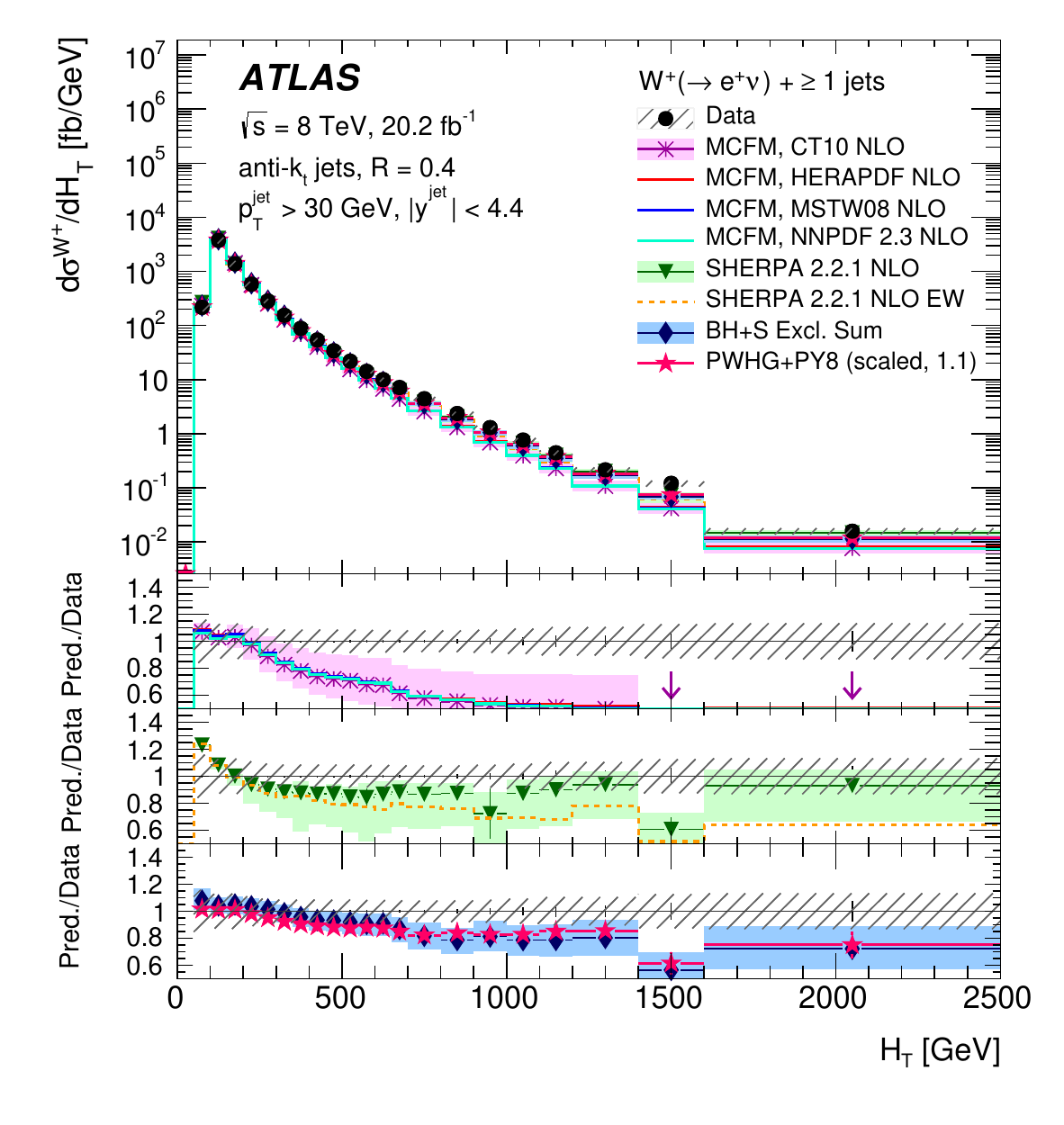} & 
\includegraphics[width=0.45\textwidth]{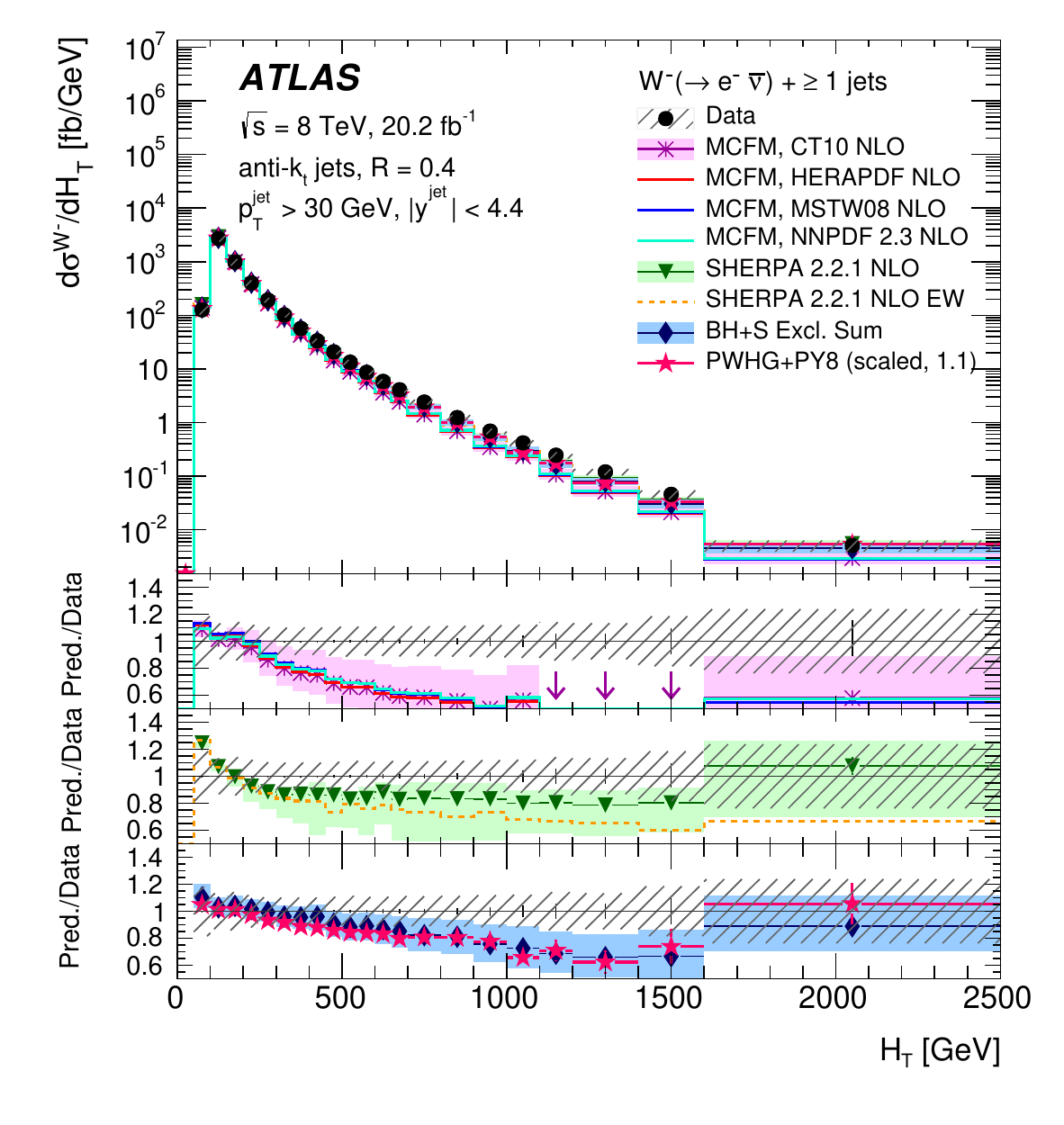} \\
\end{tabular}
\caption{Differential cross sections for the production of \Wplus\ (left) and \Wminus\ (right) as a function of the \HT for events with $\njs \ge 1$. \captionCommon }
\label{fig:res-app-HT-1jet-ch}
\end{figure*}

\begin{figure*} \centering
\begin{tabular}{cc}
\includegraphics[width=0.45\textwidth]{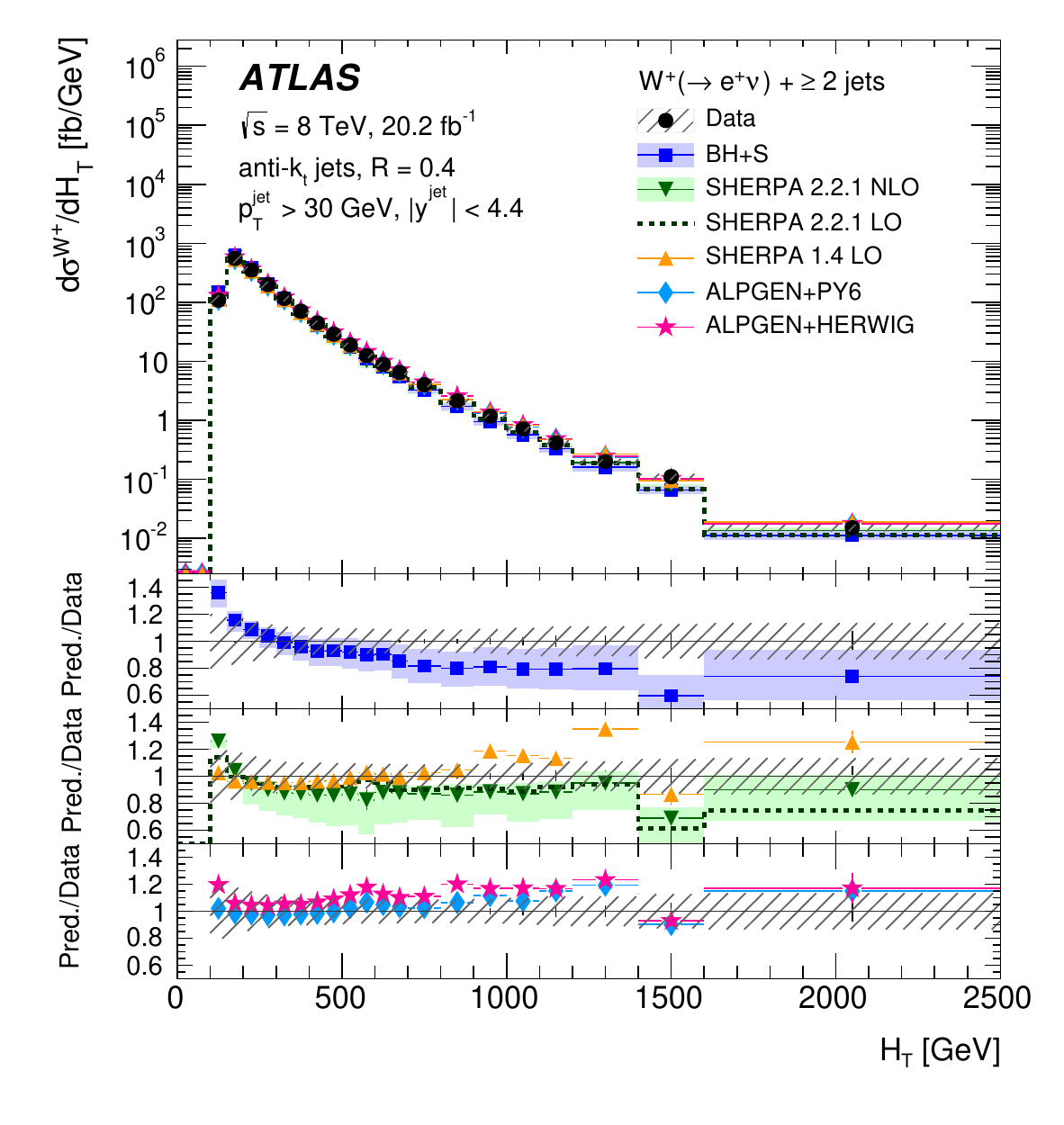} & 
\includegraphics[width=0.45\textwidth]{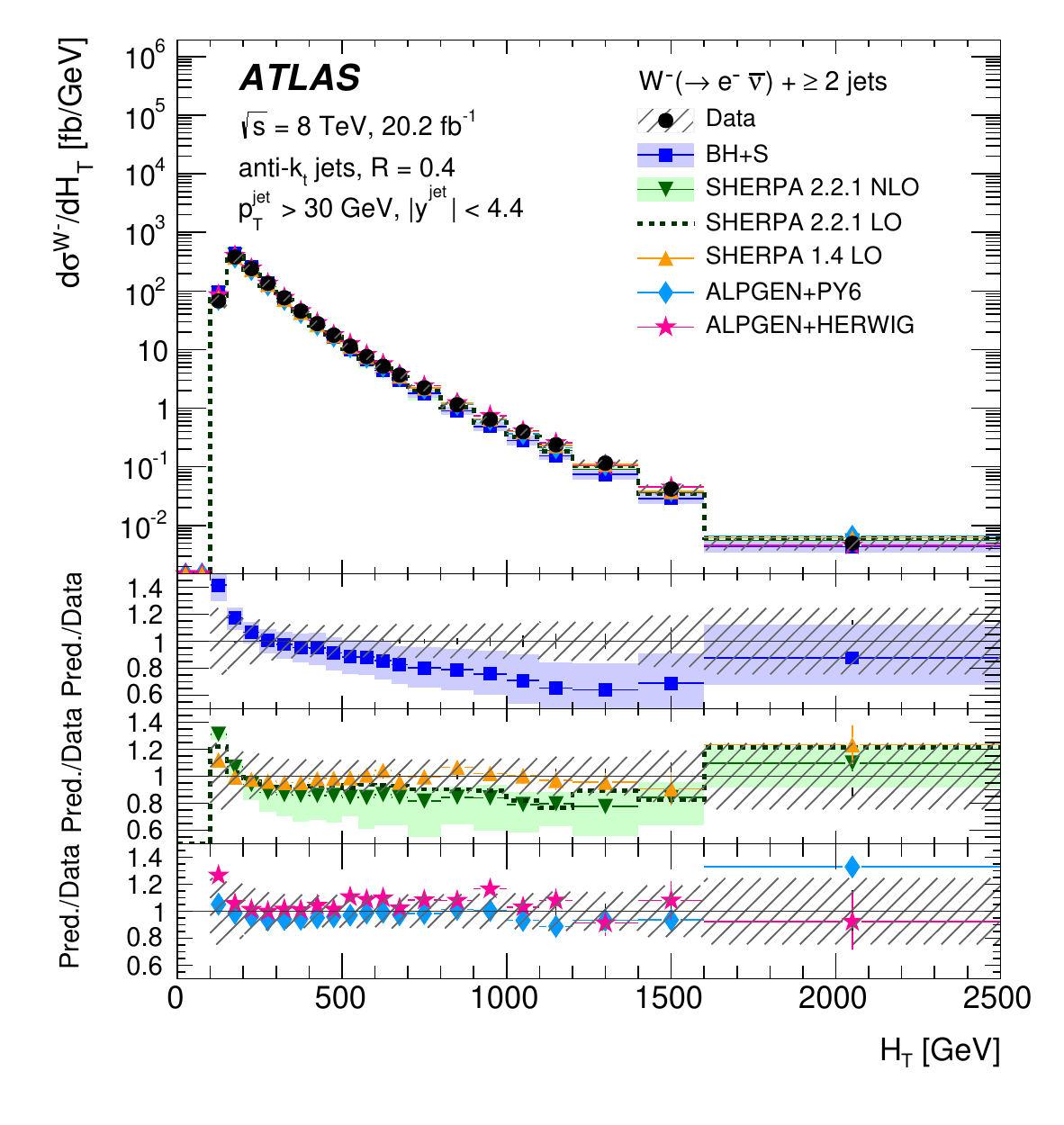} \\
\end{tabular}
\caption{Differential cross sections for the production of \Wplus\ (left) and \Wminus\ (right) as a function of the \htj\ for events with $\njs \ge 2$. \captionCommon }
\label{fig:res-ht-2jet-ch}
\end{figure*}

The \Wplus and \Wminus cross sections, which have been used to calculate the \WpmRatio cross-section ratio distributions shown before, are given for the following jet multiplicities: 
\begin{itemize}
	\item In the presence of at least one jet, the default set of predictions (\figsref{res-njet-ch}{res-leadjetrap-1jet-ch}) and the \MCFM predictions with different PDF sets (\figsref{res-app-HT-1jet-ch}{res-app-leadjetrap-1jet-ch}), corresponding to the figures shown in \secsref{Wresults-njet}{Wresults-1jet}, and
	\item in the presence of at least two jets, the default set of predictions (\figsref{res-ht-2jet-ch}{res-leadjetpt-2jet-ch}), corresponding to \figsref{res-ht-2jet}{res-leadjetpt-2jet} shown in Appendix~\ref{app:2-jet}.
\end{itemize}

% plots for njet and 2 jets

\begin{figure*} 
\centering 
\begin{tabular}{cc} 
\includegraphics[width=0.45\textwidth]{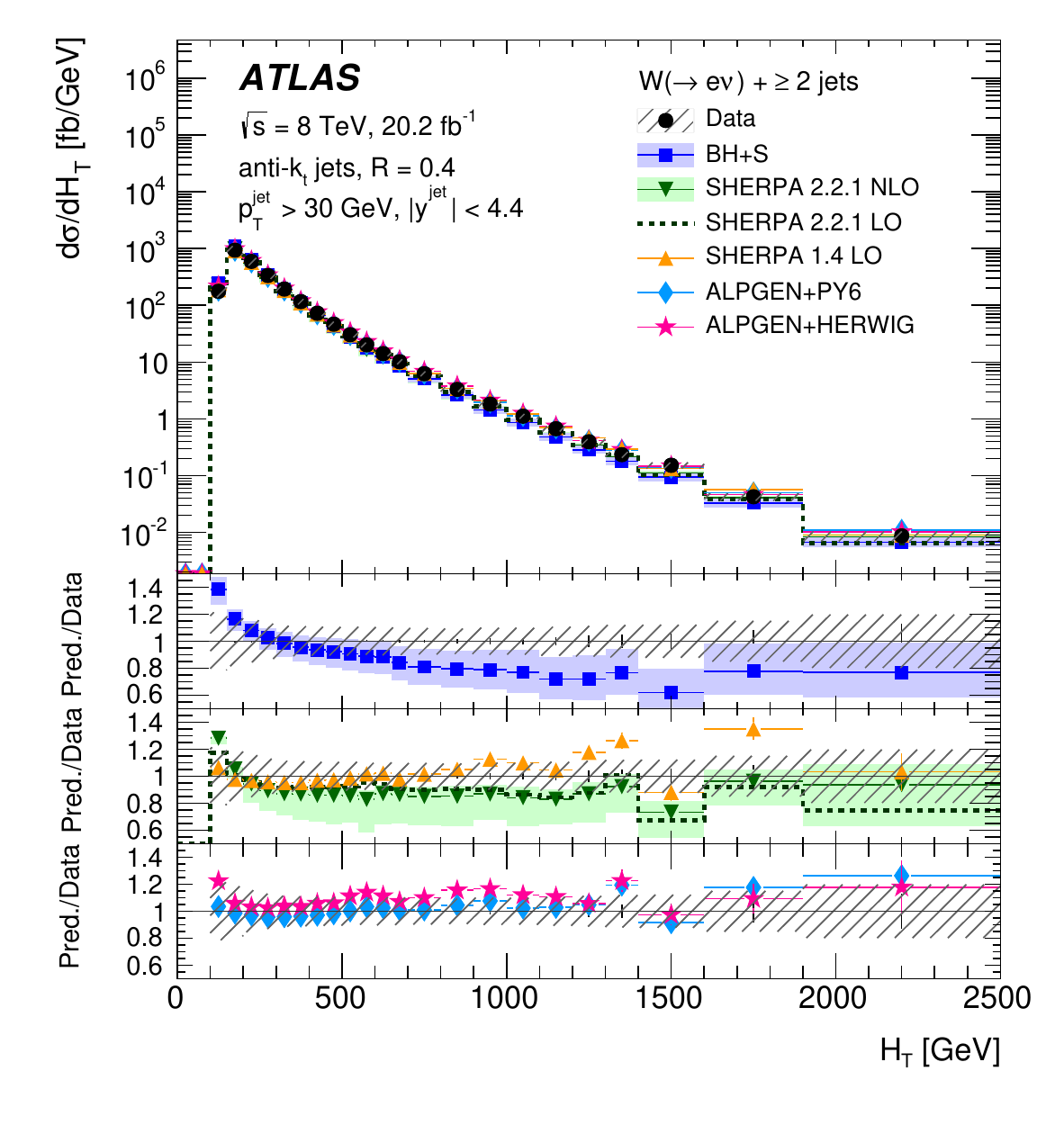} & 
\includegraphics[width=0.45\textwidth]{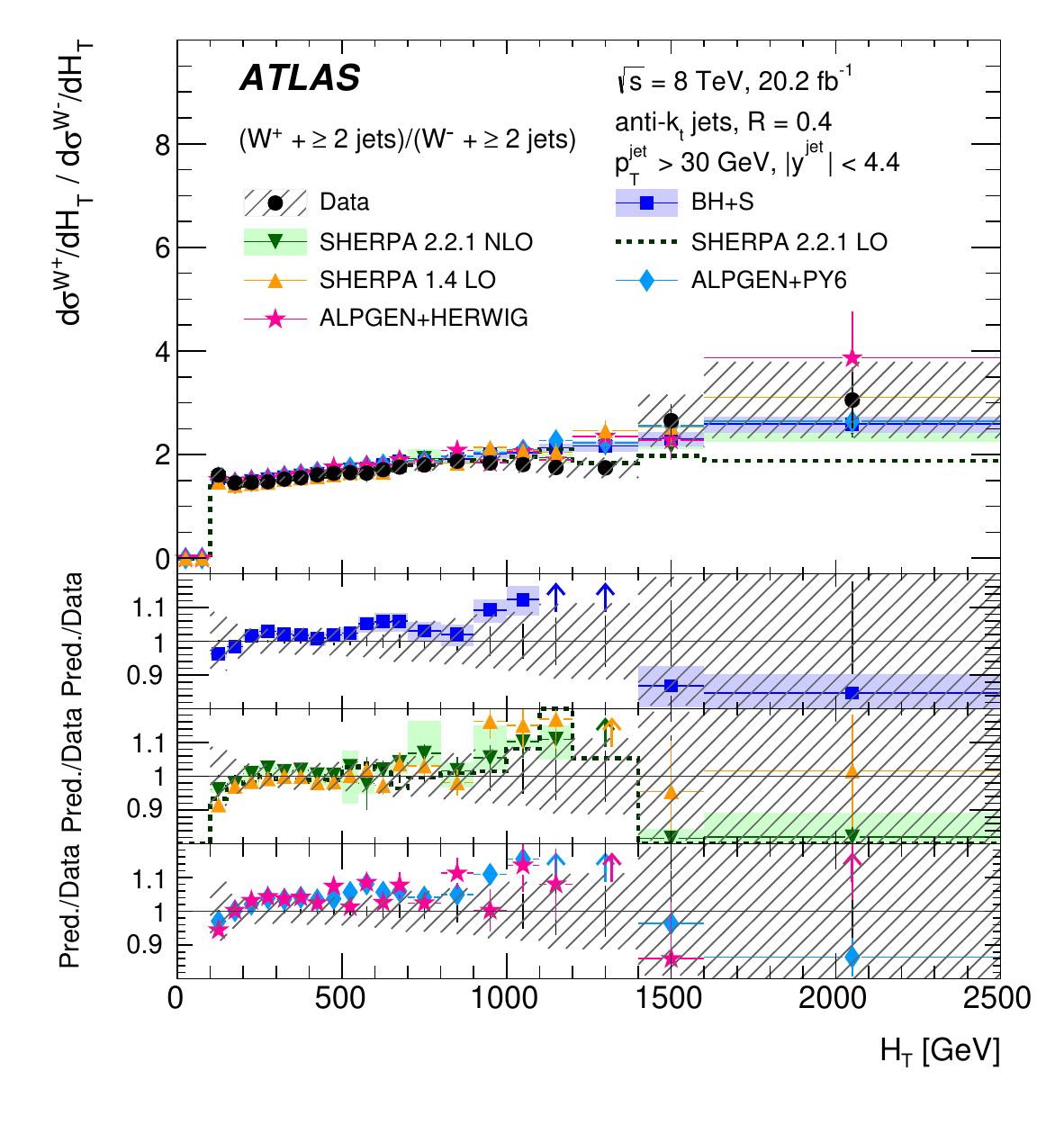} \\ 
\end{tabular} 
\caption{Differential cross sections for the production of \Wboson bosons (left) and the \WpmRatio\ ratio (right) as a function of the \htj\ for events with $\njs \ge 2$. \captionOverflow\ \captionCommon} 
\label{fig:res-ht-2jet} 
\end{figure*} 

% electron eta plots

\begin{figure*} \centering
\begin{tabular}{cc}
\includegraphics[width=0.45\textwidth]{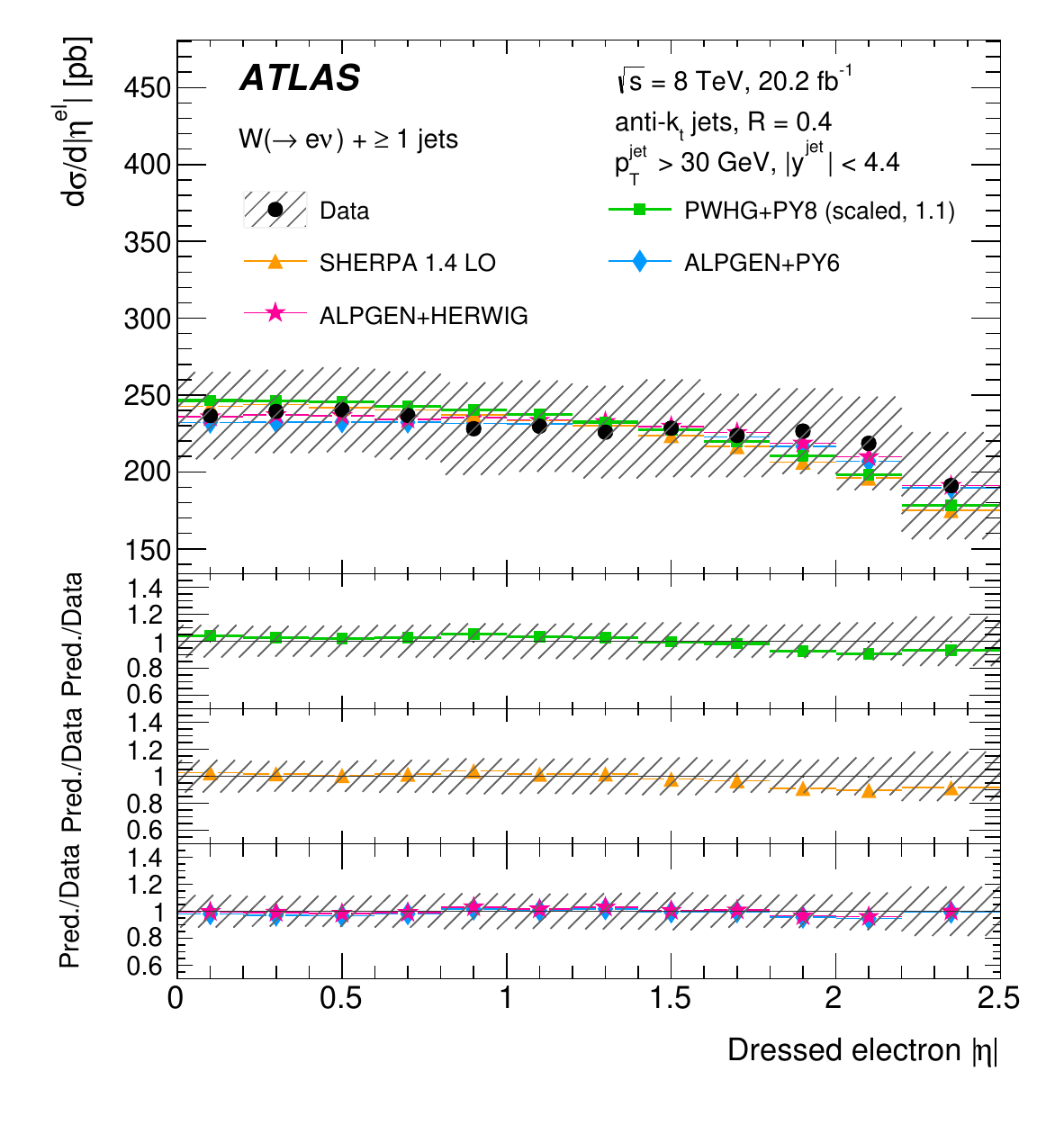} & 
\includegraphics[width=0.45\textwidth]{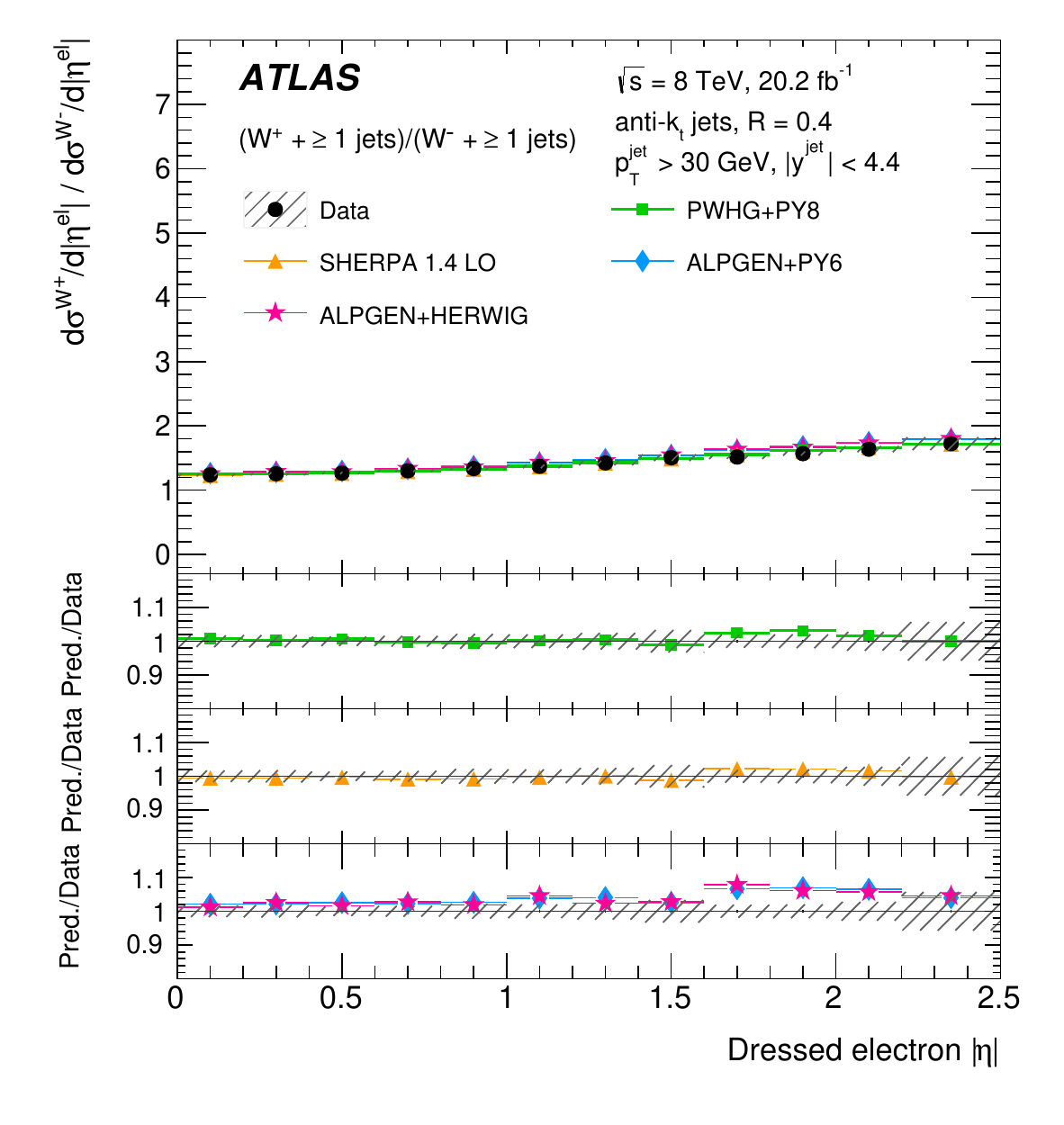} \\
\includegraphics[width=0.45\textwidth]{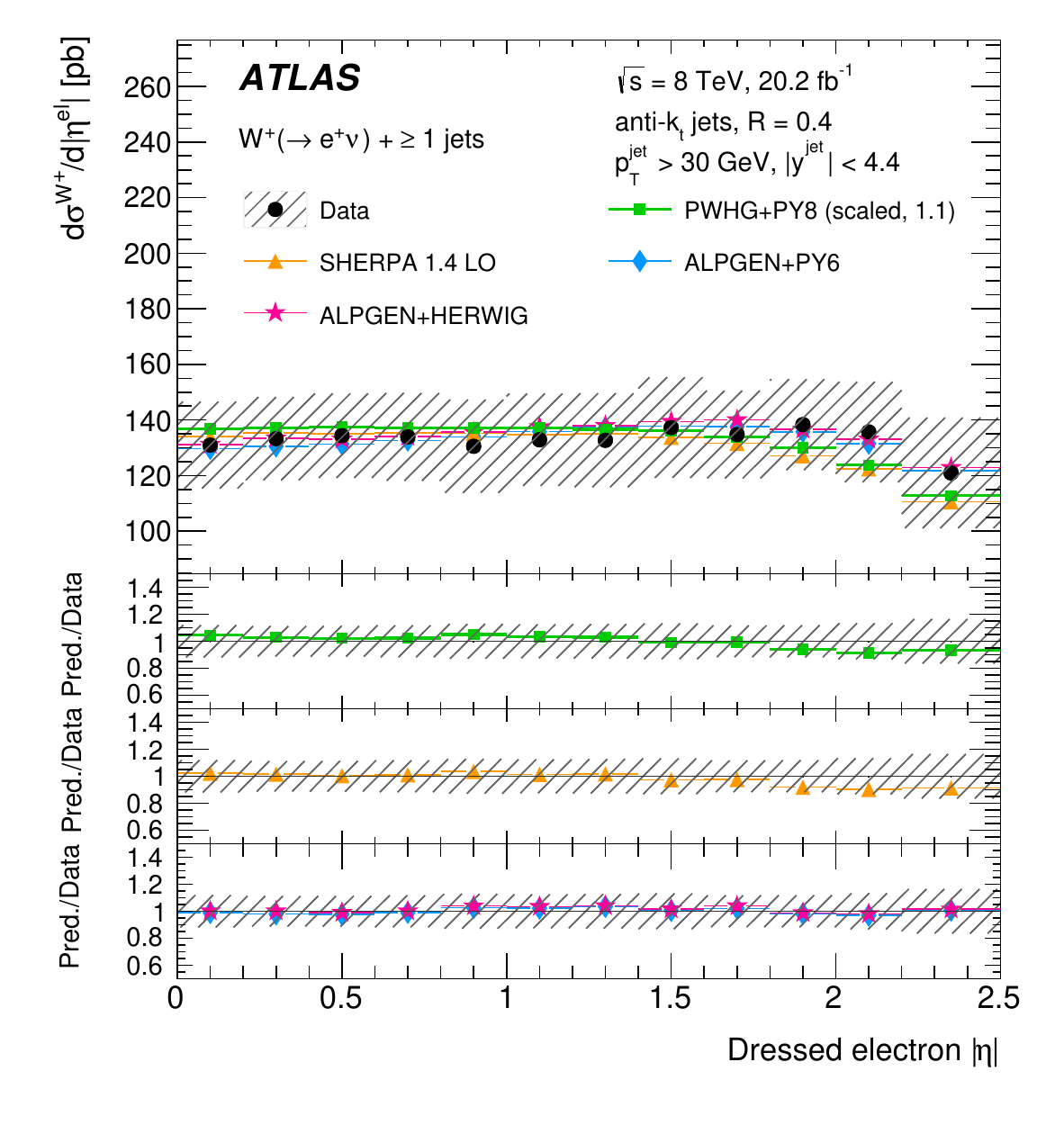} & 
\includegraphics[width=0.45\textwidth]{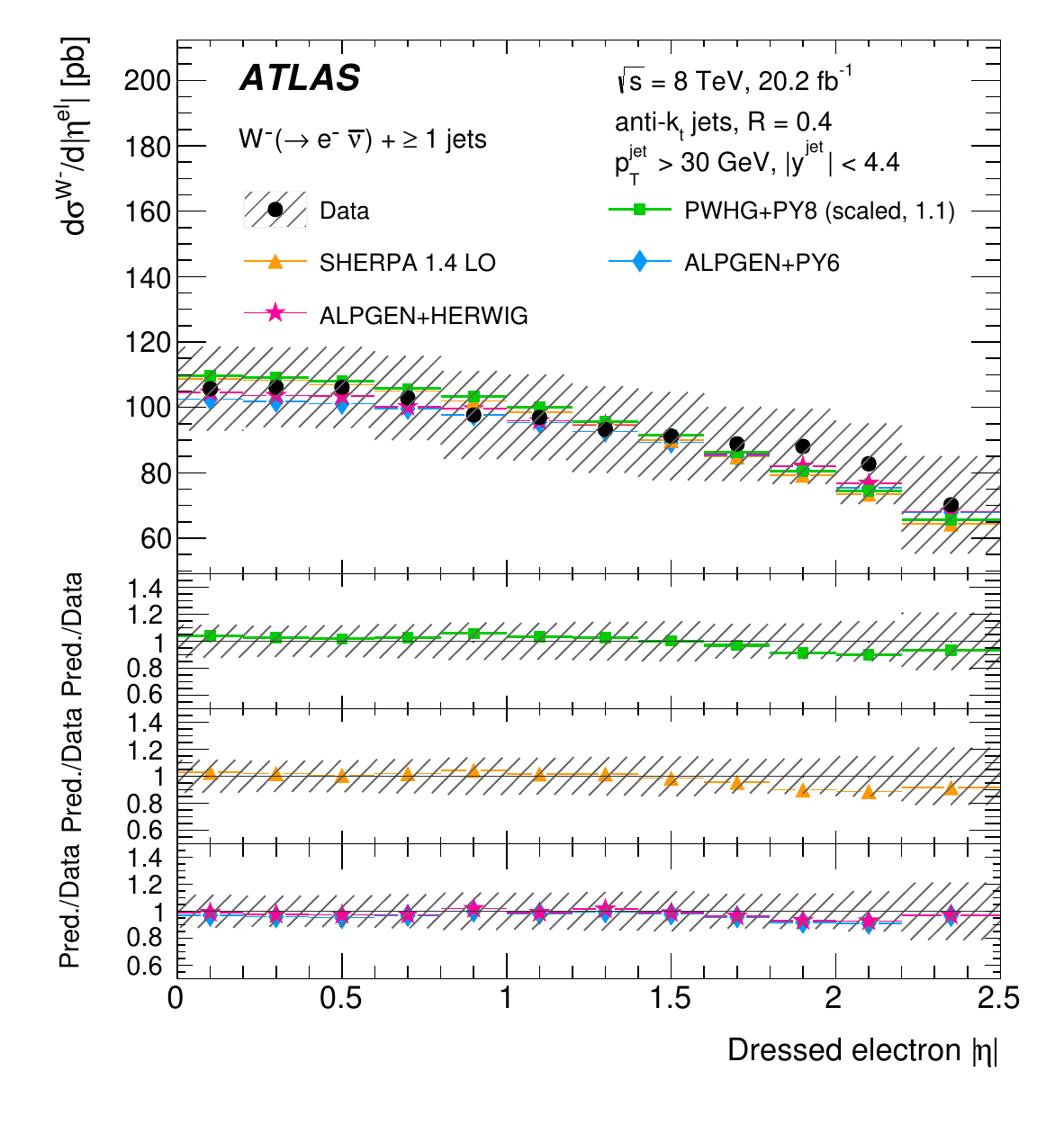} \\
\end{tabular}
\caption{Differential cross sections for the production of \Wboson bosons (top left), \Wplus\ (bottom left), \Wminus\ (bottom right) and the \WpmRatio\ ratio (top right) as a function of the electron $\eta$ for events with $\njs \ge 1$. \captionCommon}
\label{fig:res-eleta-1jet}
\end{figure*}

% W+, W- for main body plots

\begin{figure*} \centering
\begin{tabular}{cc}
\includegraphics[width=0.45\textwidth]{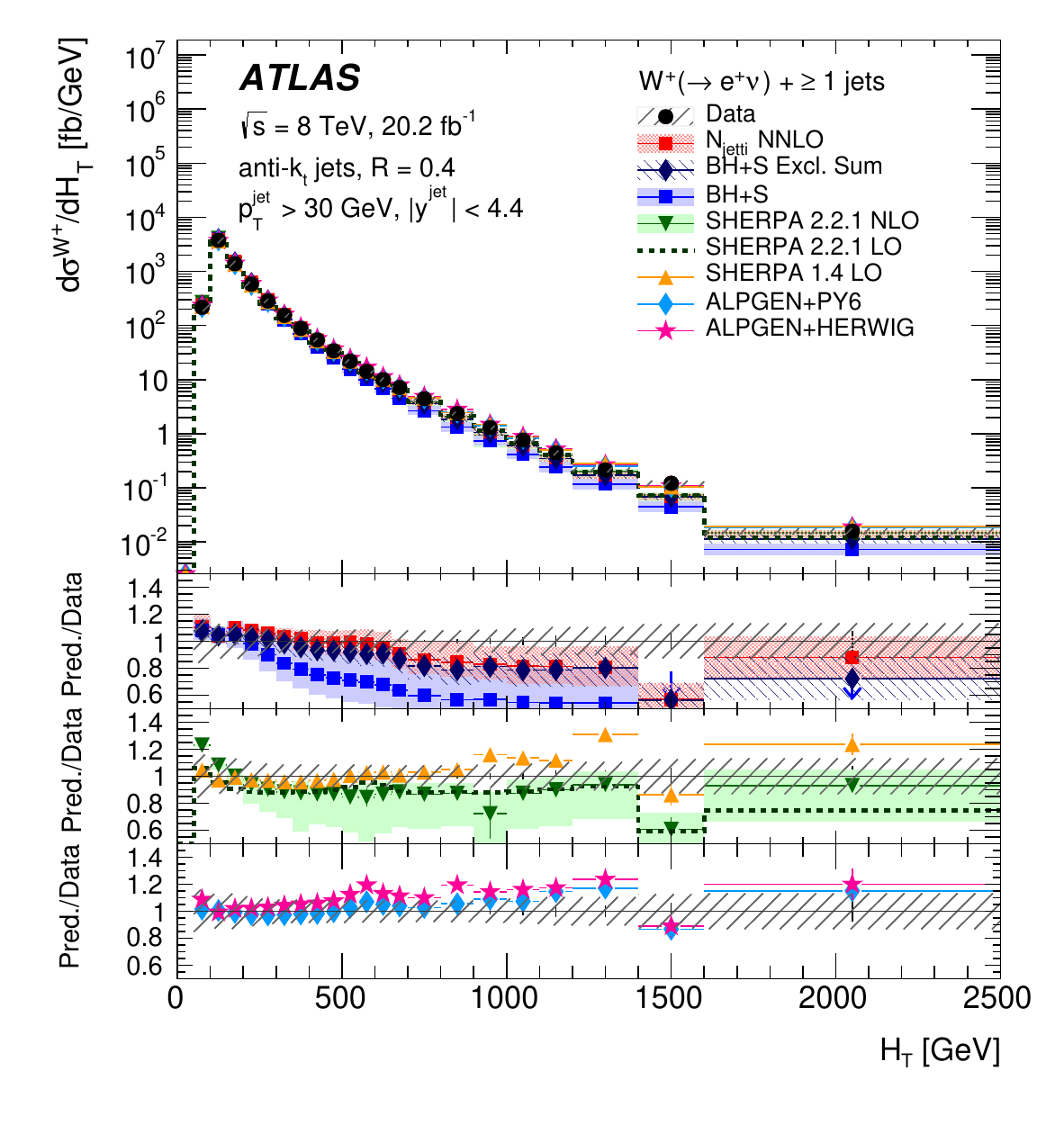} & 
\includegraphics[width=0.45\textwidth]{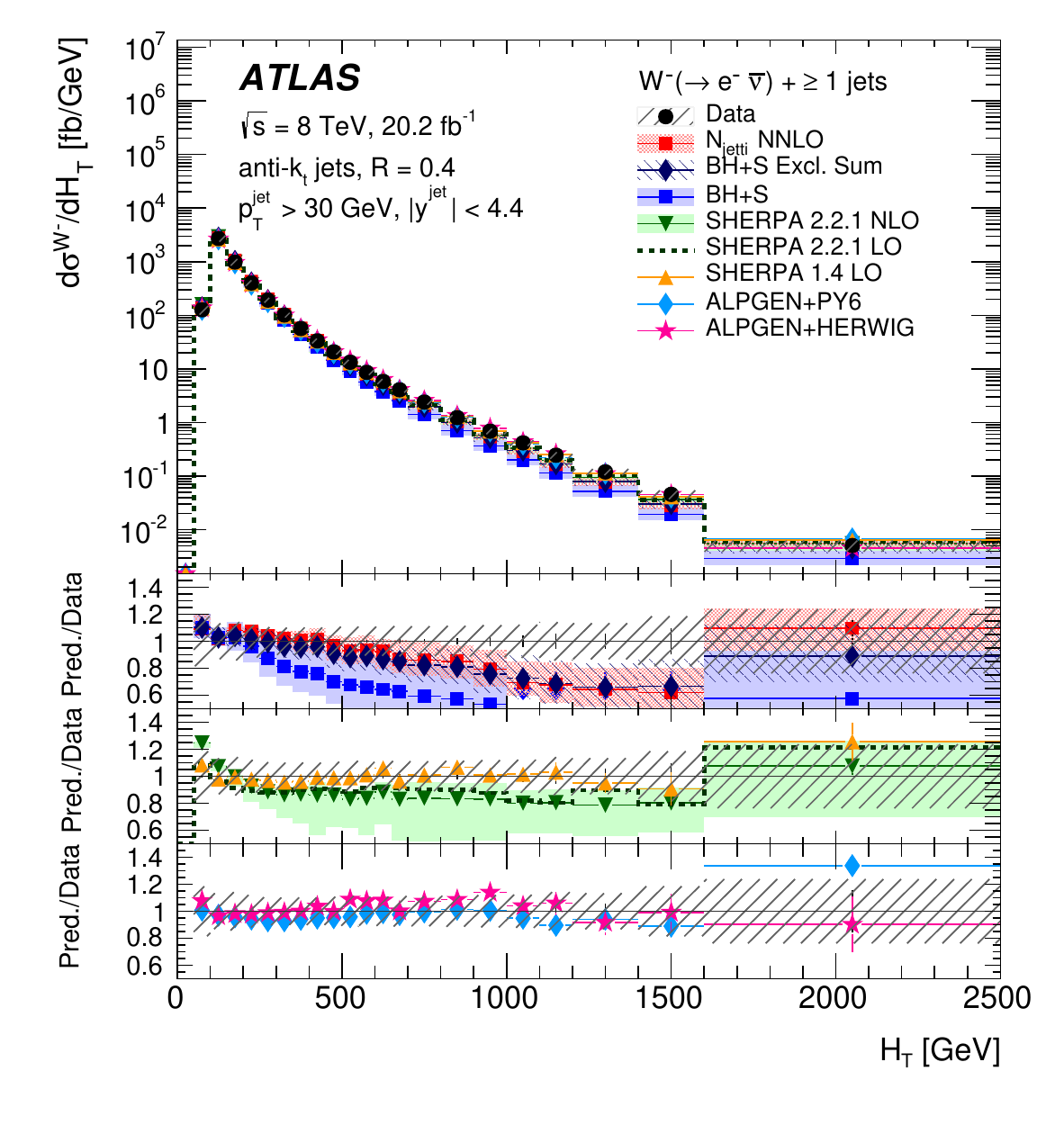} \\
\end{tabular}
\caption{Differential cross sections for the production of \Wplus\ (left) and \Wminus\ (right) as a function of the \htj\ for events with $\njs \ge 1$. \captionCommon }
\label{fig:res-ht-1jet-ch}
\end{figure*}

\begin{figure*} \centering
\begin{tabular}{cc}
\includegraphics[width=0.45\textwidth]{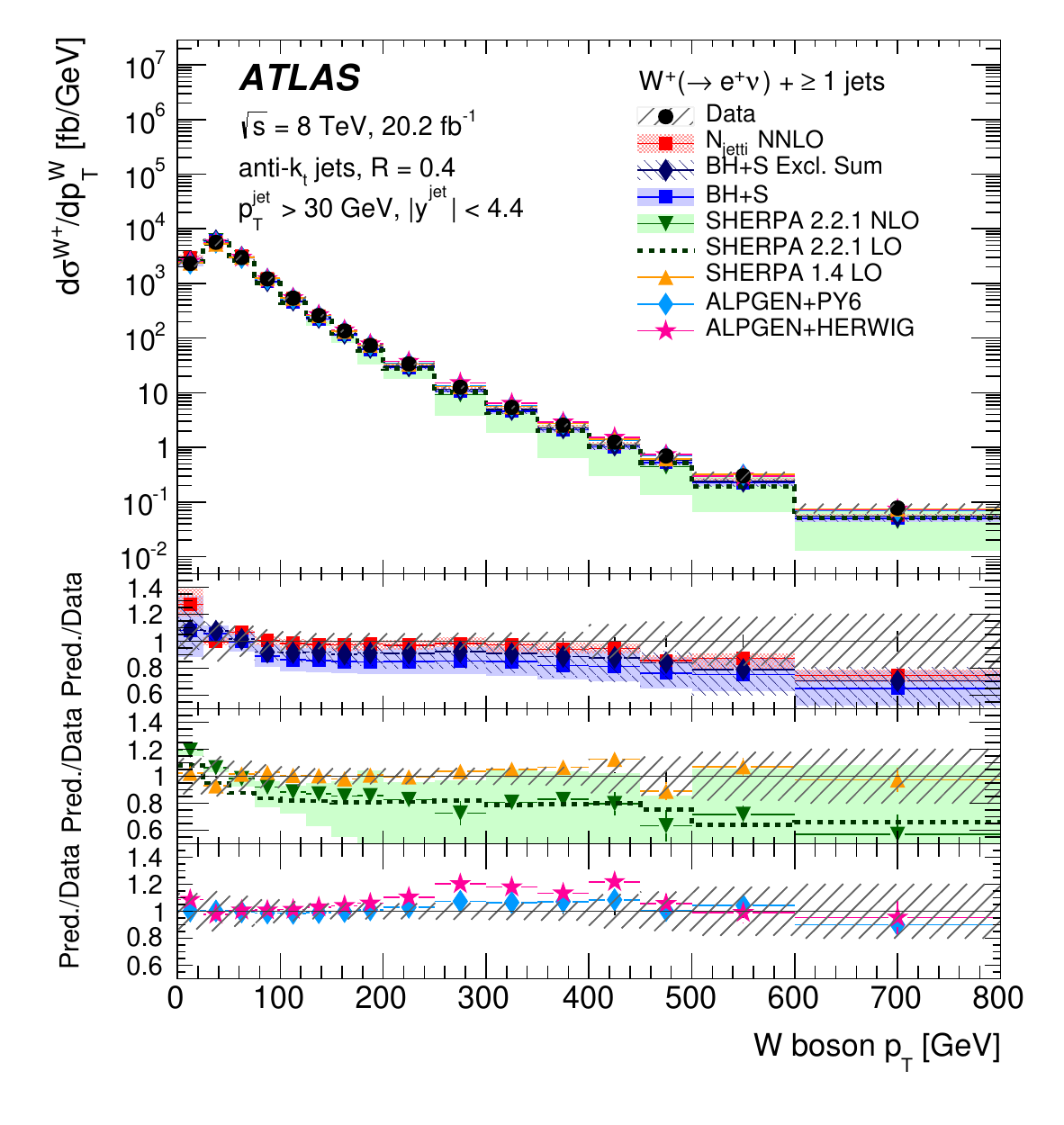} & 
\includegraphics[width=0.45\textwidth]{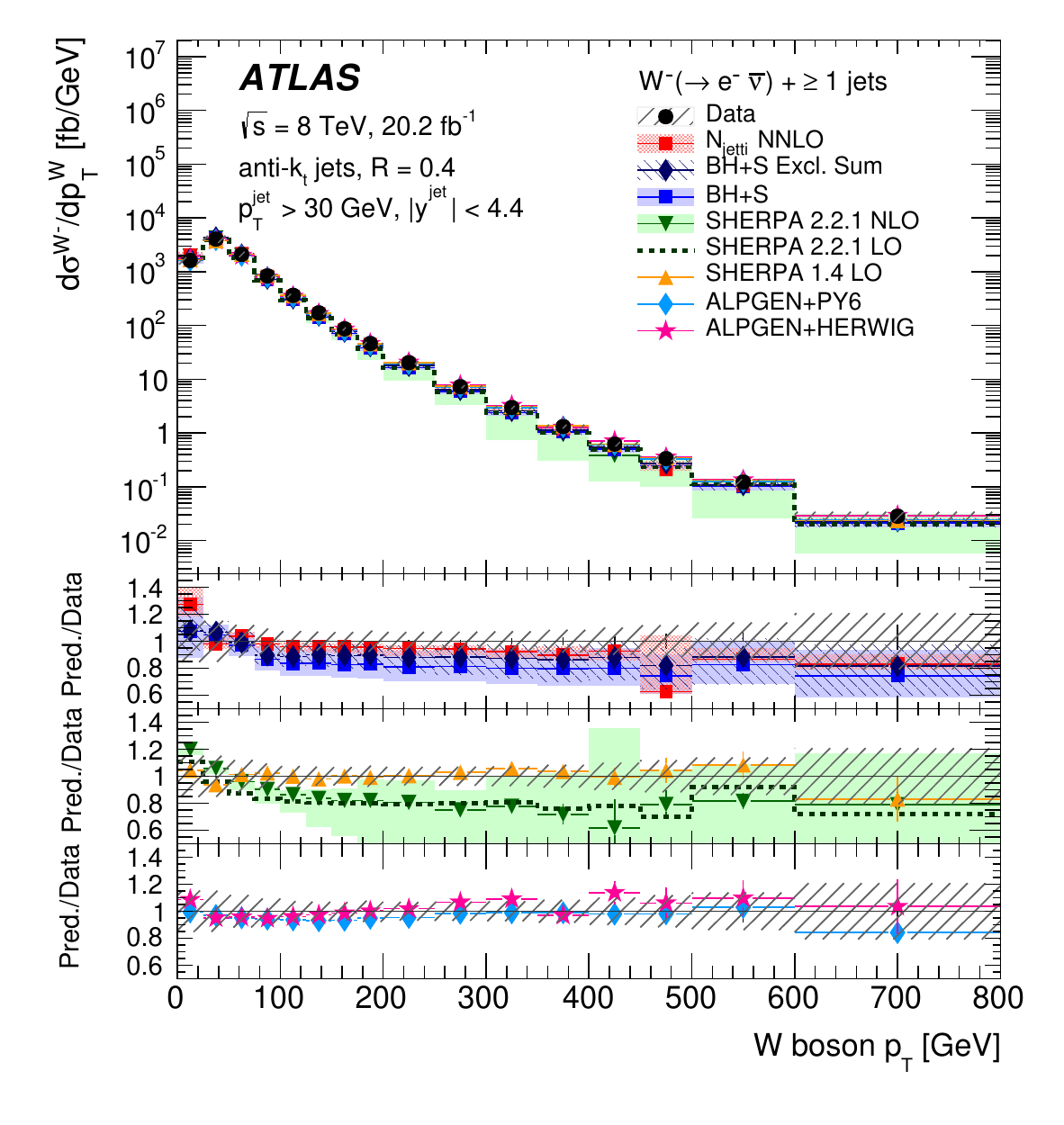} \\
\end{tabular}
\caption{Differential cross sections for the production of \Wplus\ (left) and \Wminus\ (right) as a function of the \Wboson \pt for events with $\njs \ge 1$. \captionCommon }
\label{fig:res-wpt-1jet-ch}
\end{figure*}

\begin{figure*} \centering
\begin{tabular}{cc}
\includegraphics[width=0.45\textwidth]{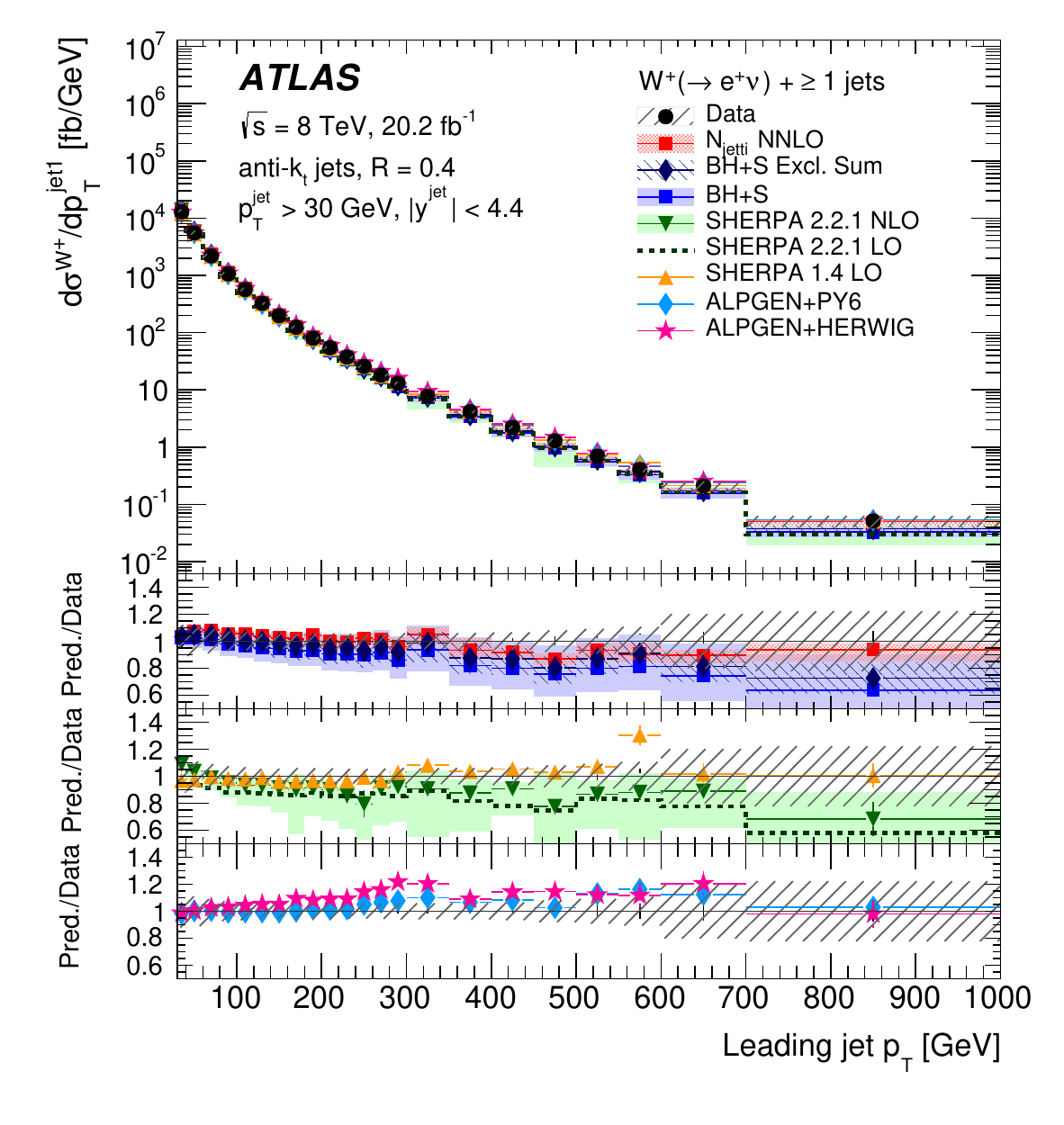} & 
\includegraphics[width=0.45\textwidth]{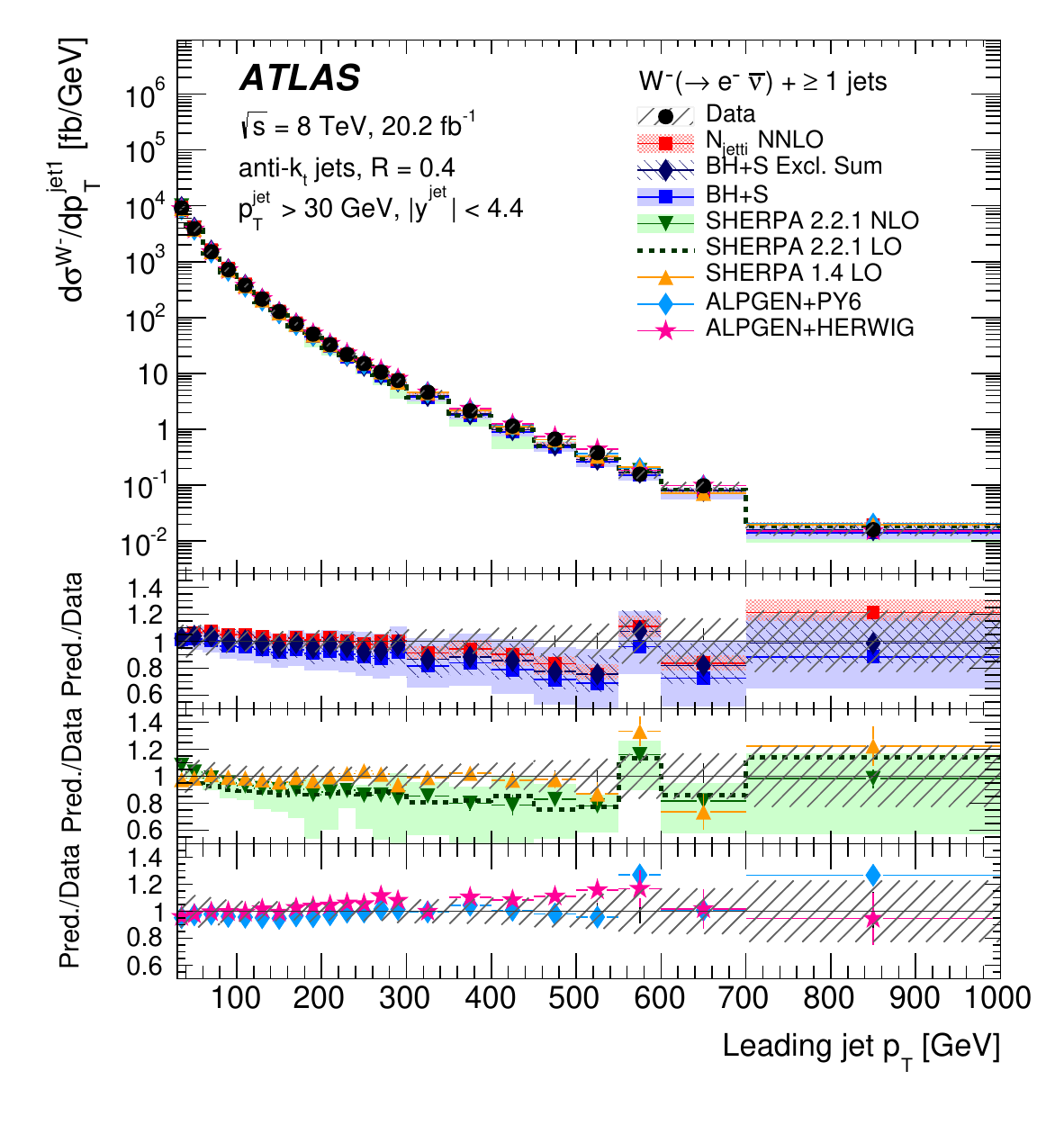} \\
\end{tabular}
\caption{Differential cross sections for the production of \Wplus\ (left) and \Wminus\ (right) as a function of the leading jet \pt for events with $\njs \ge 1$. \captionCommon }
\label{fig:res-leadjetpt-1jet-ch}
\end{figure*}

\begin{figure*} \centering
\begin{tabular}{cc}
\includegraphics[width=0.45\textwidth]{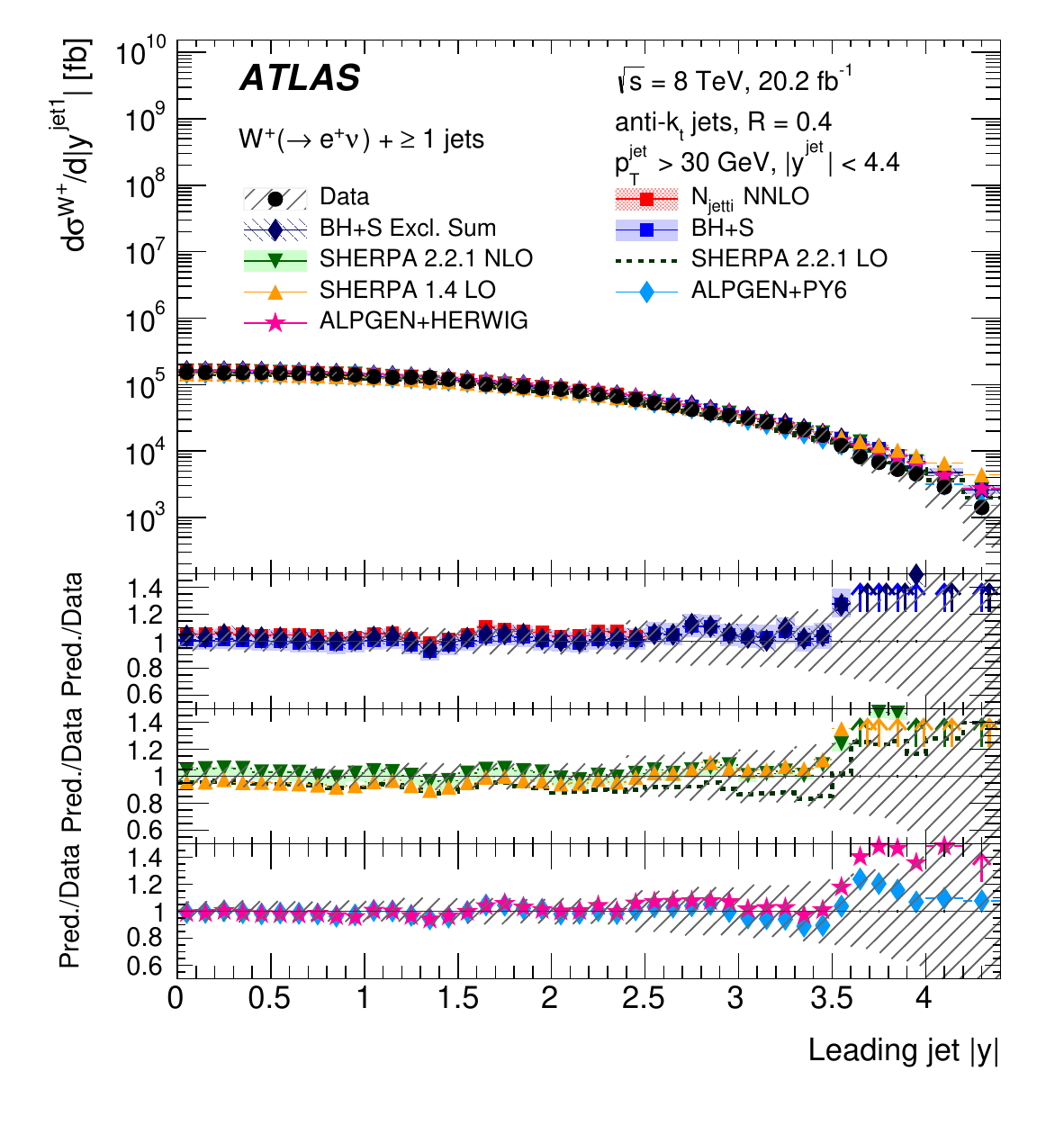} & 
\includegraphics[width=0.45\textwidth]{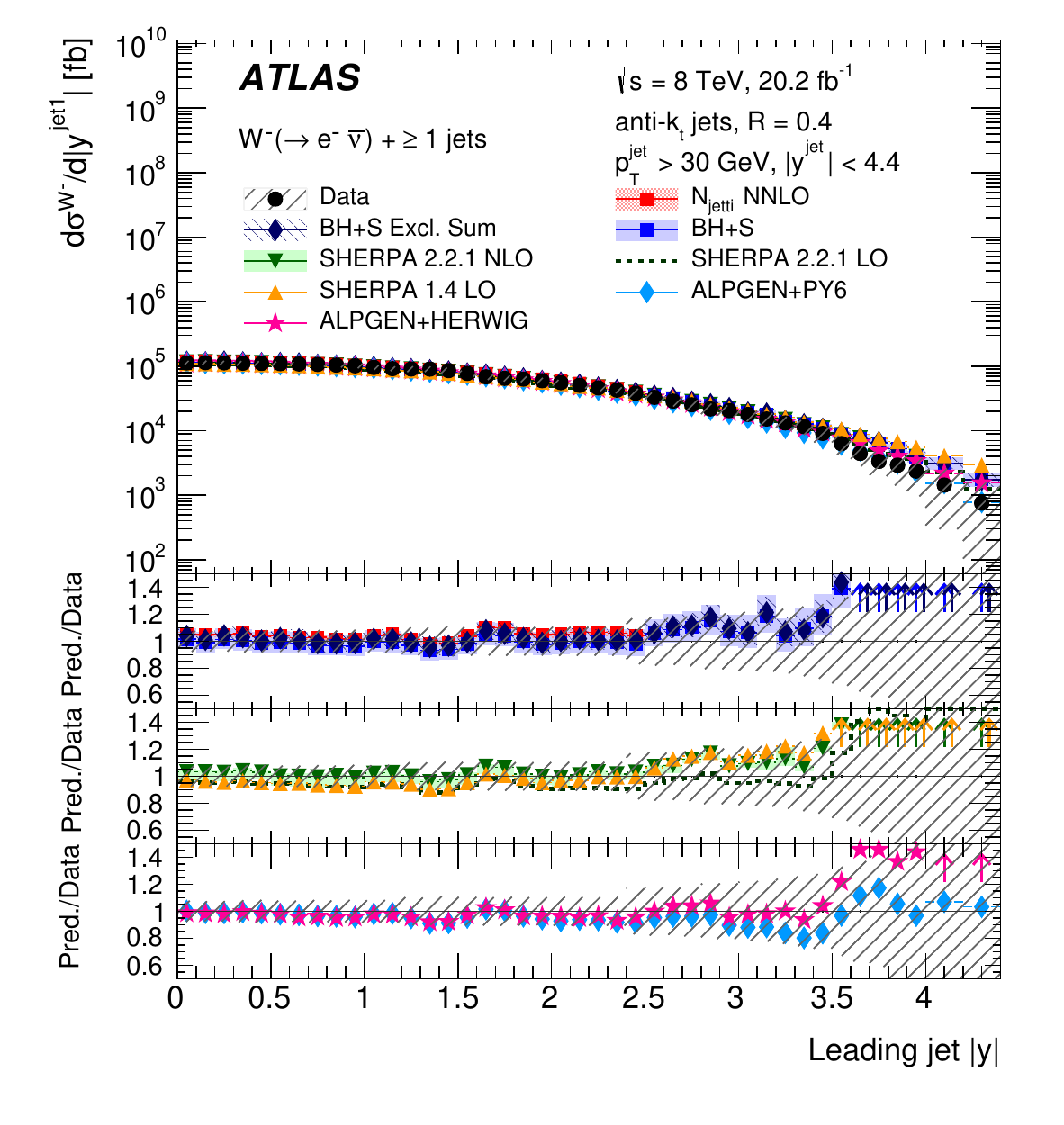} \\
\end{tabular}
\caption{Differential cross sections for the production of \Wplus\ (left) and \Wminus\ (right) as a function of the leading jet rapidity for events with $\njs \ge 1$. \captionCommon }
\label{fig:res-leadjetrap-1jet-ch}
\end{figure*}

% W+, W- with MCFM

\begin{figure*} \centering
\begin{tabular}{cc}
\includegraphics[width=0.45\textwidth]{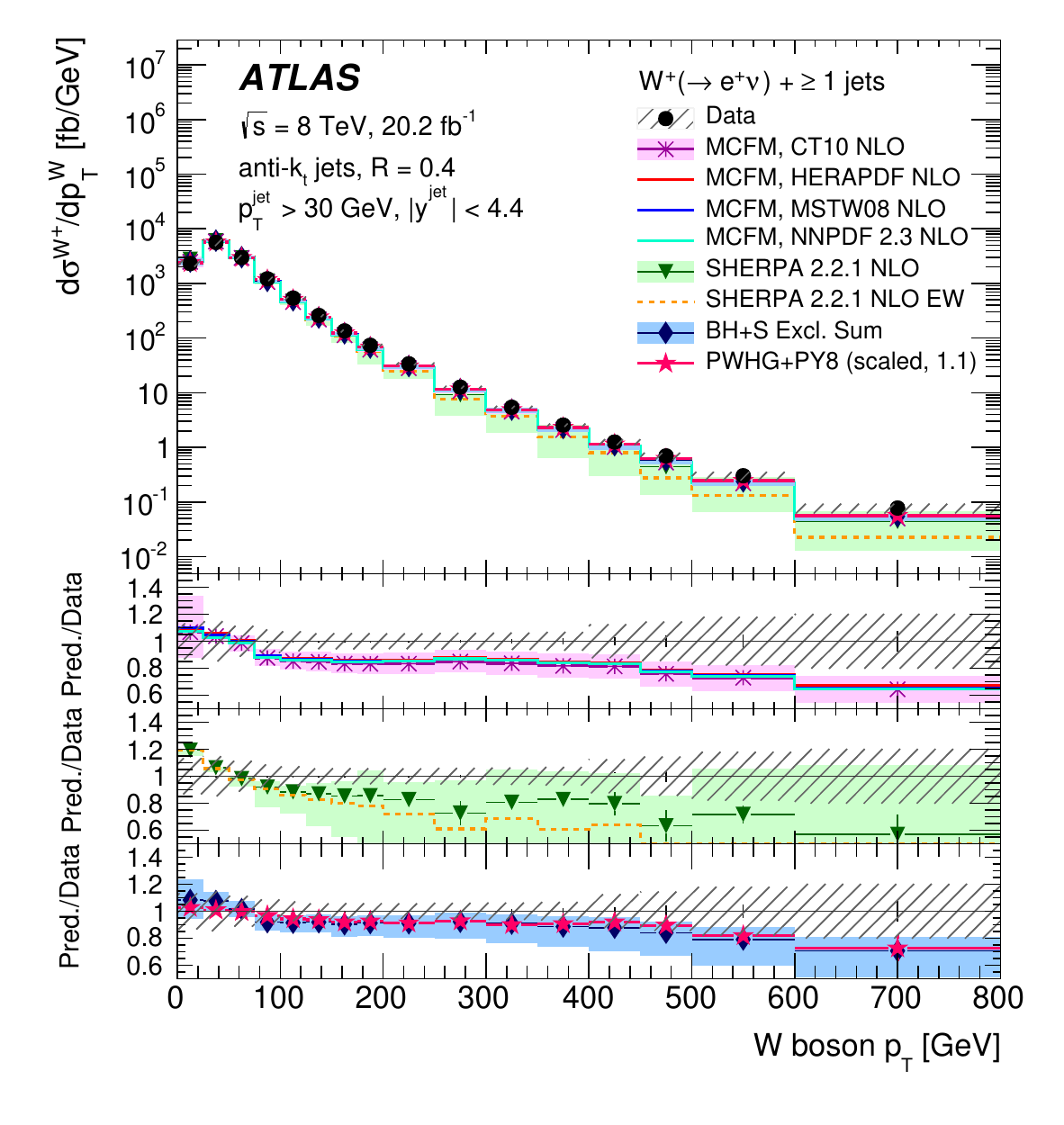} & 
\includegraphics[width=0.45\textwidth]{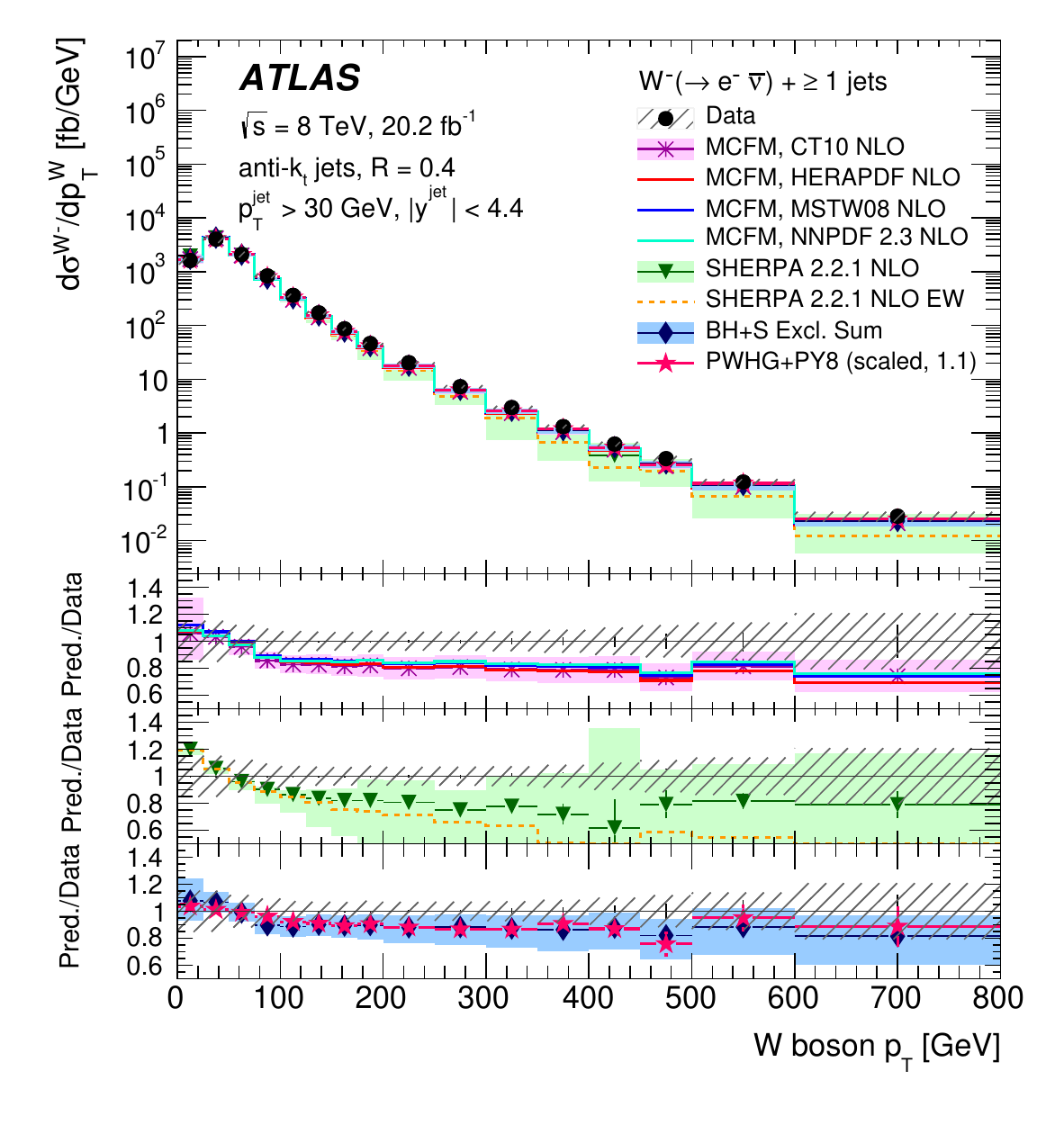} \\
\end{tabular}
\caption{Differential cross sections for the production of \Wplus\ (left) and \Wminus\ (right) as a function of the \Wboson\pt for events with $\njs \ge 1$. \captionCommon }
\label{fig:res-app-wpt-1jet-ch}
\end{figure*}

\begin{figure*} \centering
\begin{tabular}{cc}
\includegraphics[width=0.45\textwidth]{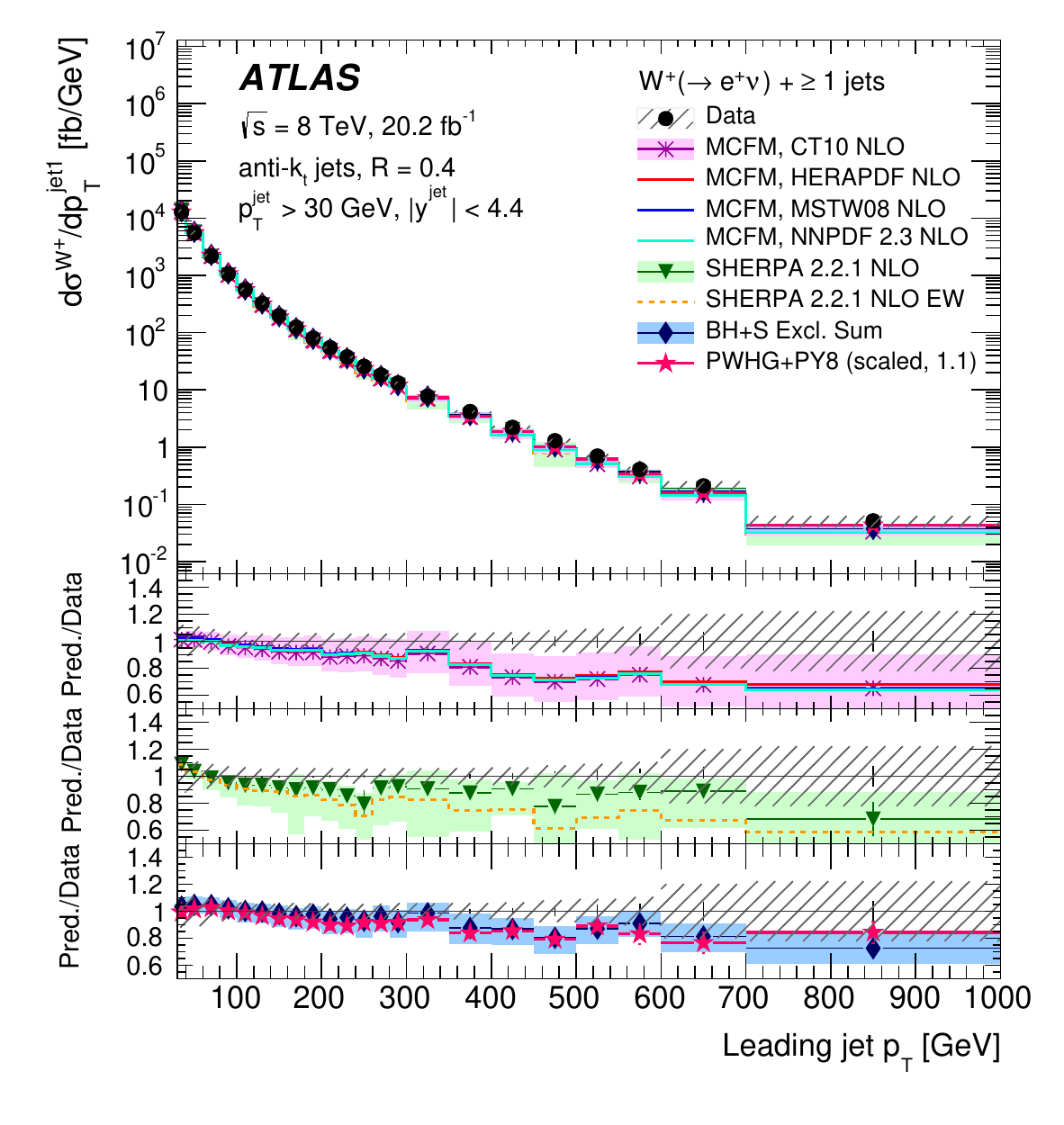} & 
\includegraphics[width=0.45\textwidth]{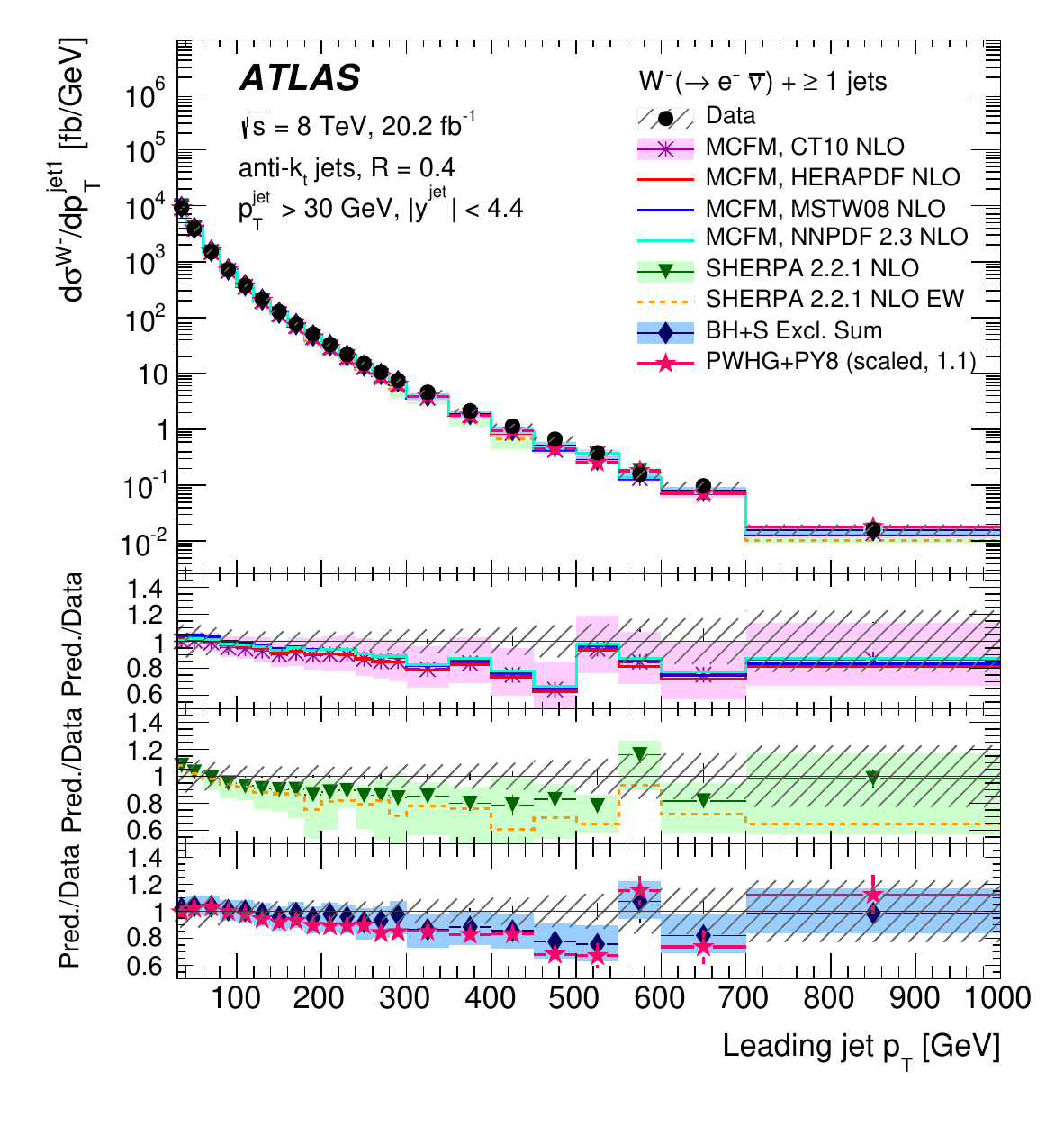} \\
\end{tabular}
\caption{Differential cross sections for the production of \Wplus\ (left) and \Wminus\ (right) as a function of the leading jet \pt for events with $\njs \ge 1$. \captionCommon }
\label{fig:res-app-leadjetpt-1jet-ch}
\end{figure*}

\begin{figure*} \centering
\begin{tabular}{cc}
\includegraphics[width=0.45\textwidth]{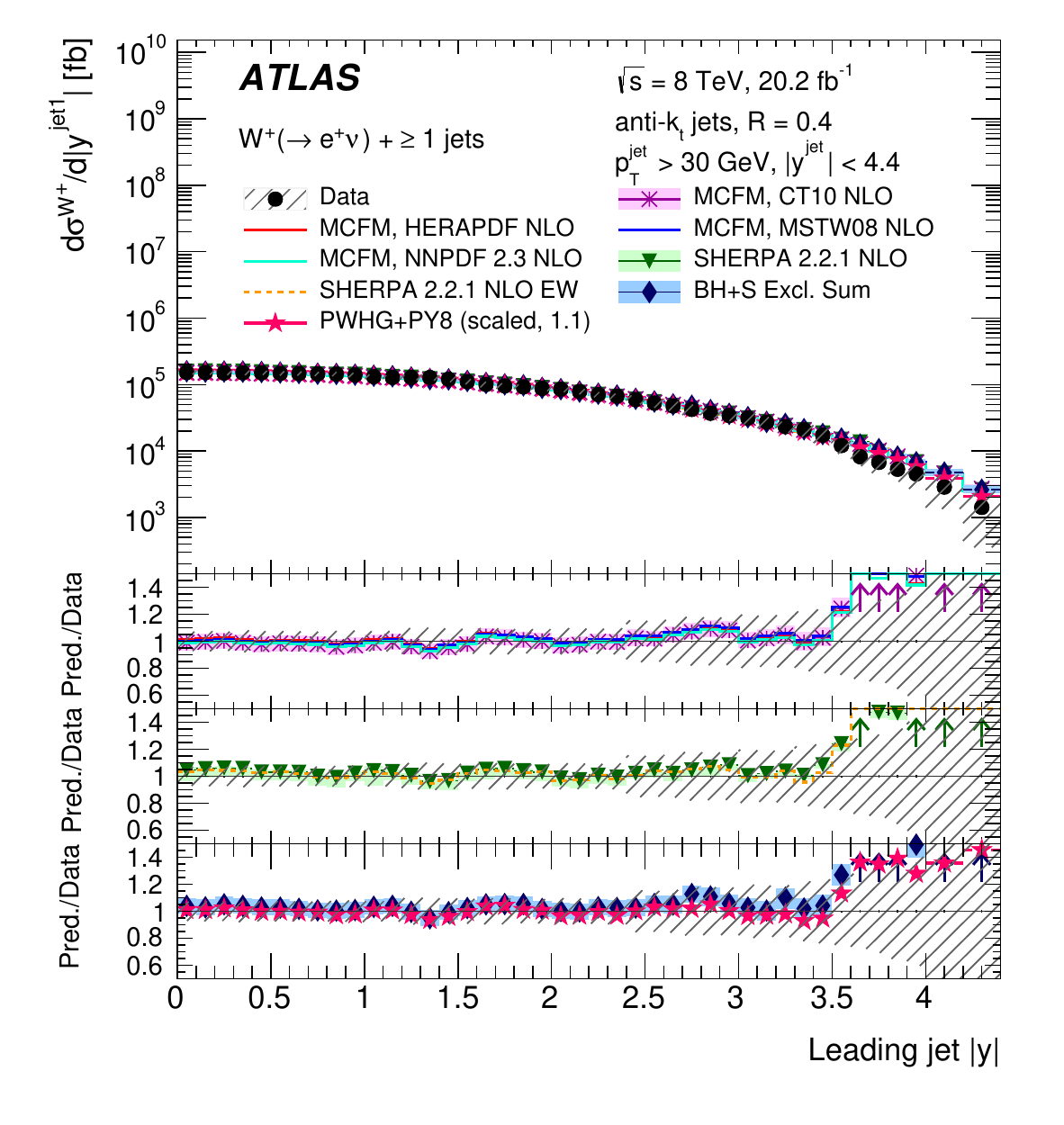} & 
\includegraphics[width=0.45\textwidth]{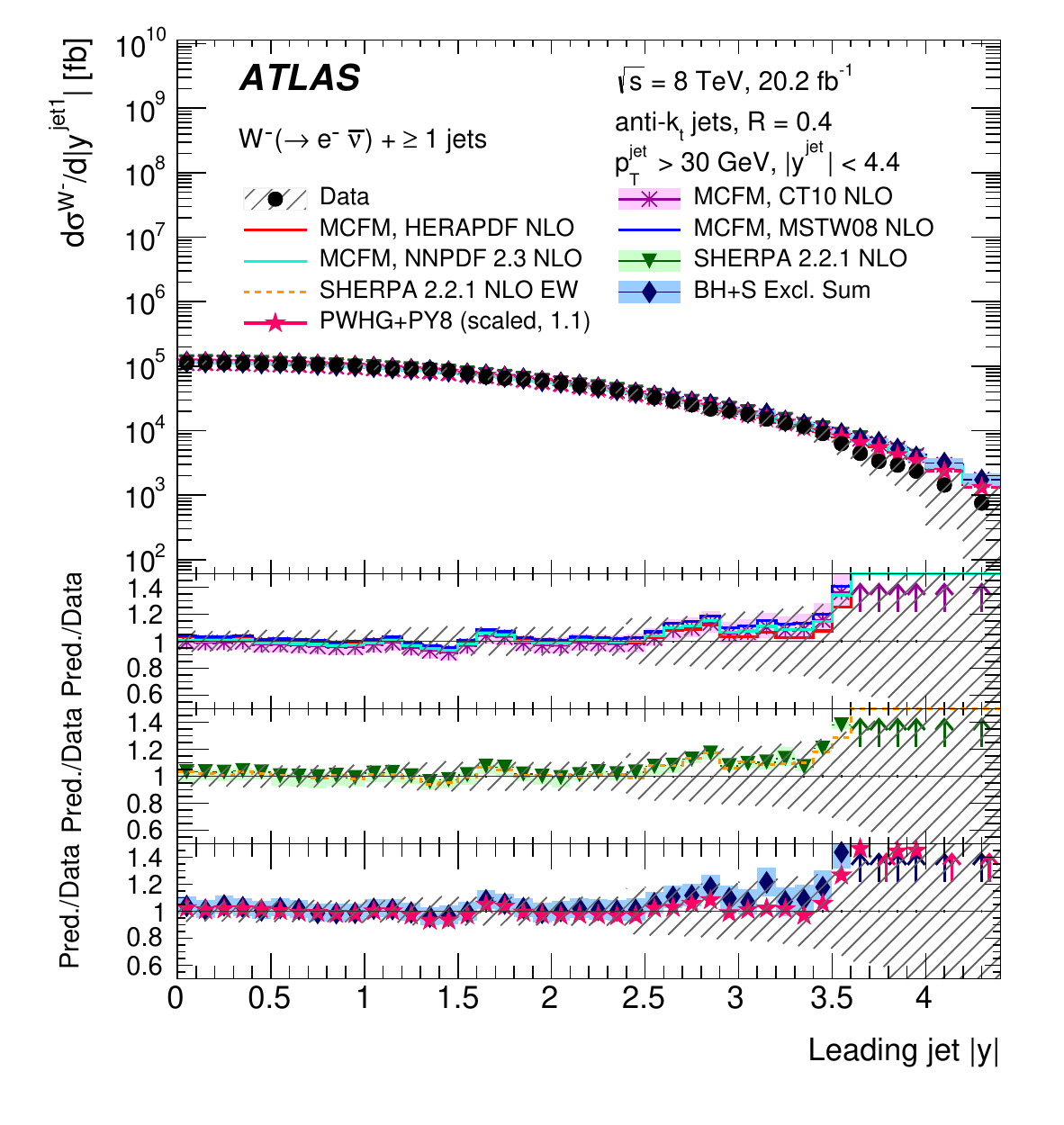} \\
\end{tabular}
\caption{Differential cross sections for the production of \Wplus\ (left) and \Wminus\ (right) as a function of the leading jet rapidity for events with $\njs \ge 1$. \captionCommon }
\label{fig:res-app-leadjetrap-1jet-ch}
\end{figure*}

% W+, W- for appendix body plots

\begin{figure*} \centering
\begin{tabular}{cc}
\includegraphics[width=0.45\textwidth]{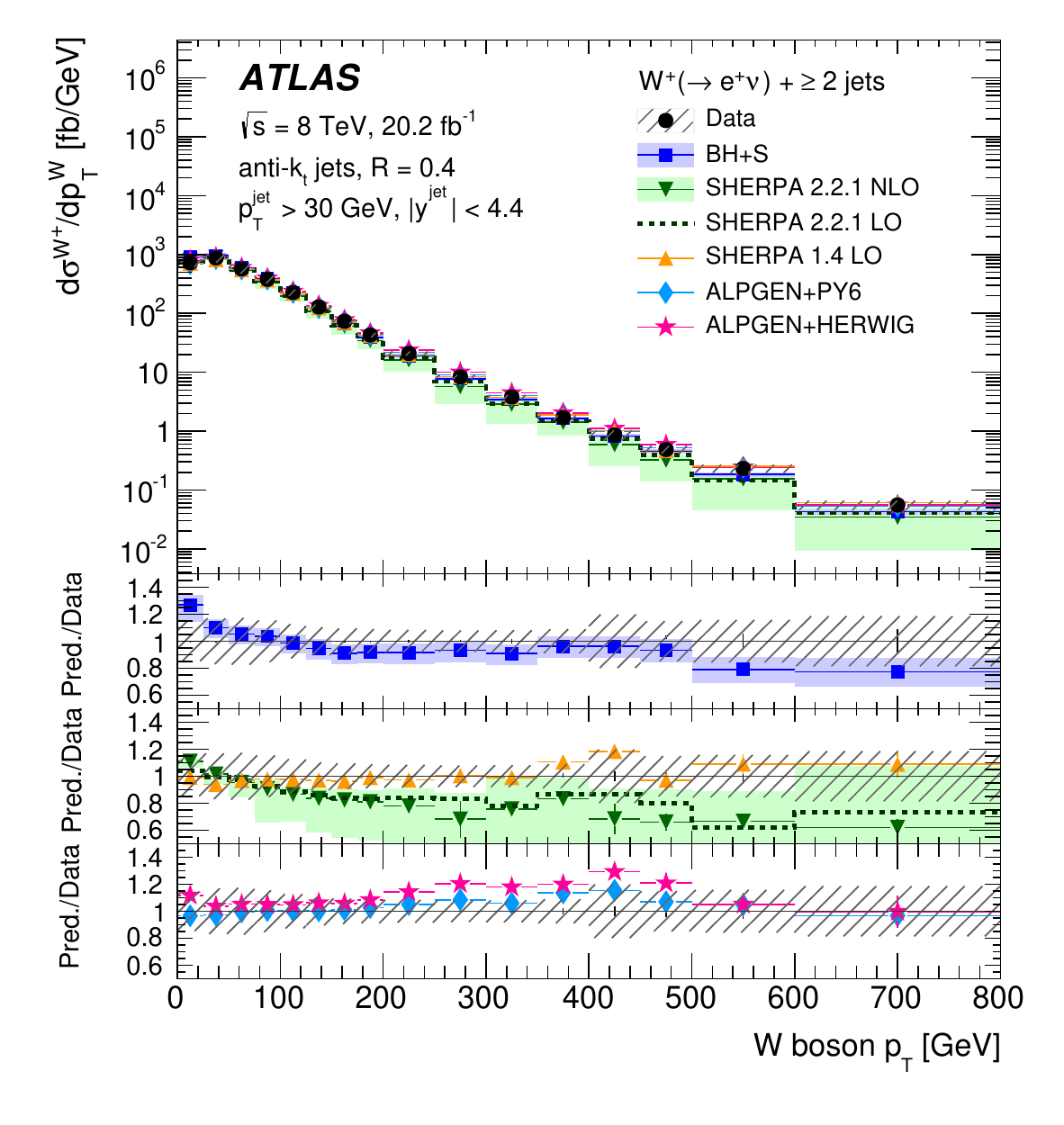} & 
\includegraphics[width=0.45\textwidth]{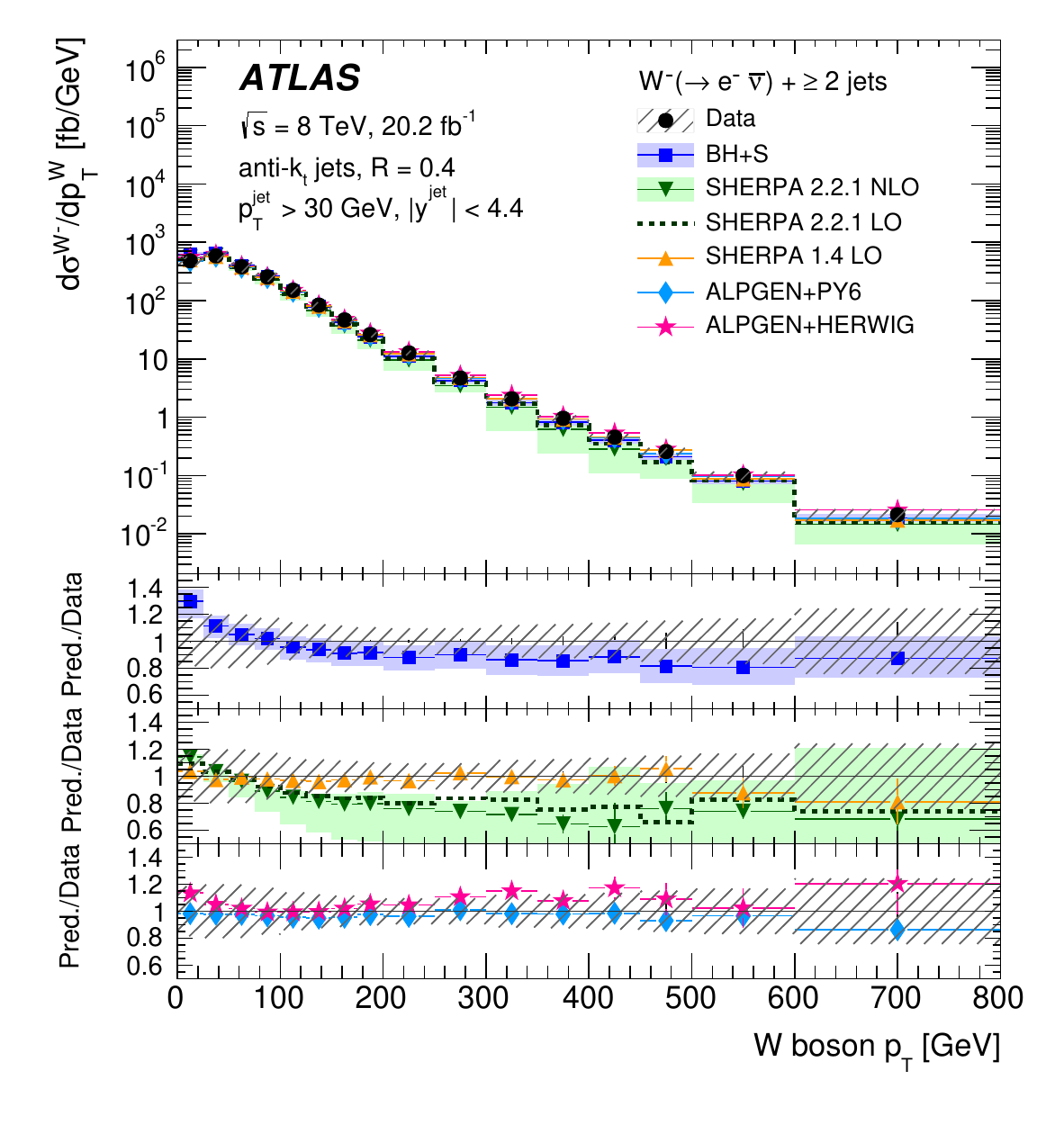} \\
\end{tabular}
\caption{Differential cross sections for the production of \Wplus\ (left) and \Wminus\ (right) as a function of the \Wboson \pt for events with $\njs \ge 2$. \captionCommon}
\label{fig:res-wpt-2jet-ch}
\end{figure*}

\begin{figure*} \centering
\begin{tabular}{cc}
\includegraphics[width=0.45\textwidth]{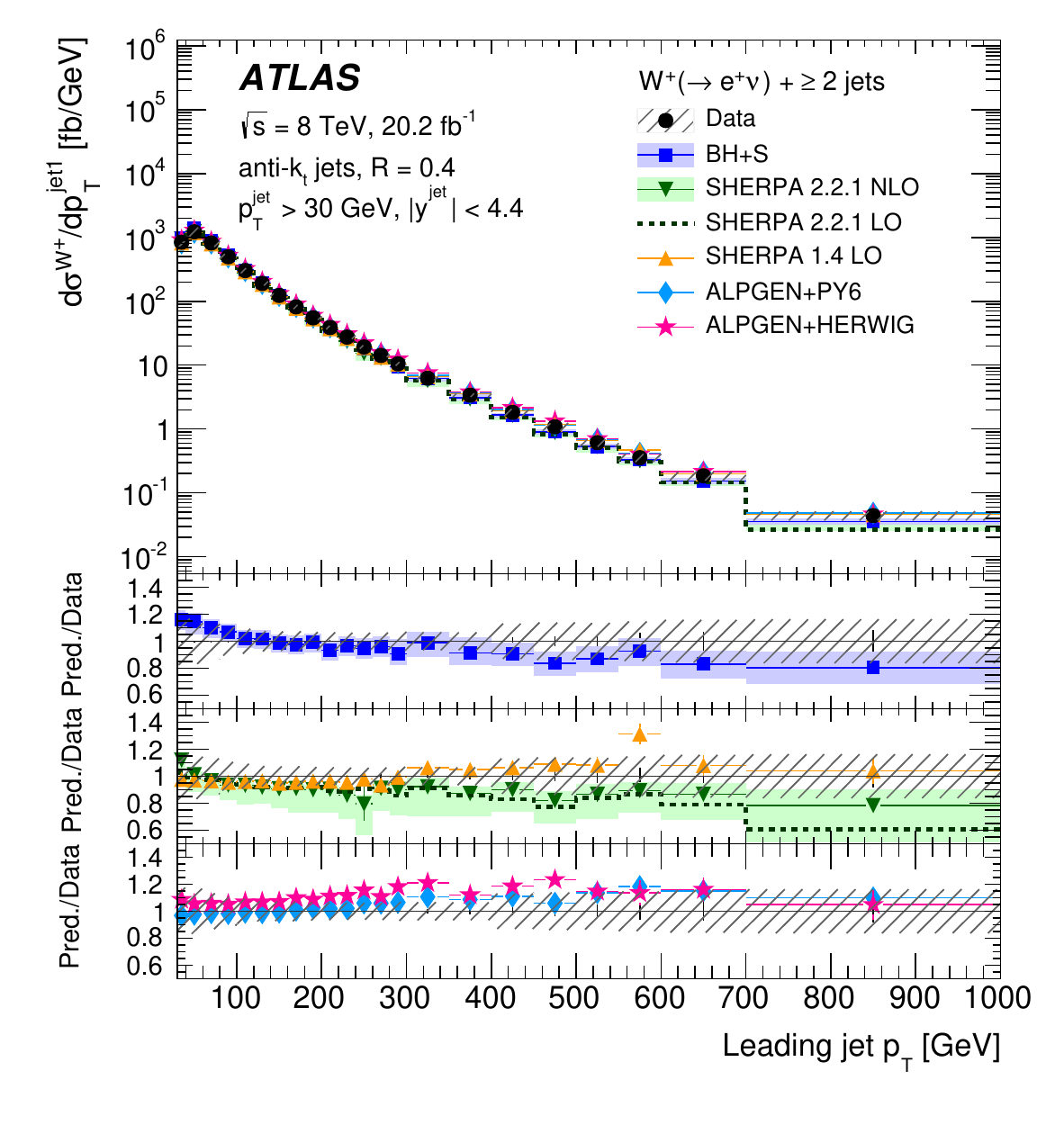} & 
\includegraphics[width=0.45\textwidth]{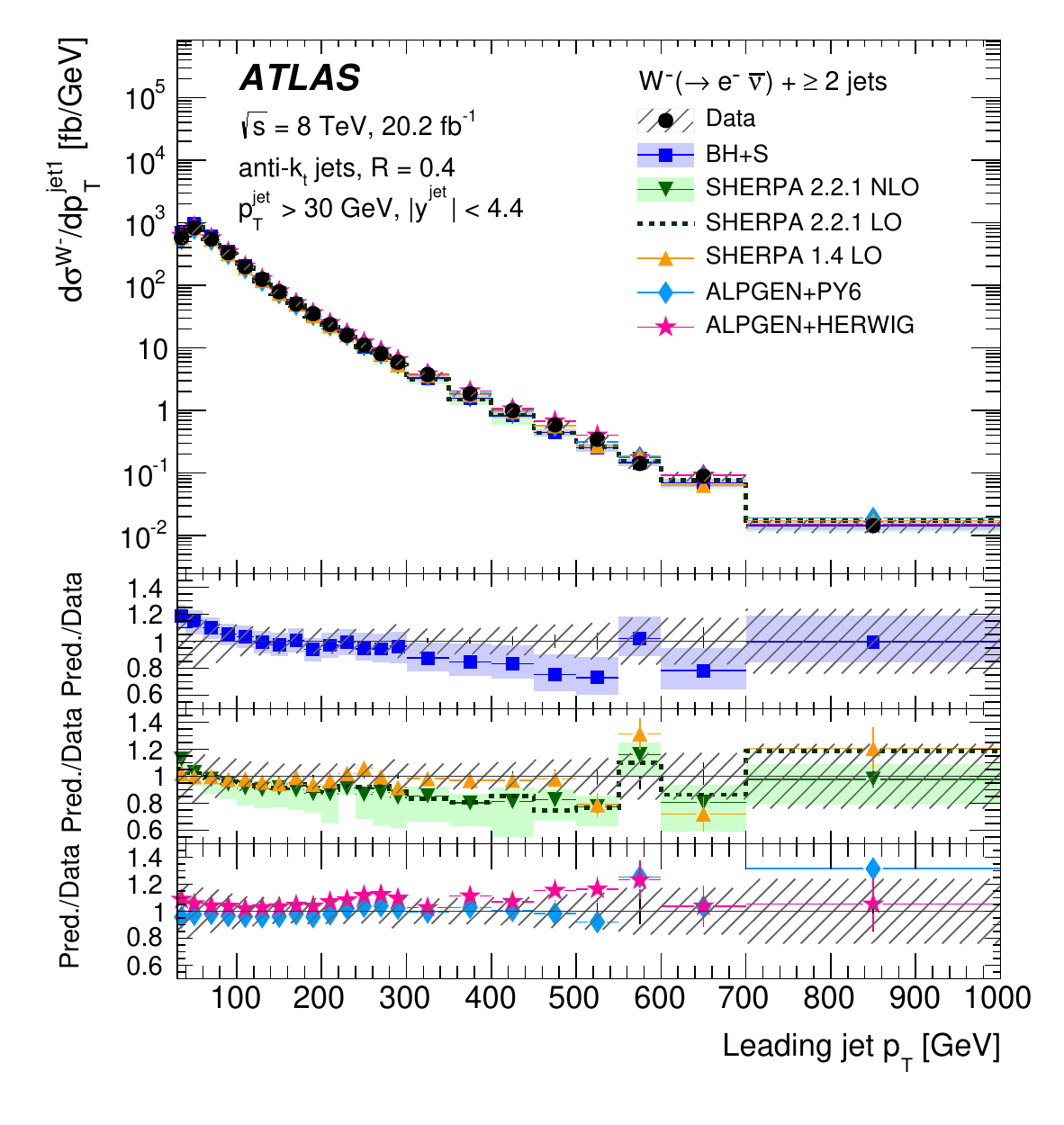} \\
\end{tabular}
\caption{Differential cross sections for the production of \Wplus\ (left) and \Wminus\ (right) as a function of the leading jet \pt for events with $\njs \ge 2$. \captionCommon }
\label{fig:res-leadjetpt-2jet-ch}
\end{figure*}

\FloatBarrier
\clearpage

%-------------------------------------------------------------------------------
% If you use biblatex and either biber or bibtex to process the bibliography
% just say \printbibliography here
\printbibliography
% If you want to use the traditional BibTeX you need to use the syntax below.
%\bibliographystyle{bib/bst/atlasBibStyleWoTitle}
%\bibliography{wxsec_2012,bibtex/bib/ATLAS}
%-------------------------------------------------------------------------------

%-------------------------------------------------------------------------------
% Print the list of contributors to the analysis
% The argument gives the fraction of the text width used for the names
%-------------------------------------------------------------------------------
%\clearpage
%\PrintAtlasContribute{0.30}
\clearpage

\newpage
% ATLAS Collaboration author list
% Data extracted on 19-Aug-2017 for paper reference STDM-2016-14
%\documentclass[11pt]{article}
%\usepackage{a4wide}%\begin{document}
\begin{flushleft}
{\Large The ATLAS Collaboration}

\bigskip

M.~Aaboud$^\textrm{\scriptsize 137d}$,
G.~Aad$^\textrm{\scriptsize 88}$,
B.~Abbott$^\textrm{\scriptsize 115}$,
O.~Abdinov$^\textrm{\scriptsize 12}$$^{,*}$,
B.~Abeloos$^\textrm{\scriptsize 119}$,
S.H.~Abidi$^\textrm{\scriptsize 161}$,
O.S.~AbouZeid$^\textrm{\scriptsize 139}$,
N.L.~Abraham$^\textrm{\scriptsize 151}$,
H.~Abramowicz$^\textrm{\scriptsize 155}$,
H.~Abreu$^\textrm{\scriptsize 154}$
% ...
\bigskip
\\
% ...
$^{12}$ Institute of Physics, Azerbaijan Academy of Sciences, Baku, Azerbaijan\\
% ...
$^{88}$ CPPM, Aix-Marseille Universit{\'e} and CNRS/IN2P3, Marseille, France\\
% ...
$^{115}$ Homer L. Dodge Department of Physics and Astronomy, University of Oklahoma, Norman OK, United States of America\\
% ...
$^{119}$ LAL, Univ. Paris-Sud, CNRS/IN2P3, Universit{\'e} Paris-Saclay, Orsay, France\\
% ...
$^{137}$ $^{(a)}$ Facult{\'e} des Sciences Ain Chock, R{\'e}seau Universitaire de Physique des Hautes Energies - Universit{\'e} Hassan II, Casablanca; $^{(b)}$ Centre National de l'Energie des Sciences Techniques Nucleaires, Rabat; $^{(c)}$ Facult{\'e} des Sciences Semlalia, Universit{\'e} Cadi Ayyad, LPHEA-Marrakech; $^{(d)}$ Facult{\'e} des Sciences, Universit{\'e} Mohamed Premier and LPTPM, Oujda; $^{(e)}$ Facult{\'e} des sciences, Universit{\'e} Mohammed V, Rabat, Morocco\\
% ...
$^{139}$ Santa Cruz Institute for Particle Physics, University of California Santa Cruz, Santa Cruz CA, United States of America\\
% ...
$^{151}$ Department of Physics and Astronomy, University of Sussex, Brighton, United Kingdom\\
% ...
$^{154}$ Department of Physics, Technion: Israel Institute of Technology, Haifa, Israel\\
$^{155}$ Raymond and Beverly Sackler School of Physics and Astronomy, Tel Aviv University, Tel Aviv, Israel\\
% ...
$^{161}$ Department of Physics, University of Toronto, Toronto ON, Canada\\
% ...
$^{a}$ Also at Department of Physics, King's College London, London, United Kingdom\\
$^{b}$ Also at Institute of Physics, Azerbaijan Academy of Sciences, Baku, Azerbaijan\\
$^{c}$ Also at Novosibirsk State University, Novosibirsk, Russia\\
$^{d}$ Also at TRIUMF, Vancouver BC, Canada\\
$^{e}$ Also at Department of Physics {\&} Astronomy, University of Louisville, Louisville, KY, United States of America\\
% ...
$^{*}$ Deceased
\end{flushleft}

%\end{document}
% Created with xml2latex.py

%-------------------------------------------------------------------------------
%\appendix
%\input{AuxMaterial}
%\part*{Auxiliary material}
%\addcontentsline{toc}{part}{Auxiliary material}
%-------------------------------------------------------------------------------

%In an ATLAS paper, auxiliary plots and tables that are supposed to be made public 
%should be collected in an appendix that has the title \enquote{Auxiliary material}.
%This appendix should be printed after the Bibliography.
%At the end of the paper approval procedure, this information can be split into a separate document
%-- see \texttt{atlas-auxmat.tex}.

%In an ATLAS note, use the appendices to include all the technical details of your work
%that are relevant for the ATLAS Collaboration only (e.g.\ dataset details, software release used).
%This information should be printed after the Bibliography.

\end{document}